\documentclass[a4paper,11pt]{article}

\usepackage{jheppub} 

\usepackage[T1]{fontenc} 
\usepackage{empheq,mathtools}
\usepackage{cancel,slashed}
\usepackage{braket}

\usepackage{subfigure}
\usepackage{graphicx}
\usepackage{epsfig}
\usepackage{amsfonts}
\usepackage{amssymb}
\usepackage{amsmath,amssymb}
\usepackage{enumerate} 
\usepackage{xcolor}
\usepackage[normalem]{ulem}

\newcommand{\be}{\begin{equation}}
\newcommand{\ee}{\end{equation}}
\newcommand{\bea}{\setlength\arraycolsep{2pt} \begin{eqnarray}}
\newcommand{\eea}{\end{eqnarray}}
\newcommand{\nn}{\nonumber}

\def\ft#1#2{{\textstyle{\frac{\scriptstyle #1}{\scriptstyle #2} } }}

\def\arXiv#1{\href{http://arxiv.org/abs/#1}{arXiv:#1}}
\def\arXiv#1#2{\href{http://arxiv.org/abs/#1}{arXiv:#1}}

\def\doi#1#2{\href{http://doi.org/#1}{#2}}

\title{\center Einstein-scalar field solutions in
AdS spacetime:
clouds, boundary conditions, and scalar multipoles }

\author{Dumitru Astefanesei$^1$, Hyat Huang$^{2,3}$, Jutta Kunz$^2$
and Eugen Radu$^4$} 
\affiliation{ $^1$ Instituto de F\'isica, Pontificia Universidad Cat\'olica de Valpara\'iso, 
Av. Brasil 2950, Valpara\'iso, Chile
\\
 $^2$  Institute of Physics, University of Oldenburg, Postfach 2503, D-26111 Oldenburg,
Germany
\\ 
$^3$College of Physics and Communication Electronics, 
Jiangxi Normal University, Nanchang 330022, China
\\
$^4$  Centre for Research and Development  in Mathematics and Applications (CIDMA),
\\
Campus de Santiago, 3810-183 Aveiro, Portugal
}

\emailAdd{dumitru.astefanesei@pucv.cl}
\emailAdd{hyat@mail.bnu.edu.cn}
\emailAdd{jutta.kunz@uni-oldenburg.de}
\emailAdd{eugen.radu@ua.pt}

\abstract{We consider an Einstein-scalar field model  which is a consistent truncation 
of
${\cal N}=8$ $D=4$
gauged supergravity, the scalar field possessing
a potential which is unbounded from below and
a tachyonic mass above the Breitenlohner-Freedman bound. 
We investigate the spherically symmetric
asymptotically anti-de Sitter soliton and black hole 
solutions, with the aim of clarifying the 
asymptotics and the possible boundary conditions at infinity. 
The emerging picture is contrasted with that found
for an Einstein-scalar field model with the same
scalar mass and a quartic self-interaction term.
We also provide arguments for the existence of 
solitonic solutions 
which can be viewed as non-linear continuation
of the (probe) scalar multipolar clouds,
with emphasis on the dipole case. 
Apart from numerical results, 
 exact  solutions are found  
for solitons with a monopole and dipole scalar field,
 as perturbations around the AdS background.}

\begin{document} 
\maketitle
\flushbottom

\section{Introduction} 

The study of scalar fields in AdS spacetime can be traced back at least to the work \cite{Salam:1977hk}, \cite{Avis:1977yn}, where the massive Klein-Gordon equation has been solved in an AdS$_4$ background. More recently, this subject has been of particular interest  mainly due to the AdS/CFT duality \cite{Maldacena:1997re},  which asserts that a consistent theory of quantum gravity in $D$-dimensions has an equivalent formulation in terms
of a non-gravitational theory in $(D-1)$-dimensions.

One well understood limit of the duality is when, in the AdS bulk, it is sufficient to consider the low energy limit of the superstring theory, namely, supergravity 
(for a review, see Ref.~\cite{Aharony:1999ti}). The supergravity models usually 
contain tachyonic scalar fields and, in some cases, there exist consistent truncations such that the matter content consists only of scalar fields.
 The solutions of these models  are particularly interesting
 due to non-trivial boundary conditions
satisfied by
  the scalar field(s) \cite{Hertog:2004ns}
	that are relevant for the dual theory \cite{Witten:2001ua}. 
For example, the scalar fields can break the conformal symmetry on the boundary 
\cite{Henneaux:2006hk},
and exact hairy black hole (BH) solutions with this 
 property were presented in \cite{Lu:2013ura}. 
Also, the lowest energy solitonic solutions can be considered 
 the true ground state of the theory \cite{Hertog:2004ns}, \cite{Hertog:2005hm},
while the study of the
thermodynamic properties of AdS  BHs
offers the possibility to better understand 
the non-perturbative aspects of certain dual field theories.

\medskip

Of interest in this context is the ${\cal N}=8$ $D=4$ gauged supergravity
model that can be obtained as a compactification of $D = 11$ supergravity on $S^7$ \cite{deWit:1986oxb}.  As discussed in \cite{Duff:1999gh}
(see  also 
 \cite{Cvetic:1999xp} and \cite{Gibbons:2005vp}),
 this model possesses a consistent  truncation with  
 three real scalar fields of equal mass, coupled to four U(1) fields, which 
can be set to zero.
The scalar fields can also vanish, a case which results
in Einstein gravity with a negative cosmological constant $\Lambda=-3/L^2$
(with $L$ the AdS radius). The main solution of interest is
the AdS$_4$ spacetime in global coordinates
\begin{eqnarray}
\label{ads}
ds^2 = 
-N(r) dt^2+ \frac{dr^2}{N(r)}
+r^2(d\theta^2+\sin^2 \theta d\varphi^2),
 ~~{\rm where}~~N(r)=1+\frac{r^2}{L^2} ,
 \end{eqnarray}  
 with $r,t$
the radial and time coordinate, respectively,
while
$\theta,\varphi$
are the usual coordinates on $S^2$. 
%
Otherwise, as discussed in Appendix A, 
we can (consistently) 
take only one scalar field to be nonzero ($n=1$),
or two of them  ($n=2$) (in which case, they are equal);
finally, all scalars can be taken nonzero and equal, $n=3$,
the resulting Einstein-(single, real) scalar field model being presented in 
Section 2.
For any nonzero $n$,
the scalar field possess a tachyonic mass $\mu^2=-2/L^2$.

\medskip

In this work, we are interested in static and localized solutions 
of the considered Einstein-scalar field model, which are also regular and have  a
finite mass. They may possess an event horizon of spherical topology 
(and then correspond to  BHs with scalar hair) or just correspond to solitonic deformations of the globally AdS spacetime (\ref{ads}).
In both cases, 
the generic expression of the (static) 
scalar field as $r\to \infty$
(which is found by considering the linearized Klein-Gordon equation 
	in a fixed AdS background)
is the sum of two modes 
\begin{eqnarray}
\label{inf}
\phi=\frac{\alpha (\theta,\varphi )}{r}+\frac{\beta (\theta,\varphi )}{r^2}+\dots,
 \end{eqnarray} 
$\alpha$ and $\beta$ being two real functions.
For a well defined theory, one has to specify a 
boundary condition on $\alpha,\beta$, 
the natural choice corresponding  to either 
$\alpha=0$ or $\beta=0$. 
However, as shown in \cite{Hertog:2004ns}
\cite{Henneaux:2002wm},
\cite{Henneaux:2004zi},
\cite{Hertog:2004dr},
one may consider a larger class of mixed boundary conditions,
with nonzero $\alpha$ and $\beta$, 
for which the conserved  global  charges are still well
defined and finite.
 The boundary conditions in this case are
defined by an essentially arbitrary function $W$  connecting
$\alpha$ and $\beta$,
with
\begin{eqnarray} 
\label{W}
\beta=\frac{d W(\alpha)}{d\alpha}~.
 \end{eqnarray} 
  Since their properties depend significantly on the choice of 
	${  W}$, this type of models have been called {\it designer gravity theories} \cite{Hertog:2004ns}.

\medskip

The expressions of $\alpha$ and $\beta$ are not determined a priori,
various boundary conditions being possible, that lead to different properties of the solutions. 
Also, $\alpha$ and $\beta$ in
the large-$r$ expansion
 (\ref{inf})
are not arbitrary, 
being determined by the imposed data at the origin
(for solitons),
 or at the horizon (for BHs).
For example,  a priori one expects
the existence of solutions with $\alpha=0$ or $\beta=0$.
However, to our best of knowledge, 
no systematic study of this aspect has been presented in the literature.
For $n=1$, we mention the study in 
Refs.~\cite{Hertog:2004dr},
	\cite{Hertog:2004rz},
	\cite{Hertog:2004bb},
	where spherically symmetric solutions with $\beta=f \alpha^2$
	are studied (with $f$ a negative constant). 
 
The first goal of the present work is
to consider 
a systematic study of 
the spherically symmetric solitonic and BH
solutions of the considered consistent truncation
of the ${\cal N}=8$ $D=4$ model.
The aim is to clarify the  dependence
of the data at infinity
on the data at the origin/horizon,
\begin{eqnarray}
\alpha \equiv \alpha(r_{0},\phi(r_{0})),~~
\beta \equiv \beta(r_{0},\phi(r_{0})),~~M_0 \equiv M_0(r_{0},\phi(r_{0})),
\end{eqnarray}
(with $r_0=0$ or $r_h$ for solitons and BHs,  
while $M_0$ is an extra-constant
which enters the far field expression of the metric), 
without imposing any  relation between $\alpha$ and $\beta$.\footnote{A similar analysis in five dimensions, but with a different goal, was presented in \cite{Anabalon:2017eri}.}
 The main results can be summarized as follows.
Firstly, both perturbative and non-perturbative solutions are considered.
The perturbative results are found for solitons,
the perturbation parameter being $\phi(0)$, the value of the scalar field 
at the origin.
The non-perturbative solutions are found by solving numerically
the field equations,
in which case we aim  for a systematic scan of the parameter space
for a large range of $r_{h},\phi(r_{h})$.
%
Secondly, our results provide strong evidence for the 
$absence$ of (soliton or BH) solutions satisfying 
the `standard' conditions
$\alpha=0$ or $\beta=0$.
That is, the ${\cal N}=8$ $D=4$ model possesses {\it designer gravity}
solutions only.

A natural question which arises in this context
 concerns the generality of these results.
In particular, is it possible to find solutions
with $\alpha=0$ or $\beta=0$ in (\ref{inf})
 for a different choice 
of the scalar field potential?
To address this aspect, 
 we study also solutions of a
model in which the scalar field still possesses
the same mass as in the ${\cal N}=8$ case;
however, the self-interaction is given by a quartic term.
As a result, 
the asymptotic behaviour of the scalar field is less constrained
and one finds $e.g.$ spherically symmetric solutions with 
$\alpha=0$
or 
$\beta=0$
in the expansion (\ref{inf}).

\medskip
Another goal of this work is
motivated by the observation that, 
to our best knowledge,
 all 
Einstein-(real) scalar field 
solutions reported in the literature
correspond to spherically symmetric configurations.
However, the (static)
solution of the linearized Klein-Gordon equation
in a fixed AdS background
possesses a general solution,
the scalar field
being a superposition of  modes,
 $\phi= \sum_{\ell m}Y_{\ell m}(\theta,\varphi)R_{\ell}(r)$ 
(with $Y_{\ell m}$   the real spherical harmonics
and  $R_{\ell}(r)$ the radial amplitude).
From this perspective, no value of $(\ell,m)$
is  privileged, 
with the same asymptotic decay (\ref{inf})
for all modes (see, also, \cite{Horowitz:2000fm}). 
Also, since
the spherically symmetric Einstein-scalar field solitons can be viewed
as a non-linear continuation of the $\ell=m=0$ mode,
 one expects similar results to exist for higher modes.
In this work we consider mainly the simplest case $\ell=1,m=0$ 
and we provide evidence that the 
qualitative picture found in the spherically symmetric case
still holds.
First, one finds again
an exact, perturbative solitonic solution,
which is interpreted as a deformation of AdS spacetime 
with a dipolar scalar field.
Nonperturbative solutions
with a $1/r^2$-decay at infinity of the scalar field
 are shown to exist in a
model $\phi^4$-selfinteraction.

\medskip

This paper is organized as follows.
In the next Section we present the general framework,
while in Section \ref{Probe} we consider the probe limit 
of the problem, 
with a study of  scalar clouds in a fixed (Schwarzschild-)AdS background.
The case of spherically symmetric configurations
is discussed in Section  \ref{spherical},
while the gravitating scalar dipoles
are studied in Section \ref{dipole}.
We conclude in Section \ref{final} with a discussion and some further remarks.
The Appendix A explains how
 the considered 
 sugra-action  is obtained
starting with the general results in Ref. \cite{Duff:1999gh}.
In the Appendix B, we provide some details on
the perturbative axially symmetric solutions, including 
the Einstein-scalar field
soliton with a quadrupole scalar field.

\section{The general framework}
\label{sec_eq_motion_action}

\subsection{The action, equations of motion
and scalar field potentials} 

We consider the Einstein-(real)scalar field model
with a negative cosmological constant
\begin{eqnarray}
\label{action}
I=\int_\mathcal{M}  d^4x \sqrt{-g}
                                 \left[
 \frac{1}{4\kappa^2}
\left( R+\frac{6}{L^2} \right)
   -\frac{1}{2} g^{ab}  \phi_{, \, a} \phi_{, \, b} - U(\phi) 
                                \right] 
	 -\frac{1}{2\kappa^2}\int_{\partial\mathcal{M}} d^3 x\sqrt{-h}K,
\end{eqnarray}
where $\kappa^2\equiv 4\pi G$ (with $G$ the Newton's constant) and  $U(\phi)$ is the scalar field potential.
Also, the last term in  (\ref{action}) is the Hawking-Gibbons
surface term \cite{Gibbons:1976ue}, where $K$ is the trace
of the extrinsic curvature for the boundary $\partial\mathcal{M}$
 and $h$ is the induced
metric of the boundary.

The corresponding Einstein-scalar field  field equations, 
as obtained from the variation of the action
(\ref{action})
 with respect to the metric and scalar field, respectively,
read:
\begin{eqnarray}
\label{eqs}
 && 
  R_{ab}-\frac{1}{2}g_{ab}R-\frac{3}{L^2}g_{ab}=2\kappa^2~T_{ab}  ,
~~~~~
\nabla^2 \phi=\frac{\partial U}{\partial \phi},
\end{eqnarray}  
where
$T_{ab}$ is the stress-energy tensor  of the scalar field,
\begin{eqnarray}
 T_{ab}=
 \phi_{, a} \phi_{,b} 
-g_{ab}  
\left(
\frac{1}{2} g^{cd}  \phi_{,c}\phi_{,d} 
 +U(\phi)
\right)  . 
 \end{eqnarray} 

\medskip 
In this work we shall consider two different expressions of the
scalar field potential $U(\phi)$.
The first case is 
of main interest,
occurring in
 a consistent trucation 
of the 
${\cal N}=8$
$D=4$
gauged
supergravity
model \cite{deWit:1986oxb},
with
\begin{equation} 
\label{U}
{\bf sugra}:~~~~
 U(\phi)=-  \frac{n}{\kappa^2 L^2}  
\sinh^2 \left ( \frac{\kappa}{\sqrt{n} }\phi \right) ,~~{\rm with}~~n=1,2,3.
\end{equation} 
The Appendix A presents
some details on how 
 the action (\ref{action}) with the above potential
can be obtained 
starting with the general results in Ref.~\cite{Duff:1999gh}.
 
The small-$\phi$ expansion of the scalar field potential (\ref{U}) is
\begin{eqnarray}
\label{small}
U(\phi) \sim -\frac{\phi^2}{L^2}-\frac{\kappa^2 \phi^4}{3 n L^2}+O(\phi^6),
 \end{eqnarray} 
Therefore, for any $n$,
 the scalar field possesses a tachyonic mass, 
with
\begin{eqnarray}
\label{mu}
\mu^2=\frac{d^2U}{d\phi^2}\bigg|_{\phi=0}=-\frac{2}{L^2},
 \end{eqnarray} 
which implies the 
asymptotic behaviour (\ref{inf}).

The second case considered in this work
corresponds to a massive scalar field
with  quartic self-interaction,
\begin{eqnarray}
\label{Uphi4}
\phi^4-{\bf model}:~~~~
U(\phi) =-\frac{\phi^2}{L^2} +\lambda \phi^4,
\end{eqnarray}
where $\lambda$ is an arbitrary constant.
Note that for the particular
value
$\lambda=-{\kappa^2}/({3n L^2})$,
the potential
(\ref{Uphi4})
 can be considered as a truncation of
the sugra-potential (\ref{U}).

\subsection{Mixed boundary conditions and holographic mass} 
\label{mass}
In this section we present a general discussion of possible boundary conditions 
for a scalar field with the tachyonic  mass (\ref{mu}), 
the interpretation within AdS/CFT duality, and a concrete method to obtain the holographic mass. We are going to follow closely
the Refs.~\cite{Marolf:2006nd} and \cite{Anabalon:2015xvl}, 
where  counterterms for the scalar fields were used 
to regularize the action and to obtain the holographic mass.
 These results can be directly applied to various examples considered in the next sections.

While in theories of gravity coupled to matter the theory is usually fully determined by the action, 
in the presence of scalar fields with tachyonic mass the situation is quite different.
That is, both modes in the scalar field's fall off (\ref{inf}) 
 are normalizable and represent physically acceptable fluctuations. 
Therefore, specifying boundary conditions for the scalar field is equivalent
to fixing the boundary data $\alpha$, $\beta$ or a specific relation between them.
 It is common to denote the mixed boundary conditions on the scalar field by 
$\beta \equiv W'(\alpha)$, where $W(\alpha)$ is an arbitrary differentiable function. 
This restriction on $\alpha$ and $\beta$ can be obtained from the vanishing 
symplectic flux flow through the boundary \cite{Amsel:2006uf}
 and it is 
interpreted as an integrability condition for the mass in the Hamiltonian formalism 
\cite{Hertog:2004ns}, \cite{Anabalon:2014fla}.

The existence of various boundary conditions for the scalar fields 
fits very well in the context of AdS/CFT duality where they are interpreted 
as multitrace deformations in the dual field theory \cite{Witten:2001ua}. 
The interpretation is as follows: if $\alpha$ is identified with the source for an operator in the dual field theory, $\mathcal{O}$, 
the dual field theory action should contain a term $\int \alpha(x) \mathcal{O}(x) d^3x$ and then $\beta$ is identified with the vacuum expectation value (VEV) of the operator, 
$\beta = \braket{\mathcal{O}}$ (and the other way around with $\beta$ the source and $\alpha$ the VEV). However, for the current work, the mixed boundary condition, $\beta = W'(\alpha) $, is the relevant one with the following interpretation: these general boundary conditions are multitrace deformations of the boundary CFT, of the form $\int  W[\mathcal{O}(x)] d^3x$. 
A generic deformation can break the conformal symmetry in the boundary, but since it is still invariant under global time translations, there  exists a conserved total mass/energy. 
As we are going to explicitly show below, the mixed boundary conditions that preserve the conformal symmetry can be obtained from the vanishing trace of the dual stress tensor and correspond to triple trace deformations.

\medskip

After this brief review of mixed boundary conditions 
for the scalar field and their interpretation within the AdS/CFT duality, 
let us obtain the holographic mass by using  the `counterterm method', 
that consists in adding suitable 
additional surface terms to regularize the action (\ref{action}). 
These counterterms are usually built up with
curvature invariants on the boundary $\partial \cal{M}$ (which is sent to
infinity after the integration);
as such,  they do not alter the bulk equations of motion.
In four spacetime dimensions, the following counterterms are sufficient to cancel
divergences
for (electro-)vacuum solutions with negative cosmological constant \cite{Henningson:1998gx}, \cite{Balasubramanian:1999re}, \cite{Skenderis:2000in}
\begin{eqnarray}
\label{ct}
I_{\rm ct}^{(0)}=-\frac{1}{2\kappa^2} \int_{\partial {\mathcal M}}d^{3}x\sqrt{-h}\Biggl[
\frac{2}{L}+\frac{L}{2}\cal{R}
\Bigg]\ ,
\end{eqnarray}
where  ${\cal R}$ is the Ricci scalar
of the boundary metric $h$.
Within this approach, the mass computation goes as follows.
First step consists in constructing a divergence-free boundary stress tensor ${\rm T}_{\mu \nu}$
from the total action
$I{=}I_{\rm bulk}{+}I_{\rm surf}{+}I_{\rm ct}^{(0)}$ 
by defining
\begin{eqnarray}
\label{s1}
{\rm T}_{\mu \nu}&=& \frac{2}{\sqrt{-h}} \frac{\delta I}{ \delta h^{\mu \nu}}
=\frac{1}{2\kappa^2 }
\left(
K_{\mu \nu}-Kh_{\mu \nu}-\frac{2}{L} h_{\mu \nu}+ L E_{\mu \nu}
\right),
\end{eqnarray}
where $E_{\mu \nu}$ is the Einstein tensor of the  boundary metric,
$K_{\mu \nu}=-1/2 (\nabla_\mu n_\nu+\nabla_\mu n_\nu)$ is the extrinsic curvature,
$n^\mu$ being an outward pointing normal vector to the boundary.
Here one supposes that the boundary geometry is foliated
by spacelike surfaces $\Sigma$ with metric
$\sigma_{ij}$
\begin{eqnarray}
\label{b-AdS}
h_{\mu \nu}dx^{\mu} dx^{\nu}=-N_{\Sigma}^2dt^2
+\sigma_{i j}(dx^i+N_{\sigma}^i dt) (dx^j+N_{\sigma}^j dt).
\end{eqnarray}
Then, 
if $\xi^{\mu}$ is a Killing vector generating an isometry of the boundary geometry,
there should be an associated conserved charge.
In this approach,
 $\rho=u^{a}u^{b}{\rm T}_{a b}$ 
is the proper energy density
while $u^{a}$ is a timelike unit vector normal to $\Sigma$.
Thus the conserved charge associated with
time translation $\partial /\partial t$ is the mass of the spacetime
\begin{eqnarray}
\label{mass-ct}
M=\int_{ \Sigma}d^{2}x\sqrt{\sigma}N_{\Sigma} \rho.
\end{eqnarray}

The presence of the scalar field in
the bulk action (\ref{action})
brings the potential danger of having divergent contributions
coming from both, the gravitational and matter actions 
\cite{Taylor-Robinson:2000xw}.
This is the case for a scalar field 
which behaves asymptotically as $O(1/r)$
($i.e.$ $\alpha\neq 0$ in (\ref{inf})),
and then
the counterterms  (\ref{ct}) will not yield 
a finite  mass.
However, 
it is still  possible to obtain a finite mass by
allowing the boundary counterterms
 to depend not only on
the boundary metric $h _{\mu \nu}$, but
also on the scalar field.
This means that the
quasilocal stress-energy tensor (\ref{s1}) also
acquires a contribution coming from the matter field. The counterterm that regularizes the action and has a valid variational principle compatible with 
the boundary condition $\beta \equiv W'(\alpha)$ is \cite{Anabalon:2015xvl}
\begin{equation}
\label{ct-M}
I_{\phi}^{ct} = -\int_{\partial M}d^{3}x\sqrt{-h}
\Big[
 \frac{1}{2 L}\phi^2
+\frac{1}{L}\frac{W(\alpha)}{\alpha ^3}\phi ^3
\Big]
\end{equation}
that
yields a supplementary contribution to the
 boundary stress tensor 
 (\ref{s1}),
\begin{eqnarray}
\label{Tsup}
{\rm T}_{\mu \nu}^{(\phi)}=-\frac{1}{L} h_{\mu \nu}
\left(
 \phi^2
+
\frac{W(\alpha)}{\alpha ^3}\phi ^3
\right ),
\end{eqnarray}
which should be taken into account 
when obtaining a finite mass, eq. (\ref{mass-ct}).

We also mention that
the background metric upon
which the dual field theory resides is
$\gamma_{\mu \nu}=\lim_{r \rightarrow \infty} \frac{L^2}{r^2}h_{\mu \nu}$
(with $r$ the radial coordinate).
Then, the expectation value of the dual CFT stress-tensor
can be calculated using the  relation \cite{Myers:1999qn}
\begin{eqnarray}
\label{r1}
\sqrt{-\gamma}\gamma^{ \mu  \lambda}<\tau_{\lambda \nu}>=
\lim_{r \rightarrow \infty} \sqrt{-h} h^{\mu  \lambda}{\rm T}_{\lambda  \nu},
\end{eqnarray}
with 
the trace
\cite{Anabalon:2015xvl}
\begin{equation}
<\tau_{\nu}^\nu> = -\frac{3}{L^4}\Big(W-\frac{\alpha\beta}{3}\Big)  .
\end{equation}
It follows that
the only mixed boundary condition, which preserves the conformal symmetry
corresponds to a triple trace deformation in dual field theory, 
$\beta\sim \alpha^2$. 
In this case, the trace of the dual stress tensor vanishes.
%

In this approach, 
given some data at infinity $(\alpha,\beta)$,
one can assign a well
defined mass 
$M$ only {\it after} defining the function $W(\alpha)$, $cf.$ eq. (\ref{W}).
In principle, this result can be circumvented by using
the counterterm in Refs.
\cite{Lu:2013ura},
\cite{Gegenberg:2003jr},
\cite{Radu:2004xp},
\begin{eqnarray}
\label{Ict}
I_{ct}^{(\phi) }= 
\frac{1}{3}
 \int_{\partial M}d^{3}x\sqrt{-h}
\Big(\phi  {n}^{\nu}\partial _{\nu}\phi  -\frac{1}{2 L}\phi^2\Big)~,
\end{eqnarray} 
which does not require to specify a condition for $\alpha$ and $\beta$.
However, this counterterm is problematic 
because it is not intrinsic to the boundary and also,
for mixed boundary conditions, the variational principle is not satisfied.
Moreover, it implies  generically a vanishing trace of the boundary stress tensor
(although for $\beta \sim \alpha^2$ the mass expression coincides with
that found using (\ref{Tsup})).

\subsection{Remarks on numerics} 

While it was possible to find some partial analytical results,
the non-perturbative solutions  
are constructed numerically, 
by integrating the system of Einstein-scalar 
field equations
(\ref{eqs})
 subject to suitable boundary conditions. 
Both solitons and BHs will be considered. However, note that in order
to simplify the problem, in the BH case  we restrict the study to the region outside the event horizon. For spherically symmetric solutions,
we use a standard Runge-Kutta ordinary differential equation solver.
All numerical calculations in the axially symmetric case have been
performed by using a professional package,
which uses a  Newton-Raphson finite difference method with
an arbitrary grid and arbitrary consistency order \cite{schoen}.

The numerics are done working with a
scaled scalar field and a scaled radial coordinate, 
\begin{eqnarray}
\label{scale}
\phi \to \phi /\kappa,~~r \to r/L,
 \end{eqnarray} 
that results in the following  Lagrangian  of the 
considered models
(note that for the $\phi^4$ case, 
we supplement (\ref{scale})
with $\lambda \to \lambda \kappa^2/L^2$)
\begin{eqnarray} 
\label{L}
{\cal L}=
  R+6 
   -\frac{1}{2} g^{ab}  \phi_{, \, a} \phi_{, \, b} 
	-U(\phi)
	 \end{eqnarray} 
	with
	\begin{eqnarray} 
{\bf sugra}:~~U(\phi)=n \sinh^2 \left ( \frac{\phi }{\sqrt{n} }\right) ,~~~~
\phi^4-{\bf model}:~~U(\phi)=-\phi^2+\lambda \phi^4,
	 \end{eqnarray} 
However, for the sake of clarity,
all equations displayed in what follows 
are given in terms of dimensionful variables.

\section{Probe limit: static scalar clouds in AdS} 
\label{Probe}

Before considering the full problem, it is 
interesting to consider first
the probe limit, 
and to study
solutions of 
the Klein-Gordon equation
in a fixed geometry,
while
neglecting the backreaction
of the scalar field.
The background can be the AdS
spacetime  or the 
Schwarzschild-AdS (SAdS) BH,
which possesses a  line element of the form (\ref{ads}),
with\footnote{The expression (\ref{SAdS})
results from the usual SAdS expression,
$N=1-2M/r+r^2/L^2$
with $M=r_h(1+{r_h^2}/{L^2} )/2$.}
\begin{eqnarray}
\label{SAdS}
 N(r)=(1-\frac{r_h}{r})
                                \big(
1+\frac{r^2}{L^2}+\frac{r r_h}{L^2}+\frac{r_h^2}{L^2} 
                                   \big),
 \end{eqnarray} 
where
 $r_h$
is  the event horizon radius
and $M$ the BH mass. 

This approximation greatly simplifies the problem
but retains some of the interesting physics.
Both {\it linear} ($i.e.$ with a mass term only in the potential $U(\phi)$)
and {\it non-linear} clouds will be considered.
In both cases, 
the mass-energy density of a configuration,
 as measured by a static observer
with $4-$velocity
$U^a\sim \delta_t^a$,
is 
$\rho=-T^t_t$.
Then one can define a   total mass of a cloud,
\begin{eqnarray}
\label{mass-cloud}
M^{(cloud)}=-\int d^3 x T_t^t=
-\int_{r_0}^\infty dr \int_0^\pi d\theta \int_0^{2 \pi}d\varphi
r^2 \sin \theta ~T_t^t.
\end{eqnarray}
One can see that, in order for 
$M^{(cloud)}$
to be finite, 
$T_t^t$ should decay faster than 
$1/r^3$ as $r\to \infty$.
Then, for the scalar field asymptotics
(\ref{inf}),
the  term proportional with $1/r$ 
should be absent, $i.e.$ $\alpha(\theta,\varphi)=0$
in the large $r$-limit.

\subsection{The linear case} 
\label{linear-clouds}

Let us start with the case
 of a massive scalar field with no self-interaction
($i.e.$ with 
$U=\frac{1}{2}\mu^2 \phi^2$)
in a fixed AdS background (\ref{ads}),
and
consider solutions of the (linear)
KG equation 
		\begin{eqnarray}
 \nabla^2 \phi=\mu^2 \phi,
\end{eqnarray}
The scalar field  can be decomposed in a sum of modes	
	\begin{eqnarray}
\phi= \sum_{\ell m} \phi_{\ell m}(r,\theta,\varphi),
~~~~{\rm with}~~~\phi_{\ell m}=Y_{\ell m}(\theta,\varphi) 
R_{\ell} (r),
\end{eqnarray}
where 
$Y_{\ell m}(\theta,\varphi) $
are the real spherical harmonics
(with
$\ell=0,1,\dots$
and
$-\ell\leq m\leq \ell$),
while the radial amplitude
$R_{\ell}(r) $ is a solution of the equation
\begin{eqnarray}
\label{eql}
\frac{1}{r^2} (r^2 N R_{\ell})'= (\mu^2+ \frac{\ell(\ell+1)}{r^2} )R_{\ell},
\end{eqnarray}
where a prime denotes the derivative $w.r.t.$
the radial coordinate $r$.
 
\begin{figure}[t] 
\begin{center}
\includegraphics[height=.34\textwidth, angle =0 ]{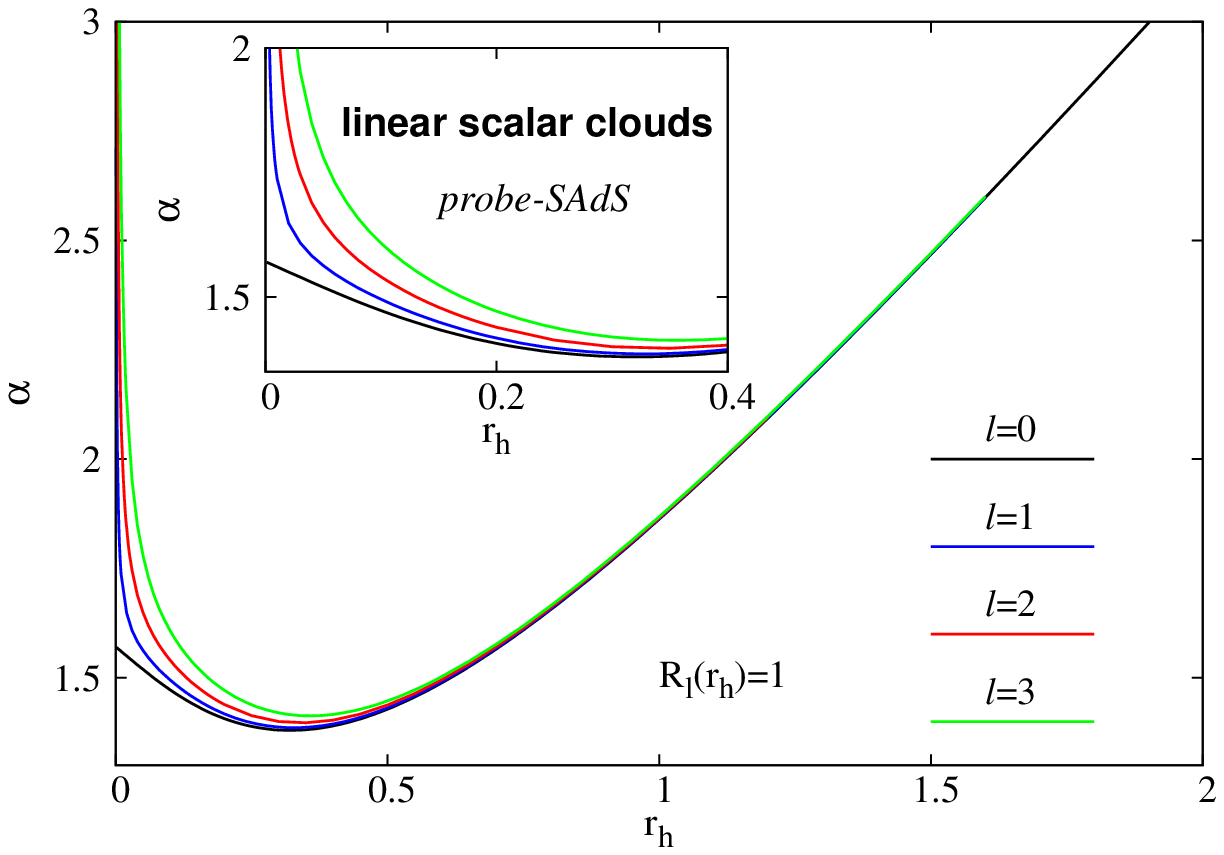}
\includegraphics[height=.34\textwidth, angle =0 ]{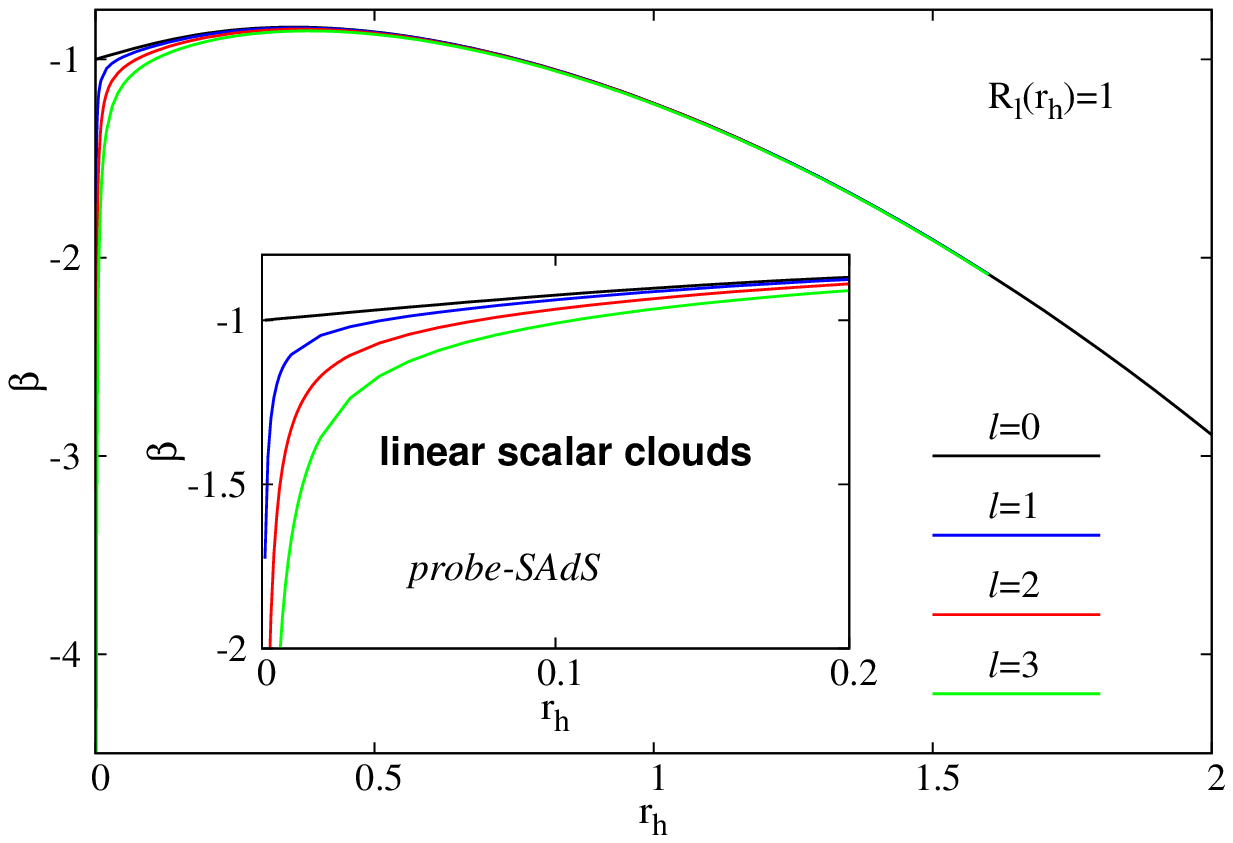}
\end{center}
\caption{
The parameters 
$\alpha$
and 
$\beta$
which enter the far field expansion of the scalar field
are shown as a function of the event horizon radius
for several values of the harmonic index $\ell$. 
The results are found for linear 
scalar clouds in a Schwarzschild-AdS background.
}
\label{ab-phi2}
\end{figure}
%
\begin{figure}[ht!] 
\begin{center}
\includegraphics[height=.34\textwidth, angle =0 ]{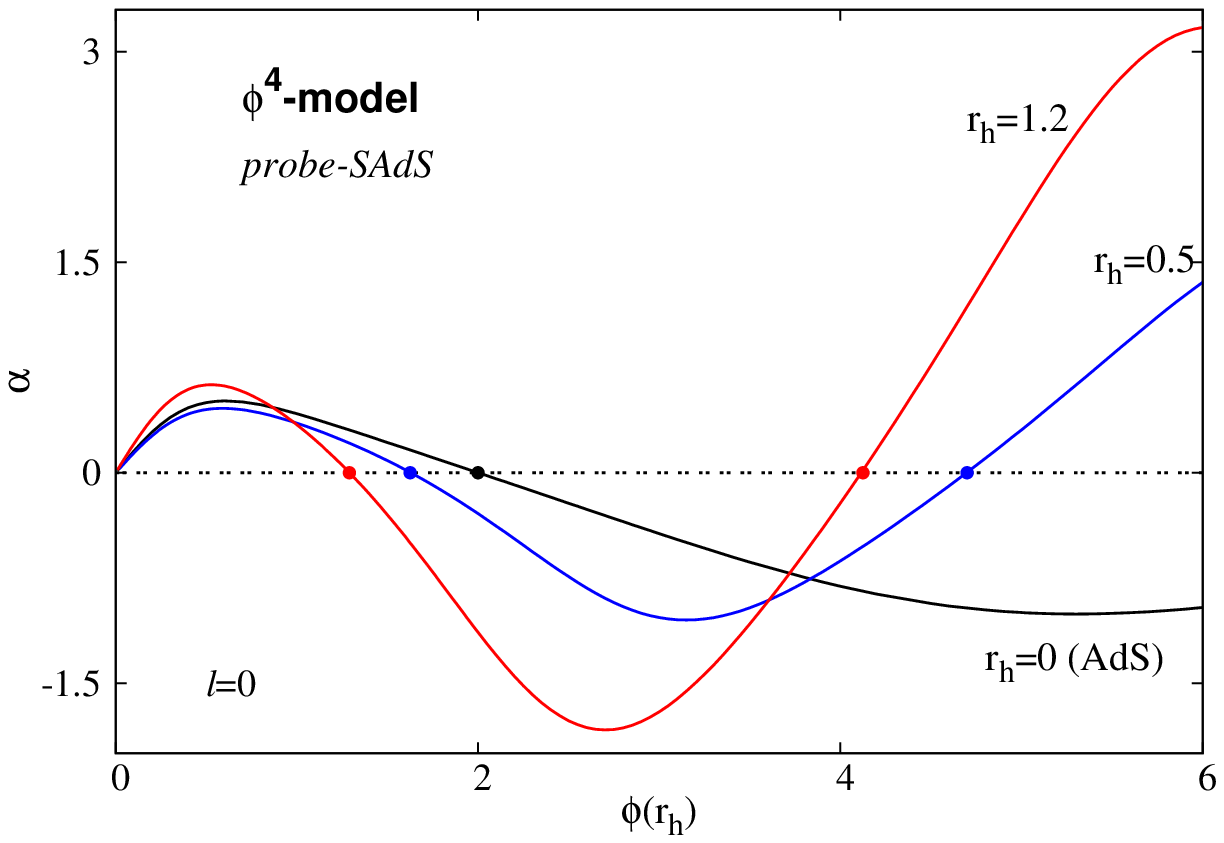}
\includegraphics[height=.34\textwidth, angle =0 ]{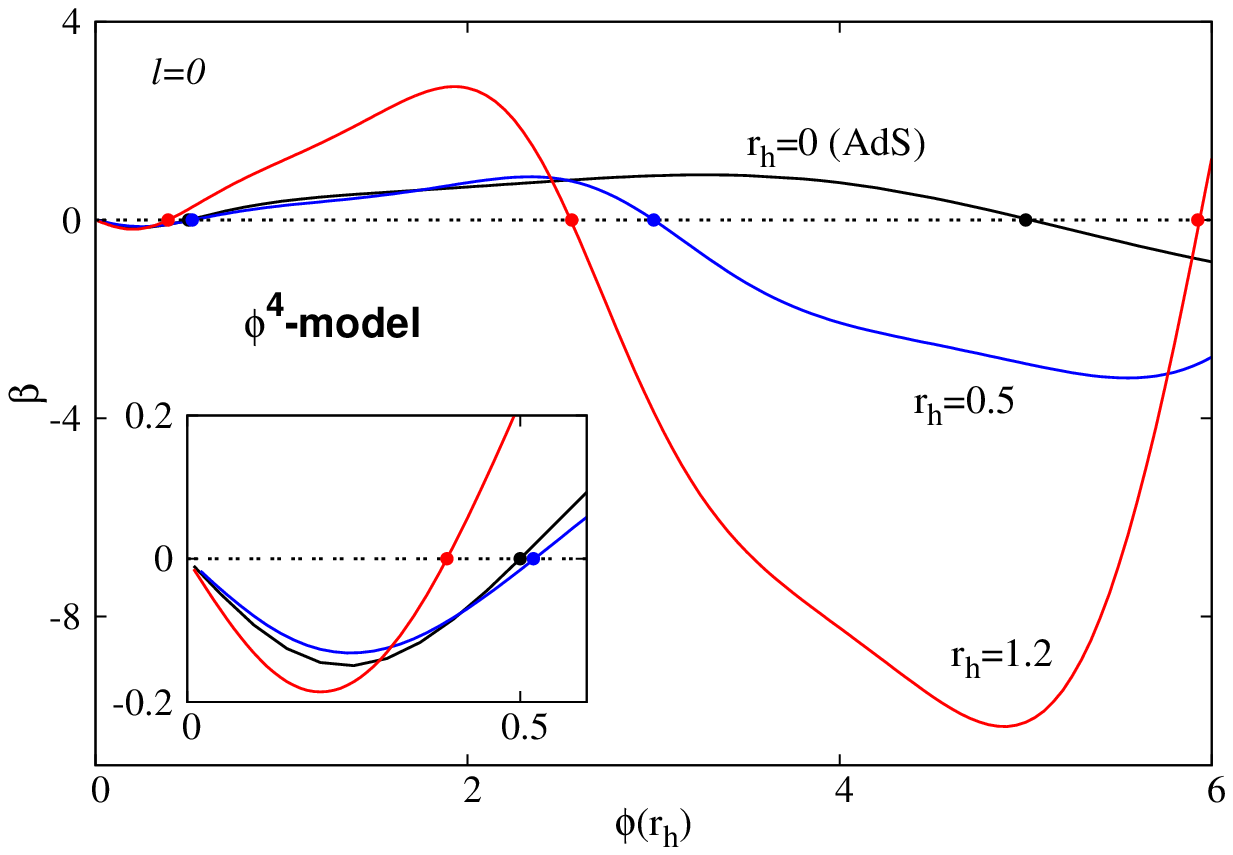}
\end{center}
\caption{
The parameters 
$\alpha$
and 
$\beta$
which enter the far field expansion (\ref{inf})
are shown as a function of the scalar field at the horizon
for solutions of a $\phi^4$-model
in a fixed background.
Note the existence of configurations with $\alpha=0$
or $\beta=0$ (marked with dots). 
}
\label{ab-phi4}
\end{figure}

For the case of interest in this work
with $\mu^2=-2/L^2$,
the general solution of the above equation reads
(with $_2F_1$  the hypergeometric function) 
\bea
R_\ell(r)&=&
 c_1
 \left(\frac{r}{L} \right)^\ell {}_2F_1
                                        \left(
\frac{1+\ell}{2} , \frac{2+ \ell}{2}   ;
  \frac{3}{2}+\ell  ;  - \frac{r^2}{L^2}
	                                       \right)\nn\\
&+&
	 c_2
 \left (\frac{L}{r} \right)^{\ell+1} {}_2F_1
	 \left( 
	 \frac{1-\ell}{2} , -\frac{\ell}{2}   ;
   \frac{1}{2}-\ell  ;  - \frac{r^2}{L^2}
	 \right) 
\eea
being the  
sum of two modes
  (with $c_1,c_2$ arbitrary constants).
However, the second term in the above relation  diverges as $r\to 0$
and thus we set $c_2=0$ (also, in what follows, 
 we take $c_1=1$). 
The explicit form of the solution for the first three values of $\ell$ reads
%
%
\begin{figure}[ht!] 
\begin{center}
\includegraphics[height=.34\textwidth, angle =0 ]{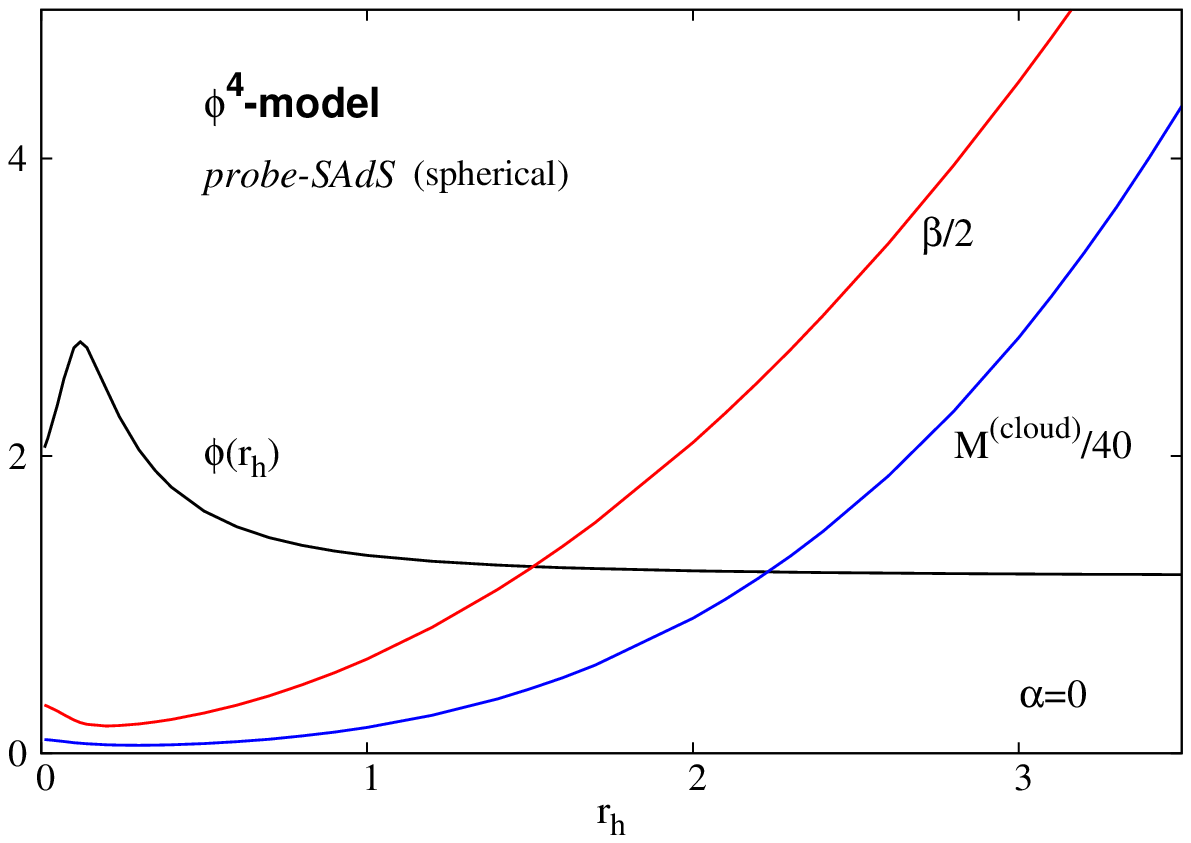} 
\includegraphics[height=.34\textwidth, angle =0 ]{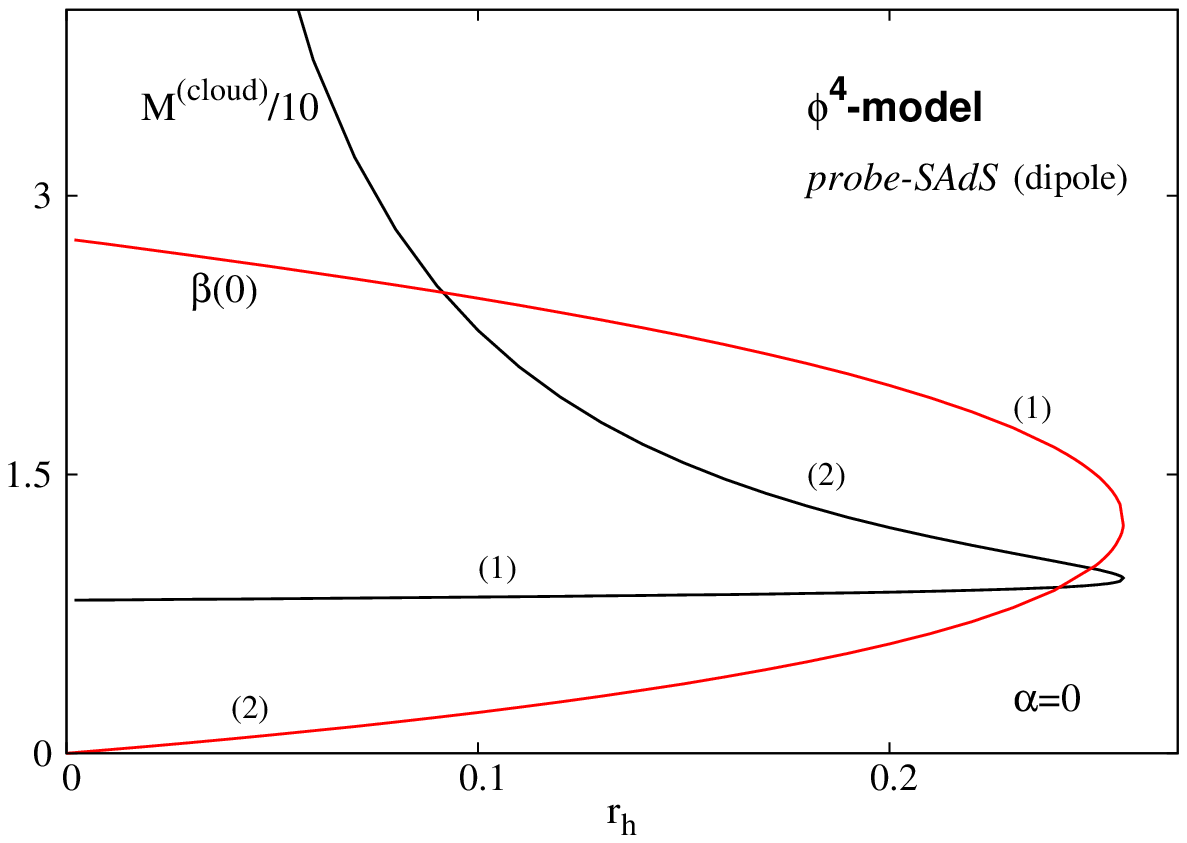} 
\end{center}
\caption{
Several quantities of interest 
are shown as a function of the event horizon radius
for $\ell=0,~1$
 solutions of the $\phi^4$-model
in a fixed Schwarzschild-AdS background.
The scalar field here decays asymptotically as $1/r^2$.
}
\label{BH-phi4-probe-var-rh}
\end{figure}
%
\bea
&&R_{0}(r)= \frac{L}{r}\arctan(\frac{r}{L}),~~
R_{1}(r)= \frac{3L}{r}-\frac{3L^2}{r^2}\arctan(\frac{r}{L}),\nn\\
&&R_{2}(r)=\frac{6L^2}{\pi r^2} 
\left(
-1+(1+\frac{r^2}{3L^2})\arctan(\frac{r}{L}
\right)
\eea 
the expressions for higher $\ell$ becoming
increasingly complicated.

As $r\to 0$, the (regular) solution has the following form
\begin{eqnarray}
 R_{\ell}(r)= (\frac{r}{L})^\ell
  -\frac{(\ell+1)(\ell+2)}{2(2\ell+3)}(\frac{r}{L})^{\ell+2}+\dots.~~
\end{eqnarray}
As spatial infinity is approached,
all multipoles decay
according to (\ref{inf}),
such that the approximate form of a $(\ell,m)$-mode reads
\be
\nn
\phi_{\ell m}(r,\theta,\varphi)= 
\frac{\alpha (\theta,\varphi)}{r}
+\frac{\beta(\theta,\varphi)}{r^2} 
+\dots,
\ee
where
\be
\nn
\alpha(\theta,\varphi)=\frac{\sqrt{\pi}\Gamma(\ell+\frac{3}{2}) L}{(\Gamma(\frac{\ell}{2}+1))^2}Y_{\ell m}(\theta,\varphi),~
\beta(\theta,\varphi)=\frac{\sqrt{\pi}\Gamma(\ell+\frac{3}{2})L^2}{(\Gamma(\frac{\ell+1}{2}))^2}Y_{\ell m}(\theta,\varphi).\nn
\ee
This behaviour strongly contrasts
with that found for a Minkowski spacetime background,
where the scalar mode which is regular at $r=0$
diverges as $r\to \infty.$
This feature
can be traced back to the ``box"-like behaviour of the AdS spacetime,
and is present also  for a Maxwell field  \cite{Herdeiro:2015vaa}.

The above solution contains already several features that
will also be found in the (self-gravitating) sugra-case.
First, 
one notices that
all multipoles share  the same far field decay.
Second,
the radial amplitude $R_{\ell}(r)=$ is always nodeless.
Moreover, both parameters  $\alpha$
and $\beta$
are non-zero.
As such, the linear cloud mass,
as computed according to (\ref{mass-cloud}) diverges,
although the energy density
$\rho$
 is finite everywhere.

One may ask if the situation is different
when considering a SAdS BH background.
Although no exact solutions of  the radial equation
appear to exist,
the eq.~(\ref{eql}) can 
  be solved numerically. 
The approximate expansion of the radial amplitude as $r\to r_h$
reads
\begin{eqnarray}
 R_{\ell}(r)= R_{\ell}(r_h)
\left(
1+\frac{ \ell(\ell+1)L^2-2r_h^2}{r_h(L^2+3r_h^2)} (r-r_h)
\right)+\dots,~~
\end{eqnarray}
while the asymptotic expansion of a mode is given by (\ref{inf}),
with 
\begin{eqnarray}
\alpha=p_1 Y_{\ell m}(\theta,\varphi) ,~~~
\beta=p_2 Y_{\ell m}(\theta,\varphi) ,
\end{eqnarray}
where the constants $p_1,p_2$
depend on the value of $r_h$.

As seen in Figure \ref{ab-phi2},
the presence of a horizon does not change the 
picture found for solitons.
In particular, there are no scalar clouds\footnote{The apparent divergence 
for $\ell>0$
of $\alpha$, $\beta$
as $r_h\to 0$ is an artifact of solving the
(linear) 
 equation (\ref{eql}) 
with the boundary condition $R_\ell(r_h)=1$,
while $R_\ell (0)=0$ in the solitonic limit.
} with
$\alpha=0$
or 
$\beta=0$.

\begin{figure}[t]
\centering
{ 
\includegraphics[height=.34\textwidth, angle =0 ]{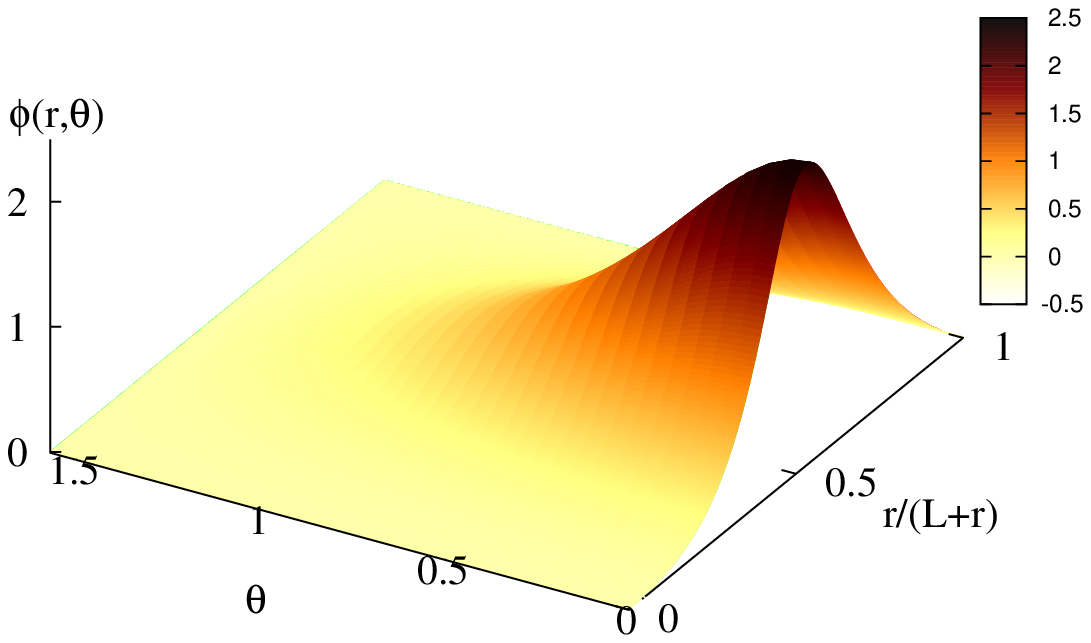}
\includegraphics[height=.34\textwidth, angle =0 ]{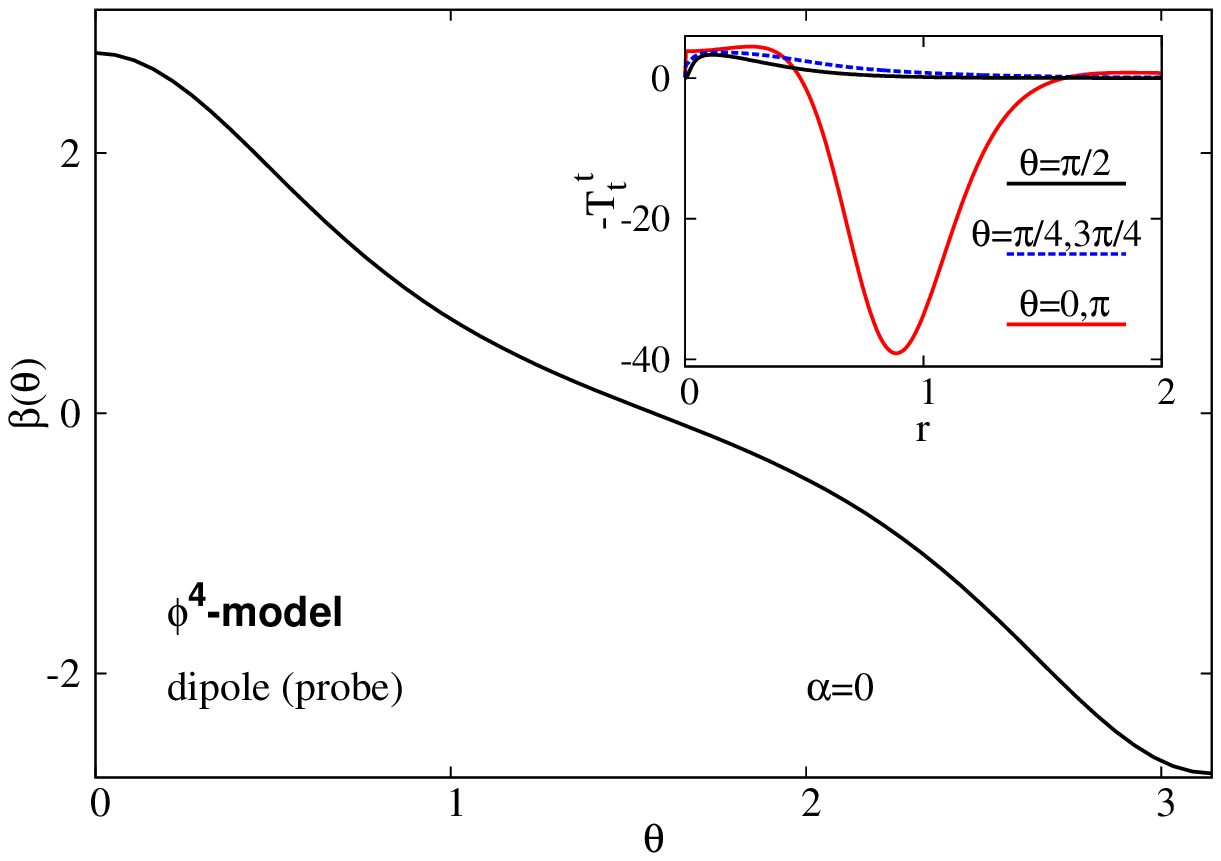}
}
\caption{ 
{\it Left panel:} The scalar field profile is shown for the
(fundamental) 
dipole solution 
with $1/r^2$ decay
in the $\phi^4$-model and a
fixed AdS background.
Only half of the space is shown here, 
with $\phi(r,\theta)=-\phi(r,\pi-\theta)$.
 {\it Right panel:} The function $\beta$
which enters the large-$r$ asymptotics (\ref{inf}) of $\phi$
is shown for the same solution. 
The inset shows the energy density $\rho=-T_t^t$ for several 
angular directions. 
}
\label{profile-phi4-probe}
\end{figure}

\subsection{Non-linear clouds in the $\phi^4$ model} 
\label{phi4-probe}

One may inquire how general are the above results and
 what are the new results induced by the scalar field
self-interaction.
In particular, 
are there scalar clouds with 
$\alpha=0$ 
(which then would possess a
finite mass)?

 While the  picture found for the 
sugra-potential (\ref{U})
appears to be qualitatively similar 
to that found for linear clouds
(and we could not find solutions with
$\alpha=0$), 
the situation 
is different for a model with a quartic
self-intercation.
An indication in this direction\footnote{An exact solution
with
$\alpha=0$ and a finite mass $M^{(cloud)}= {8\pi^2 L}/({27 \lambda^2})$ 
is found in a model with a {\it cubic}-selfinteraction
\begin{eqnarray} 
U(\phi) =-\frac{\phi^2}{L^2} +\lambda \phi^3,~~{\rm and}~~
\phi(r)=  -\frac{4}{3\lambda}\frac{1}{1+\frac{r^2}{L^1}}.
\end{eqnarray}
}   
as provided by  the existence of the 
following exact solution\footnote{Note that the solution
with real $\phi$
exists for  $\lambda<0$ only.} 
describing a spherically symmetric soliton
in the $\phi^4$-model:
\bea
\label{ex-sol}
&&\phi(r)= \frac{1}{2\sqrt{-\lambda}}\frac{1}{\sqrt{1+\frac{r^2}{L^2}}},\nn\\
&&{\rm with}~~~
\phi(r) \to \frac{1}{2 \sqrt{-\lambda}}\frac{L}{r}-\frac{L^3}{4 \sqrt{-\lambda} r^3 }+\dots
~~{\rm as}~~ r\to \infty~~i.e.~~~\beta=0.
\eea

We have studied generalizations of this exact solution,
by solving numerically the scalar field equation 
for a (S)AdS background 
and varying $\phi(r_h)$
 (the value of the scalar field at the horizon or the origin, $r_h=0$, for the soliton),
the values of the parameters 
$\alpha$
and 
$\beta$
being extracted from the numerical output.
In the numerics, we set $\lambda=-1$
without any loss of generality, via
a suitable scaling of the scalar field.

Some numerical results are shown in Figure \ref{ab-phi4}
where one can observe that the parameters 
$\alpha$
and 
$\beta$
can be zero, for a (presumably infinite)
set of discrete values of 
$\phi(r_h)$.
For example, 
in the solitonic case ($r_h=0$),
the exact solution
(\ref{ex-sol})
corresponds to
$\phi(0)=1/2$,
while
the first configuration with
$\alpha=0$
is found for  
$\phi(0)=2$
(which strongly suggests the existence of 
an exact solution also in that case).
Moreover, for $\phi(0)>2$ the solutions possess at least a node.

In Figure \ref{BH-phi4-probe-var-rh} 
(left panel)
we present a number of relevant 
quantities as a function of the horizon radius
for the subset of (finite mass) 
solutions with
$\alpha=0$.
As one can see, no restrictions seem to exist on the BH size,
while for large enough BHs, both $M$ and $\beta$
increase linearly with $r_h$.
 

\medskip

Since one has to solve numerically a (nonlinear) 
partial differential equation (PDE), the case of higher multipoles is technically more complicated. Restricting to axisymmetric configurations,
this is a particular case of the problem discussed in Section 3, 
being solved by using
a similar numerical approach.
Moreover, we have mainly studied the  
case of finite mass dipoles\footnote{However, we have confirmed
the existence of similar solutions also for $\alpha=0$ quadrupoles.}
$i.e.$
with $\alpha=0$
in the scalar far field expansion, the $1/r^2$
decay being imposed by introducing a new function 
$\psi=r \phi$, and requiring $\psi\to 0$
as $r\to \infty$.
Other boundary conditions satisfied by 
the scalar field are 
$\partial_\theta \phi=0$
at $\theta=0,\pi$
and $\phi=0$ at $r=0$,
while the field is (still) odd-parity,
$\phi(\theta)=-\phi(\pi-\theta).$
In Figure \ref{profile-phi4-probe}
(left panel)
we display 
the profile of the AdS dipole solution,
which possesses a finite mass $M^{(cloud)} \simeq 8.208$.
We note that the extrema of the field are located on the 
$z-$axis (with $z=r\cos \theta$) and are symmetric $w.r.t.$
the equatorial plane.

We shall also mention that
the $\ell-$labeling of the solutions in terms of multipoles
is ambiguous in a non-linear setup, 
since the `pure-cloud' feature of the linear solutions is lost
due to self-interaction.
As such, $\ell$ stands rather for the number of 
angular nodes of the scalar profiles
  (with $\ell=0$ for spherical solutions, 
$\ell=1$ for dipoles (one node at $\theta=\pi/2$), etc).
For example, for a dipole solution,
one can write  the following expansion of the scalar field 
\begin{eqnarray} 
\phi(r,\theta)=\sum_{k\geq 0}f_{k}(r) {\cal P}_{2k+1} (\cos \theta),
\end{eqnarray}
with ${\cal P}_n(x)$ the Legendre polynomials.
Although the $k=0$ term dominates, the
contribution of the higher order terms is also nontrivial,
as can be seen already
in the profile of the function $\beta(\theta)$
(in the right panel of Figure  \ref{profile-phi4-probe}). While for a linear cloud 
$\beta \sim \cos (\theta)$,
this is not the case when there exists a nonzero contribution 
of higher order ${\cal P}_{2k+1}$-terms.
 
As expected, 
similar solutions are found   (numerically)
in the presence of a BH horizon,
$i.e.$ for a SAdS background.
The numerics is done in terms of a new radial coordinate
$\bar r=\sqrt{r^2-r_h^2}$ in the line-element 
(\ref{ads}),
(\ref{SAdS}),
such that the horizon is located at $\bar r=0$,
where we impose 
$\partial_{\bar r} \psi=0$
(with $\phi=r\psi $).
The boundary conditions on the $z$-axis 
and at infinity are similar to those imposed for $r_h=0$. In Figure \ref{BH-phi4-probe-var-rh}. 
we have shown the mass of the solutions  and the value
of $\beta(0)$
as a function of the horizon radius.
One can see that the picture is very different as
compared to that found in the spherically symmetric
case.
As for $\ell=0$,
a branch of solutions
(label (1) in Figure \ref{BH-phi4-probe-var-rh}) 
smoothly emerges 
when adding a horizon at the center of a soliton.
Along this branch the mass increases, while
the maximal value of $\beta$ decreases.
However, for $\ell=1$, one finds the existence
of a maximal value of $r_h$,
with 
a back bending and the
occurrence of a secondary branch of solutions
(label (2)), which extends backwards in $r_h$.
As $r_h\to 0$
along this secondary branch, 
$\beta \to 0$,
while
the mass of the solutions appears to diverge.\footnote{
This behaviour can be understood by noticing that,
for the (scaled) units (\ref{scale}) we employ in numerics,
$r_h\to 0$
can also be approached as $L\to \infty$
(and thus a vanishing cosmological constant),
while the background metric becomes the Schwarzschild BH.
However, no smooth solution  exists
in  this case \cite{Herdeiro:2015waa}.
}

\section{Spherically symmetric Einstein-scalar field solutions} 
 \label{spherical}
 \subsection{The Ansatz, equations and asymptotics} 

The spherically symmetric solutions
are constructed by using the following Ansatz
  for the metric and scalar field 
\begin{eqnarray}
ds^2=-N(r)e^{-2\delta (r)}dt^2+\frac{dr^2}{N(r)}+r^2 (d\theta^2+\sin^2\theta d\varphi^2),~~
{\rm and}
~~\phi \equiv \phi(r)~,
\end{eqnarray}
where it is convenient to take
\begin{eqnarray}
N(r)=1+\frac{r^2}{L^2}-\frac{2m(r)}{r},
\end{eqnarray} 
with $m(r)$ a mass function.
From (\ref{eqs}) we find the following equations
for the metric functions and the scalar field:
\begin{eqnarray}
\label{eomeq}
m'= \kappa^2  r^2(\frac{1}{2}N\phi'^2+U(\phi)),~~
\delta'=-\kappa^2   r \phi'^2,~~
\phi''+(\frac{2}{r}+\frac{N'}{N}-\delta')\phi'-\frac{1}{N}\frac{dU(\phi)}{d\phi}=0.
\end{eqnarray}
There is also
a 2nd order constraint equation 
\be
\frac{1}{2}N''
-N\delta''
+\frac{N'}{r}
-\delta' 
\left(
\frac{N}{r}+\frac{3}{2}N'-N\delta'
\right)
-\frac{3}{L^2}
+ \kappa^2 (N\phi'^2+2 U(\phi))=0,
\ee
which, however, is a differential consequence
 of the equations for $m,\delta$ 
in (\ref{eomeq}).

The Ricci scalar $R$ and the Kretschmann scalar
 $K=R^{\mu\nu\rho\sigma}R_{{\mu\nu\rho\sigma}}$ are given by
\bea 
R&=&2N\delta''-N''+3\delta'N'-\ft{2N(r \delta'-1)^2}{r^2}-\ft{4N'}{r}+\ft{2}{r^2},\nn\\
K&=&\ft{4(N-1)^2}{r^4}+\ft{2N'^2}{r^2}+\ft{2(N'-2N\delta')^2}{r^2}+\bigg(2N(\delta'^2-\delta'')+N''-3\delta'N'\bigg)^2.
\eea
For all solutions reported in this work, both
$R$ and $K$ are regular everywhere on the considered 
domain of integration\footnote{Thus, for BHs, this holds on and outside the 
event horizon.}.

The system of equations (\ref{eomeq}) will be solved 
first perturbatively
and then numerically.  
In both cases, it is useful to find the approximate
form of the solutions at the boundaries of the domain of integration.

 \subsubsection{The small-$r$ expansion} 
Starting with the solitonic case,
the small-$r$ solution can be written 
in the form
\begin{eqnarray}
\label{r=0}
m(r)=\sum_{k\geq 3} m_{(k)}r^k,~~
\delta(r)=\delta(0)+\sum_{k\geq 1} \delta_{(k)}r^k,~~
\phi(r)=\phi(0)+\sum_{k\geq 1} \phi_{(k)}r^k,~~
\end{eqnarray}
 the series coefficients 
$m_{(k)}$,
$\delta_{(k)}$,
$\phi_{(k)}$,
being determined by the values of the functions 
$\phi$ and $\delta$
at $r=0$.
No general pattern for these coefficients appears to exist,
the first terms being
\begin{eqnarray}
\nonumber
&&
m_{(3)}=\frac{1}{3} \kappa^2 U_0 ,~
m_{(4)}=0,~
m_{(5)}=\frac{2\kappa^2 U'^2-0}{45},
m_{(6)}=0,~
m_{(7)}=\frac{5\kappa^2 U'^2_0 }{252 L^2}
\left(
-1+L^2(\frac{2}{3}\kappa^2 U_0+\frac{6}{25}U''_0
\right),
\\
&&
\nonumber
\delta_{(1)}=\delta_{(2)}=\delta_{(3)}=0,~
\delta_{(4)}=-\frac{1}{36}\kappa^2 U'^2_0,~
\delta_{(5)}=0,~
\delta_{(6)}=\frac{1}{27 L^2}\kappa^2 U'^2_0
\left(
1-L^2(\frac{2}{3}\kappa^2 U_0+\frac{1}{10}U''_0)
\right),
\\
&&
\nonumber
\phi_{(1)}=0,~
\phi_{(2)}=\frac{1}{6}U'_0,~
\phi_{(3)}=0,~
\phi_{(4)}=\frac{U'_0}{12L^2}
\left(
-1+\frac{2}{3}\kappa^2 L^2 U_0+\frac{1}{10}L^2 U''_0 
\right),
\phi_{(5)}=0,
\end{eqnarray}
with
$U_0 =U (\phi(0))$,
$U_0'=U'(\phi(0))$,
$U_0''=U''(\phi(0))$.

 \subsubsection{The near-horizon solution} 
Apart from solitons, we are also interested in BH solutions.
They possess a non-extremal horizon\footnote{We did not find any indication
for the existence of extremal BH solutions.
In fact, their absence is also suggested by the absence of an attractor
solution 
with an $AdS_2\times S^2$ geometry,
which would describe the near horizon of the extremal BHs.} 
located at $r=r_h>0$,
where $N(r_h)=0$.
Close to the  horizon, we assume the existence of a power series expansion
of the solution in $r-r_h$,
with 
\be
\label{r=rh}
m(r)=\sum_{k\geq 0} \bar m_{(k)}(r-r_h)^k,~
\delta(r)=\delta(r_h)+ \sum_{k\geq 1} \bar \delta_{(k)}(r-r_h)^k,~~
\phi(r)= \phi(r_h)+\sum_{k\geq 1} \bar \phi_{(k)}(r-r_h)^k,
\ee
the  coefficients being determined by the horizon values of 
the functions
$\phi$ and $\delta$. 
One finds $e.g.$  
\begin{eqnarray}
\nonumber
&&
\bar m_{(0)}=\frac{r_h(r_h^2+L^2)}{2L^2},~~ 
\bar m_{(1)}= \kappa^2 r_h^2 U_h,~~ 
\bar \delta_{(1)}=-\frac{\kappa^2 r_h^3 U'^2_h }
{ \left(1+\frac{3r_h^2}{L^2}-2\kappa^2 r_h^2 U_h \right)^2},~~
\bar \phi_{(1)}=\frac{r_h U'_h}{1+\frac{3r_h^2}{L^2}-2\kappa^2 r_h^2 U_h},~
\\
&&
\nonumber
\bar m_{(2)}=  \kappa^2r_h
\left(
 U_h+\frac{3r_h^2 }{4}
\frac{U'^2_h}{1+\frac{3r_h^2}{L^2}-2\kappa^2 r_h^2  U_h }
\right),~
\\
&&
\nonumber
\bar \delta_{(2)}=
-\left(
 (1+\frac{3r_h^2}{L^2})(1-\frac{3r_h^2}{L^2}-r_h^2  U_h'')
+2\kappa^2 r_h^4 \bigg((\frac{6}{L^2}- U''_h-6)U_h-2\kappa^2 U_h \bigg)
\right)
\frac{2\kappa^2 r_h^2   U'^2_h}{4(1+\frac{3r_h^2}{L^2} -2\kappa^2 r_h^2 U_h)^4},
\\
&&
\nonumber
\bar \phi_{(2)}=-
\left(
(1+\frac{3r_h^2}{L^2})(\frac{6}{L^2}-U''_h)U'_h
-2\kappa^2 U_h'
\bigg(
(2(1+\frac{6r_h^2}{L^2})-r_h^2 U''_h)U_h+r_h^2 U_h'^2-4\kappa^2 r_h^2U_h^2
\bigg)
\right)\nn\\
&&\qquad\quad\times\frac{r_h^2}{4(1+\frac{3r_h^2}{L^2} -2\kappa^2 r_h^2 U_h)^3)},
\end{eqnarray}
where we denote
$U_h =U (\phi(r_h))$,
$U_h'=U'(\phi(r_h))$,
$U_h''=U''(\phi(r_h))$.
 Also, since the model (\ref{L})
is invariant 
 when taking $\phi \to -\phi$,
it is enough to consider the case 
$\phi(r_h)>0$, only
(or $\phi(0)>0$ for solitons).
 
 \subsubsection{The large-$r$ approximate solution} 
Finally, the large-$r$ approximate expression of the solutions 
holds for both solitons and BHs.
For a generic scalar potential with 
\be
U|_{\phi=0}=0,\qquad 
\ft{\partial U}{\partial \phi} \big|_{\phi=0}=0,\qquad 
\ft{\partial^2 U}{\partial \phi^2}\big|_{\phi=0}=-\ft{2}{L^2},\qquad 
\ft{\partial^3 U}{\partial \phi^3}\big|_{\phi=0}=0,
\ee
an approximate form of the solutions\footnote{Let us remark that the equations of the model are invariant
when taking $\delta \to \delta+const.$,
a symmetry which is lost when imposing
$\delta(\infty)=0.$
}
  can be written as  series in $1/r$, 
with
\begin{eqnarray}
\label{inf-gen} 
 m(r)=M_0-\frac{\alpha^2 \kappa^2}{2L^2} r
+ \sum_{k\geq 1} \frac{\tilde m_{(k)}} {r^k} ,~~
\delta(r)= \sum_{k\geq 1} \frac{\tilde \delta_{(k)}} {r^k} ,~~
\phi(r)= \frac{\alpha}{r}+ \frac{\beta}{r^2}
+\sum_{k\geq 3} \frac{\tilde \phi_{(k)}} {r^k} ,~~
\end{eqnarray}
with
the coefficients depending on the free parameters
 $\{M_0,\alpha,\beta \}$.
One finds, $e.g.$,
\begin{eqnarray}
\nonumber
&&
\tilde m_{(1)}=-\frac{\kappa^2}{2}
\left(
\alpha^2+\frac{2\beta^2}{L^2}+\alpha^4 (\frac{ 2\kappa^2}{L^2}+\frac{1}{4} U^{(4)})
\right),\nn\\
&&
\tilde m_{(2)}=-\frac{\alpha\kappa^2}{6}
\left(
\alpha (-M_0+2\alpha \beta (\frac{ 8\kappa^2}{L^2}+U^{(4)})+\frac{1}{15}\alpha^3 U^{(5)} +4 \beta
\right),
\\
\nonumber
&&
\tilde \delta_{(1)}=0,~~
\tilde \delta_{(2)}=\frac{1}{2}\alpha^2 \kappa^2,~
\tilde \delta_{(3)}=\frac{4}{3}\alpha \beta \kappa^2,~
\tilde \delta_{(4)}=\frac{\kappa^2}{8} 
\left(
       8\beta^2+\alpha^2 L^2 
(\frac{6\kappa^2}{L^2}+U^{(4)})
\right),~
\\
\nonumber
&&
\tilde \phi_{(3)}=\frac{\alpha^2 L^2}{12}(\frac{ 6\kappa^2}{L^2}+U^{(4)}),~
\tilde \phi_{(4)}=\frac{ L^2}{12} 
\left(
\alpha (4M+ \alpha \beta (\frac{8\kappa^2}{L^2}+U^{(4)})
+\frac{1}{12}\alpha^3 U^{(5)} -4 \beta
\right),
\end{eqnarray}
where $U^{(n)}$ denotes $\ft{\partial^n U}{\partial \phi^n}|_{\phi=0}$.

Also, the leading order expression of the metric functions 
$g_{rr}$ and $g_{tt}$ reads
\begin{eqnarray}
\nonumber
&&
g_{rr}=\frac{1}{N(r)}=\left(\frac{L}{r} \right)^2
-\left(1+\frac{\alpha^2 \kappa^2}{ L^2} \right)
\left(\frac{L}{r} \right)^4+\frac{2M_0}{L}\left(\frac{L}{r}\right)^5+O(1/r^6),
\\
&&
\nonumber
-g_{tt}=N(r)e^{-2\delta(r)}=1+\frac{r^2}{L^2}-\frac{2M_0+\frac{8\alpha \beta \kappa^2}{3L^2}}{r}+O(1/r^2),~
\end{eqnarray}
such that the spacetime
is still asymptotically (locally) AdS.

 \subsubsection{Quantities of interest} 

The Hawking temperature and horizon area of the BH solutions are fixed by
the horizon data, with
\begin{eqnarray}
T_H=\frac{1}{4\pi }N'(r_h)e^{-\delta(r_h)},~~A_H=4\pi r_h^2.
\end{eqnarray}

The mass computation is a straightforward application of the general
formalism  in Section \ref{mass}.
For a given design function $W$
(as given by (\ref{W})), 
the non-vanishing components of the resulting boundary stress-tensor are
(here we choose $\partial {\mathcal M}$ to be a three surface of fixed $r$,
while $n_{\nu}=\sqrt{g_{rr}}\delta_{r \nu}=\delta_{r \nu}/\sqrt{N}$):
\begin{eqnarray}
\label{BD4}
\nonumber
{\rm T}_{\theta}^{\theta} 
={\rm T}_{\varphi}^{ \varphi}=
\Big(
 \frac{M_0 L}{2\kappa^2}
-\frac{1}{ L}(W-\alpha \beta)
\Big)\frac{1}{r^3}
+O\left(\frac{1}{r^4} \right),
~~
{\rm T}_{t}^t=\Big(
-\frac{ M_0 L}{ \kappa^2}
-\frac{1}{L}(W+\alpha \beta)-
\Big)\frac{1}{r^3}+O\left(\frac{1}{r^4}\right).
\end{eqnarray}
Then
the mass of these solutions, as computed from (\ref{mass}) is
\begin{eqnarray}
\label{Mct}
M= 4\pi 
\Big(
\frac{M_0}{\kappa^2}+\frac{ \alpha\beta+W}{ L^2}
\Big),
\end{eqnarray}
 with $W$ the function (\ref{W})
imposing a condition between $\alpha$ and $\beta$.

\subsection{ Solutions in the ${\cal N}=8$ $D=4$ model} 
\label{sugra}
\subsubsection{ Perturbative  solitons } 
\label{sugraP}

In the solitonic case,
a simple enough exact solution can  found
perturbatively in terms of the scalar amplitude $\phi(0)=\epsilon$.
The Ansatz for a perturbative approach is:
\begin{eqnarray}
m(r)=\sum_{k\geq 2}\epsilon^k m_k(r),~~
\delta(r)=\sum_{k\geq 2}\epsilon^k \delta_k(r),~~
\phi(r)=\sum_{k\geq 1}\epsilon^k \phi_k(r),~~
\end{eqnarray} 
 
The solution for the lowest order in 
$\epsilon$
is valid for any scalar selfinteraction,
since only the mass term is relevant here. 
One finds
\begin{eqnarray}
\label{lower}
&&
\phi_1(r)=\frac{L}{r}{\cal X}(r),~\phi_2(r)=0,~
m_2(r)=\frac{\kappa^2 L}{2}{\cal X}(r) (1-\frac{L}{r}{\cal X}(r)),~m_3(r)=0,
\\
\nonumber
&&
\delta_2(r)=-\frac{\kappa^2}{2}
\left(
\frac{1}{N_0(r)}
+{\cal X}(r)(\frac{2L}{r}+(1-\frac{L^2}{r^2}){\cal X}(r))
-\frac{\pi^2}{4}
\right),~\delta_3(r)=0,
\end{eqnarray} 
where we define the auxiliary functions
\begin{eqnarray}
N_0(r)=1+\frac{r^2}{L^2},~~{\cal X}(r)=\arctan(\frac{r}{L}).
\end{eqnarray} 

Although the equations can be solved to the next order in 
$\epsilon$,
the solution with a generic
parameter
 $n$
in the potential (\ref{U}) 
is exceedingly complicated.
Thus in what follows, we shall restrict our study to the special case 
$n=1$,
where the solution still possesses a simple enough form, 
with
\begin{eqnarray}
\nonumber
&&
 \phi_3(r)=-\frac{\kappa^2}{2 N_0(r)}
             \left(
1-\frac{L}{3r}{\cal X}(r)
\left(
3+N_0(r)
-\frac{L^2}{r^2}N_0^2(r){\cal X}^2(r)
\right)
            \right),~~ \phi_4(r)=0,
\\
\nonumber
&&
m_4(r)=\frac{\kappa^4 L}{24 N_0(r)}
\bigg(
-\frac{9r}{L}+(\frac{19r^2}{L^2}+25){\cal X}(r)
-\frac{2r}{L}(13+\frac{2r^2}{L^2}+\frac{11 L^2}{r^2} ){\cal X}^2(r)
\\
\nn
&&\qquad\qquad
+6(\frac{ L^2}{r^2}-\frac{r^2}{L^2}){\cal X}^3(r)  
+\frac{4L}{r}N_0^2(r) {\cal X}^4(r)  	
\bigg),
\\
\nonumber
&&
\delta_4(r)=\frac{\kappa^4}{192}
                      \bigg[
\frac{8}{N_0(r)^2}
 \bigg(
14+\frac{5r^2}{L^2}
-\frac{10r}{L}(1-\frac{ r^2}{L^2}+\frac{16L^2}{5r^2}){\cal X}(r)  
 \bigg)+(\frac{5r^2}{L^2}+\frac{16L^2}{5r^2}-3)N_0(r) {\cal X}(r)^2\nn\\
&&\qquad\qquad
-\frac{8r}{L}(1-\frac{L^4}{r^4} )N_0(r) {\cal X}(r)^3 
-2 (1+\frac{2L^2}{r^2}+\frac{3L^4}{r^4})N_0(r)^2{\cal X}(r)^4+\pi^2(\pi^2-10)
                       \bigg].
\end{eqnarray} 
\begin{figure}[t]
\centering 
{
\includegraphics[height=.34\textwidth, angle =0 ]{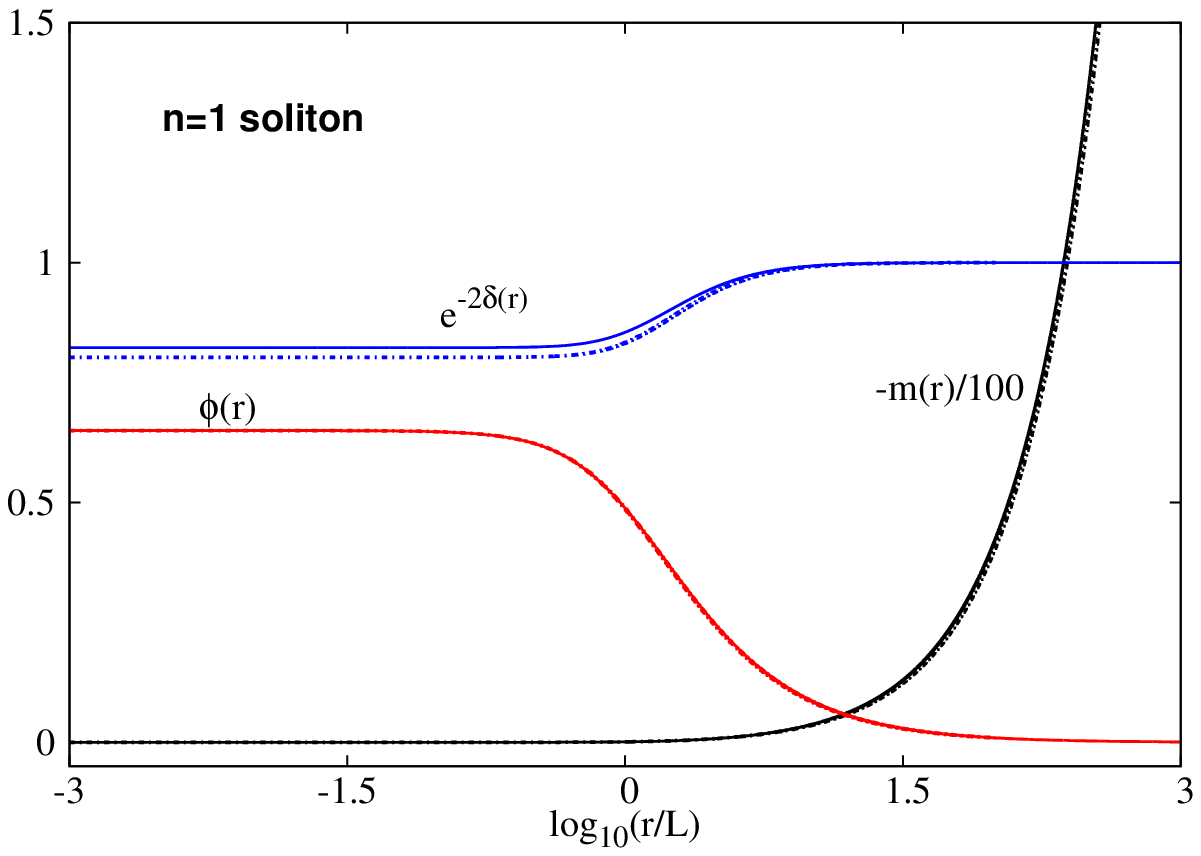}
\includegraphics[height=.34\textwidth, angle =0 ]{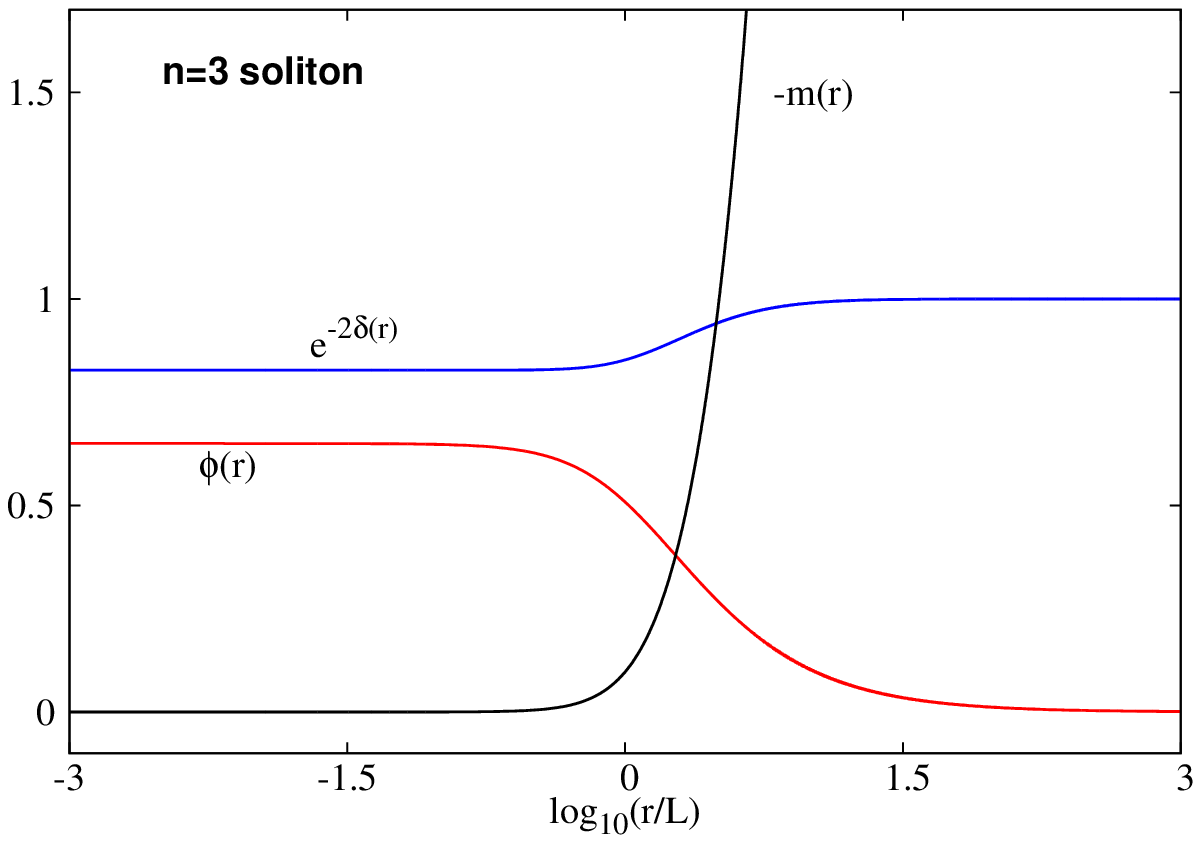}
}
\caption{ 
Typical profiles of  $n=1,3$  soliton solutions 
with the same value of the scalar field at the origin
$\phi(0)=0.65$
are shown as a function of the radial coordinate. 
 The corresponding  perturbative solution 
is also shown for the $n=1$ case (dotted curves).
.}
\label{profile-soliton-sugra13}
\end{figure}
%
\begin{figure}[ht!]
\centering
{ 
\includegraphics[height=.34\textwidth, angle =0 ]{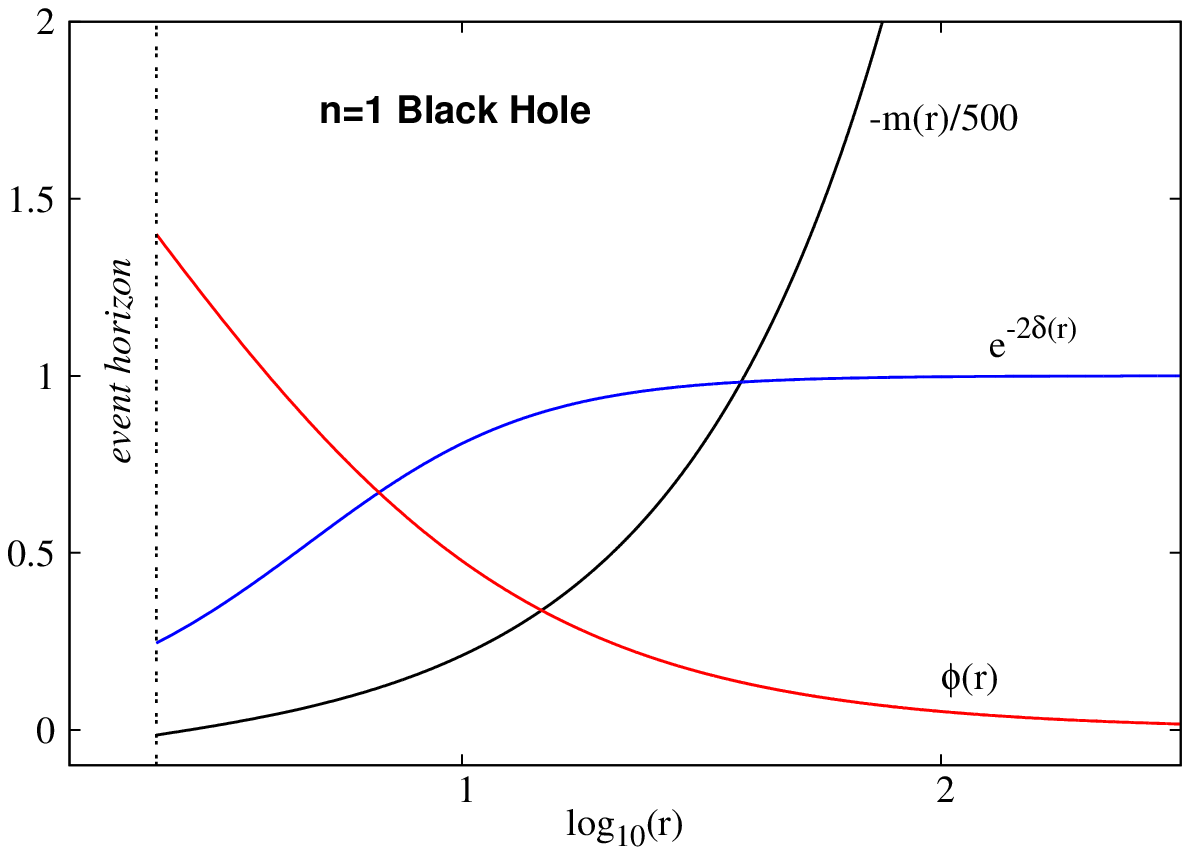}
\includegraphics[height=.34\textwidth, angle =0 ]{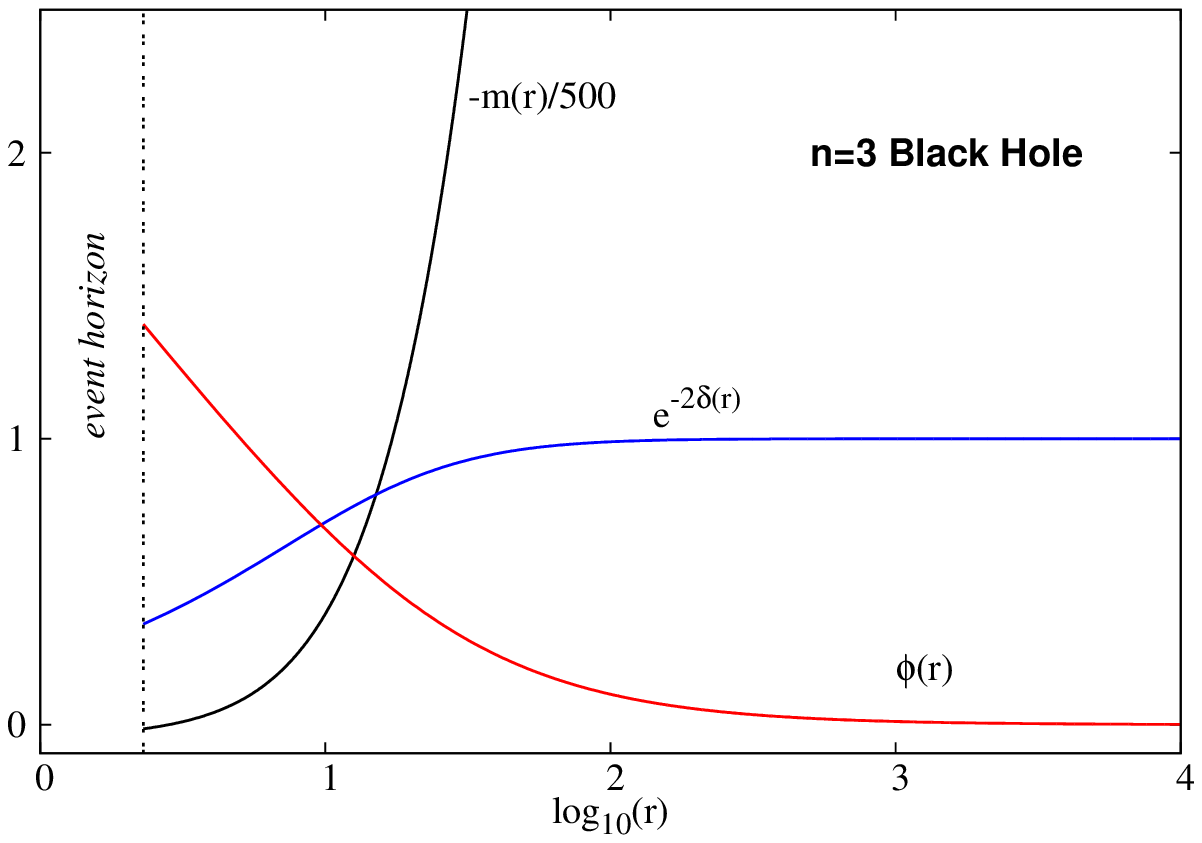}
}
\caption{ 
Typical profiles of $n=1,3$  black hole solutions 
with the same value of the scalar field at the horizon
$\phi(r_h)=1.4$
are shown as a function of the radial coordinate.
}
\label{profile-BH-sugra13}
\end{figure}
While $\phi_4(r)=0$,
the  function 
$\phi_5(r)$ 
 is more complicated,  with the presence of
  the poly-logarithm function 
$Li_n(x)$,  
\begin{eqnarray} 
\phi_5(r)=\kappa^4\sum_{k=0}^5 f_k(r) {\cal X}(r)^k,
\end{eqnarray} 
 where
\begin{eqnarray} 
\nonumber
f_0(r)&=&-\frac{1}{4N_0(r)}
                                 \left(
1+\frac{2 i \pi^4 N_0(r)}{15}
\frac{L}{r}
(1-\frac{90}{\pi^4} Li_{4}(-\frac{r+i L}{r-i L}) )
                                 \right),
\\ 
\nonumber
f_1(r)&=&
\frac{7}{40 N_0(r)^2}\frac{L}{r}
\left(
1
+\frac{2 r^4}{7L^4}
+\frac{19 r^2}{7L^2}
-\frac{40}{7} N_0(r)^2 (5 Li_{3}(-\frac{r+i L}{r-i L})+\zeta(3) )
\right),
\end{eqnarray} 

\begin{figure}[ht!]
\begin{center}
\includegraphics[height=.34\textwidth, angle =0 ]{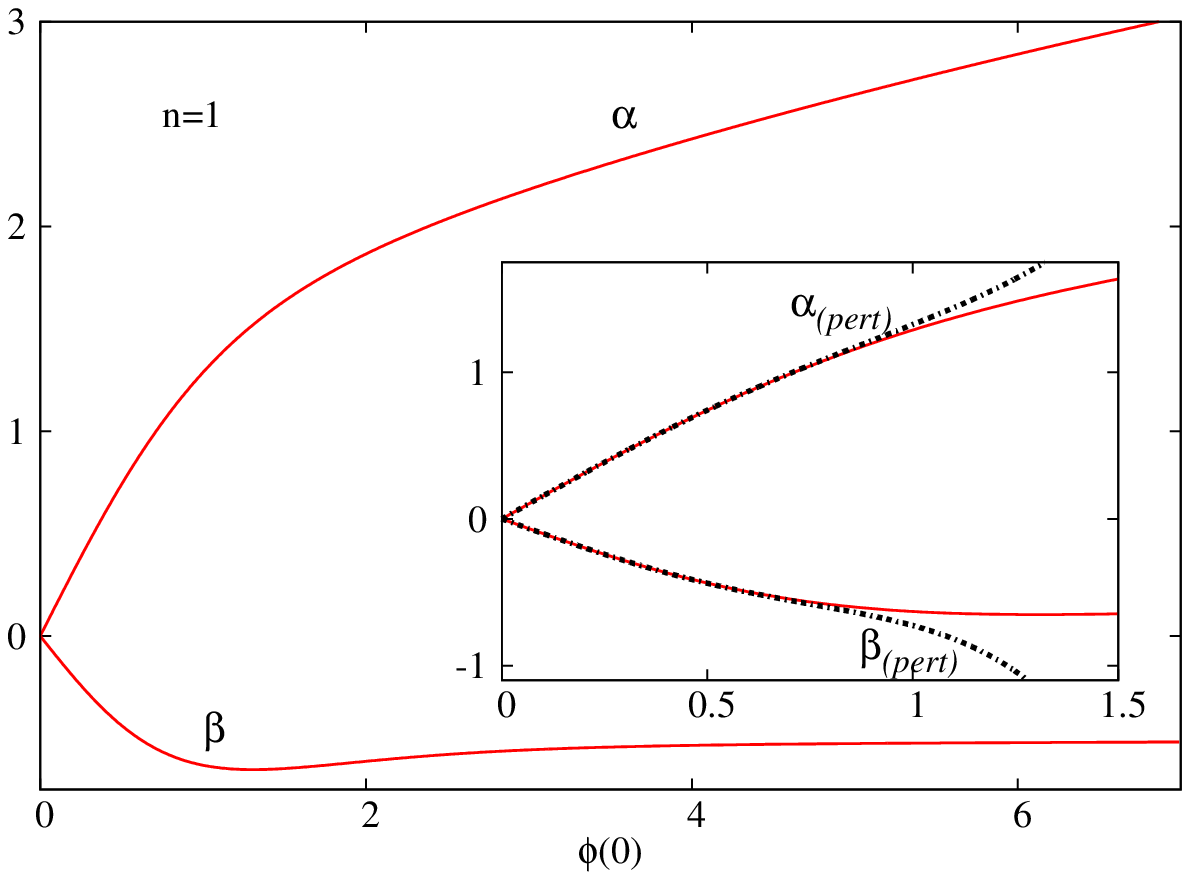}
\includegraphics[height=.34\textwidth, angle =0 ]{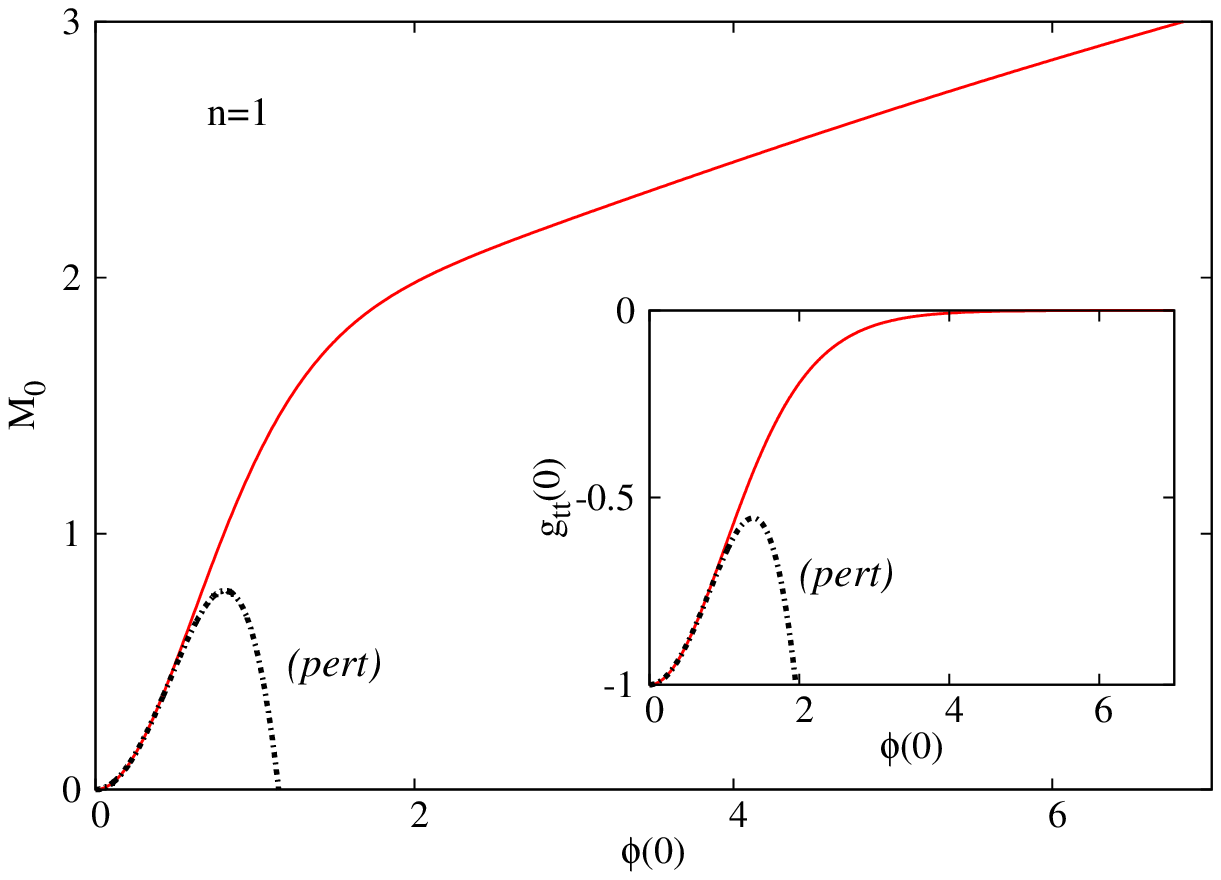}
\includegraphics[height=.34\textwidth, angle =0 ]{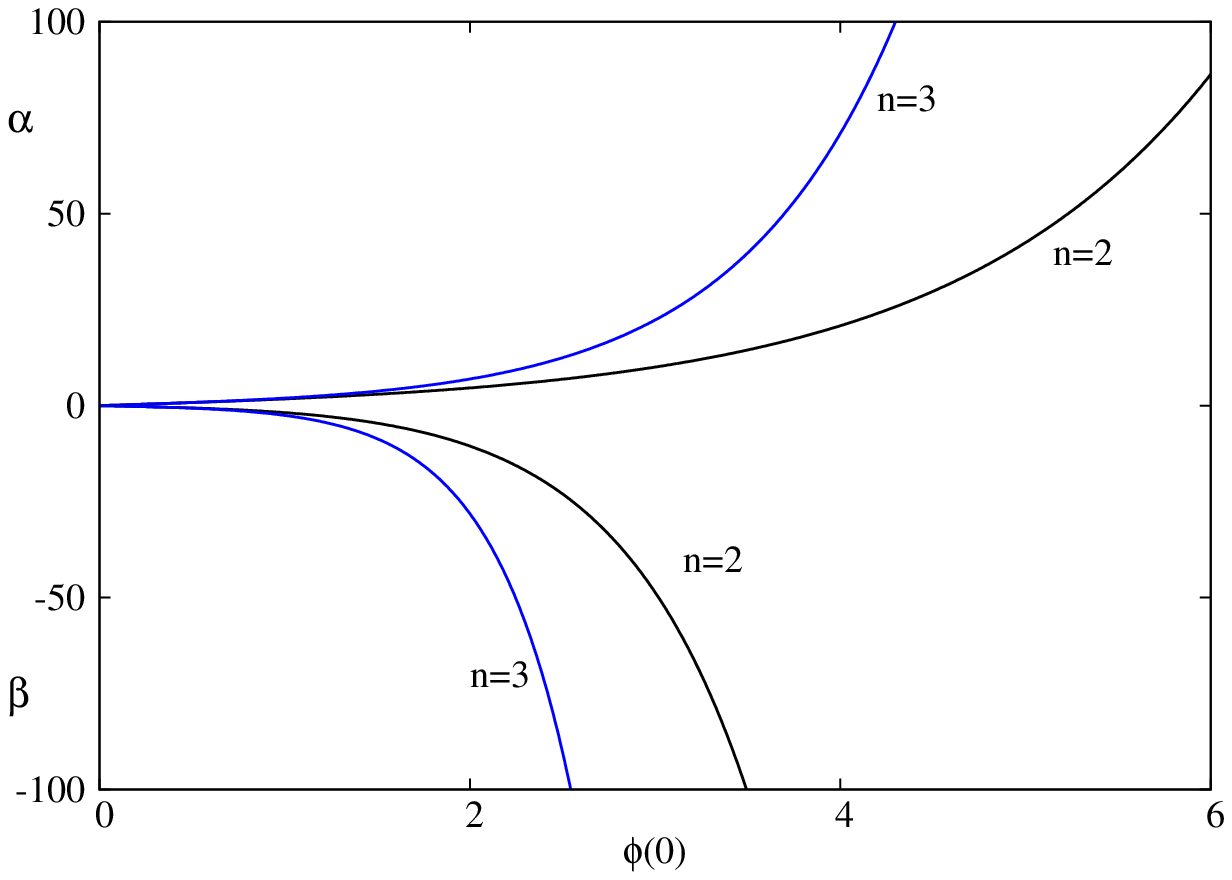}
\includegraphics[height=.34\textwidth, angle =0 ]{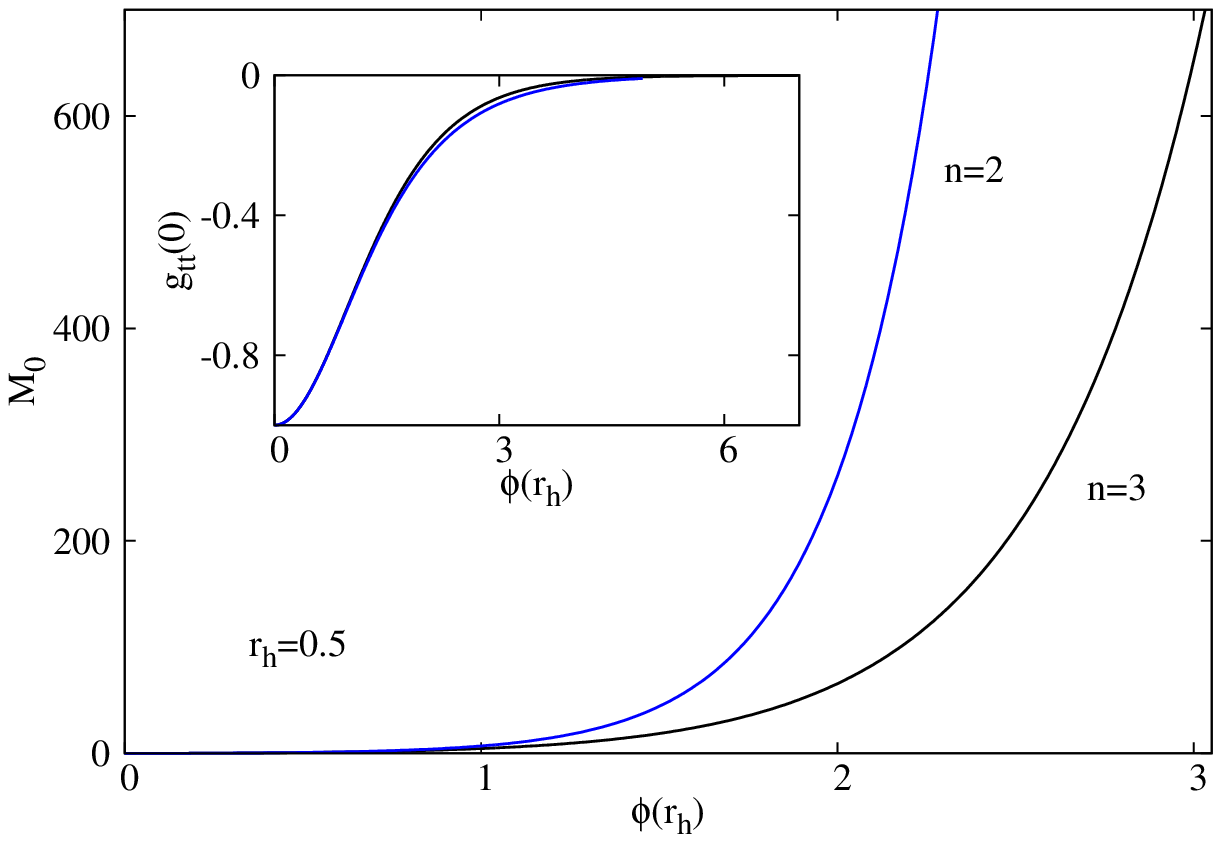}
\end{center}
\caption{
Several quantities of interest are shown as a function of the scalar 
field at the origin for $n=1$ and $n=2,3$ solitonic solutions.
The perturbative results are also shown for $n=1$  (dotted curves).
}
\label{quant123}
\end{figure}
%
\begin{figure}[ht!]
\begin{center}
 \includegraphics[height=.30\textwidth, angle =0 ]{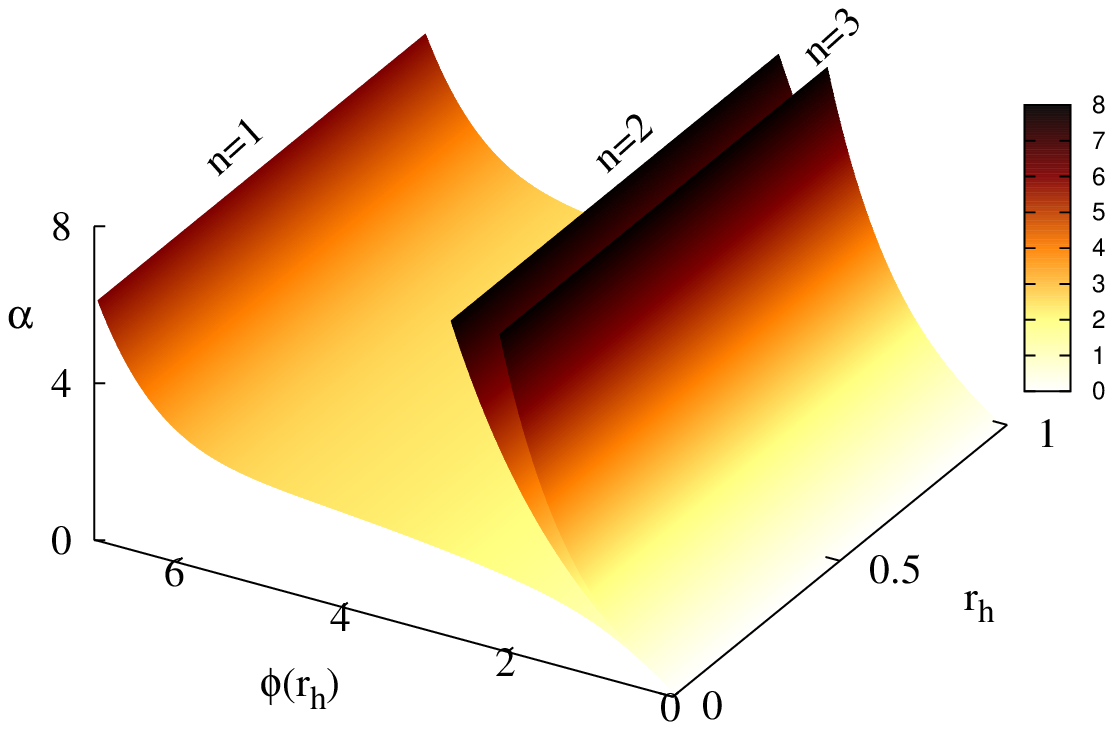}
\includegraphics[height=.30\textwidth, angle =0 ]{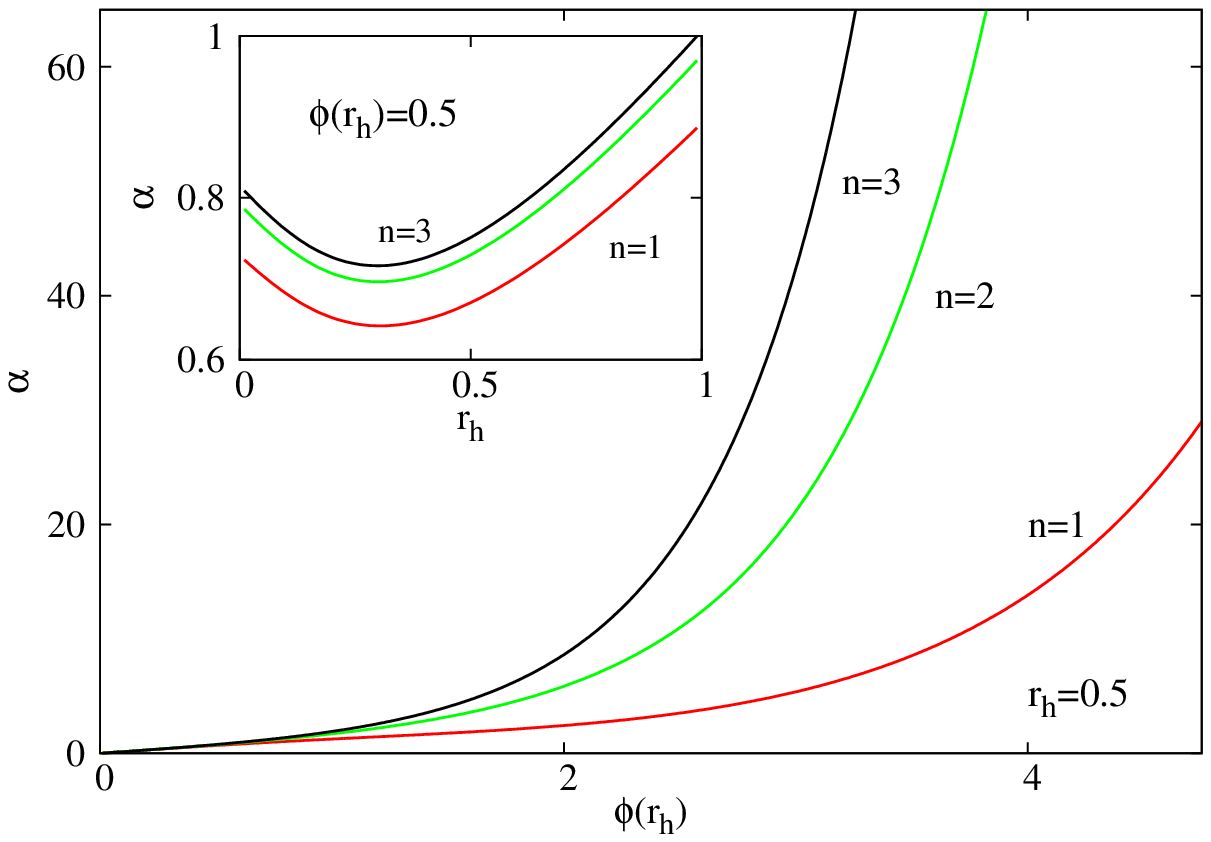}
 \includegraphics[height=.30\textwidth, angle =0 ]{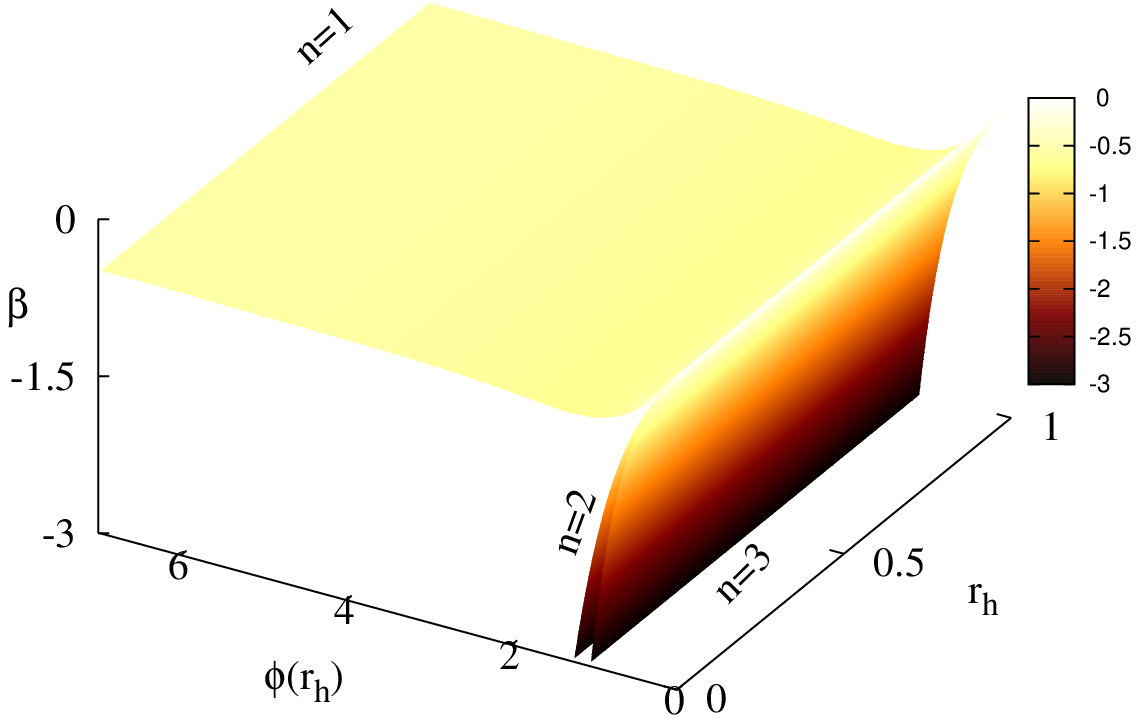}
\includegraphics[height=.30\textwidth, angle =0 ]{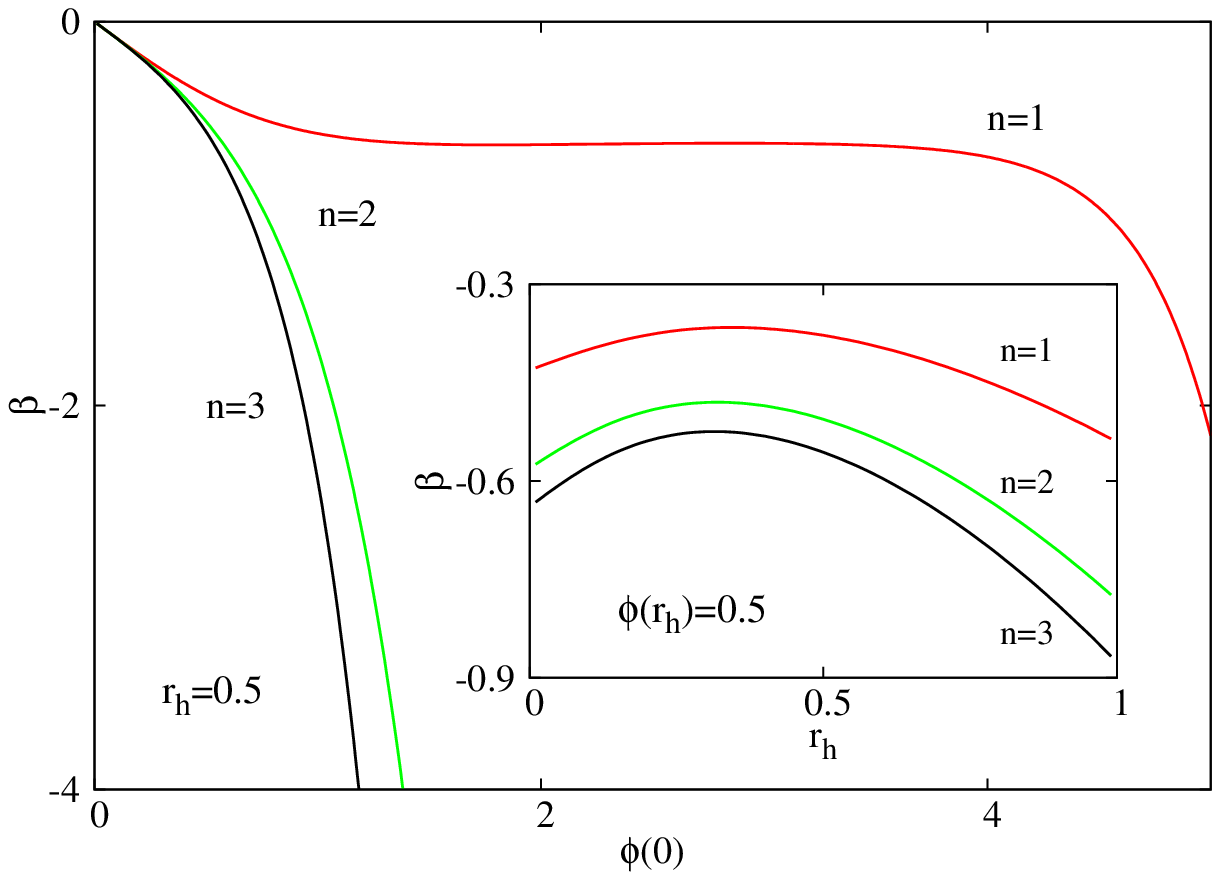}
%
\end{center}
\caption{ 
The parameters $\alpha$
and 
$\beta$
which enter the far field expansion of the scalar field 
are shown as a function of the value of 
the horizon radius and of 
the scalar field at the horizon
for families of $n=1,2,3$ black hole solutions.
The right panel shows different slices of these plots.
}
\label{alpha-beta-sugra123}
\end{figure}
%
\begin{eqnarray}  
\nonumber
f_2(r)&=&
(1+\frac{  r^2}{L^2}+\frac{2 r^4}{3L^4})\frac{45 L}{r}
-480 i N_0(r)^2  Li_{2}(-\frac{r+i L}{r-i L}),
\\
\nonumber
f_3(r)&=&
\frac{L}{24 r} 
\left(
1-24 i\pi-\frac{8L^2}{r^2}+48\log(\frac{L+i r}{2r} )
\right),
\\ 
\nonumber
f_4(r)&=&\frac{L}{24 r}(-8 i+\frac{5 L}{r}-\frac{  L^3}{r^3}),~~~~
f_5(r)=\frac{N_0(r)}{120}\frac{L^3}{r^3}
(1+\frac{9L^2}{r^2}),
\end{eqnarray} 
where 
  $\zeta(x)$ 
is the Riemann zeta function\footnote{Note that 
$\phi_5(r)$
is a real function, despite the presence of $i$ in its expression.}.

The above expressions 
allow for a discussion of some  basic properties 
of the solitonic solution. 
For example,   the small-$r$
expansion of the scalar field reads
\begin{eqnarray} 
\phi(r)=\epsilon
       - \big (
\epsilon
+\frac{2\kappa^2}{3}\epsilon^3
+\frac{2\kappa^4}{15} \epsilon^5
         \big)
		 \frac{r^2}{3L^2}+O(r^4),
															~~~{\rm thus}~~\phi(0)=\epsilon.
\end{eqnarray}  

In the context of this work, 
the coefficients
 $\alpha,\beta,M_0$
which enter the far field asymptotics  
are of special interest, with
\begin{eqnarray}
\nonumber
&&
 \alpha=\frac{\pi L}{2}\epsilon
+\frac{\pi L}{36} \kappa^2
\left(
1-\frac{1}{16}(12+\pi^2)
\right)
  \epsilon^3 
	+\frac{\pi L}{3840} \kappa^4 a_5   \epsilon^5 +\dots,
	\\
	&&
	\label{abs}
 \beta=- L^2 \epsilon
+ \frac{\pi^2 L^2 \kappa^2}{8} (1-\frac{16}{3\pi^2} ) \epsilon^3
+ \frac{  L^2 \kappa^4}{480} b_5 \epsilon^5+\dots,
\\
&&
	\nonumber
M_0= \frac{3\pi L}{4}\epsilon^2 \kappa^2 
\left(
1+\frac{1}{72}\epsilon^2 \kappa^2(54-11\pi^2)
\right)+\dots,
\end{eqnarray}
(where we denote
$
a_5=96+\pi^4+\pi^2(20-960 \log 2 )+5280 \zeta(3),
$
$
b_5=-144+5 \pi^2(3+\pi^2-48 \log 2)+840 \zeta(3))
$).
One notices that
$\alpha$ is positive 
and
$\beta$ negative 
  to  order ${\cal O}(\epsilon)^5$,
which suggests the absence of solutions with
$\alpha \leq 0 $ and $\beta \geq 0$ to all orders,
a conjecture which is confirmed by the nonperturbative 
results in the next Subsection. 
Also,
one should remark that 
$(\alpha,\beta,M_0)$
are not independent, being parameterized by 
$\phi(0)$.
The choice of the function
$W$ 
(cf. eq. (\ref{W}))
fixes this parameter; 
for example, 
$ \beta=-\alpha^2$
for $\epsilon=\phi(0)\simeq 0.4$,
while $ \beta=-\alpha^3$
for
 $\epsilon=\phi(0)\simeq 0.519$).

The expression of the metric potential
at the origin (which
provides a measure on how strong are the gravity effects)
is also of interest, with
\begin{eqnarray} 
-g_{tt}(0)=
1-\frac{1}{4}\kappa^2(\pi^2-8)\epsilon^2 
+\frac{1}{48}\kappa^4 (96-19\pi^2+\pi^4)\epsilon^4+\dots,
\end{eqnarray} 

Finally, we mention that given a design function $W$
(which would fix the parameter $\phi(0)$),
the mass $M$
of the solitons results directly from  
the eqs.~(\ref{Mct}), (\ref{abs}).

\begin{figure}[ht!]
\begin{center}
\includegraphics[height=.34\textwidth, angle =0 ]{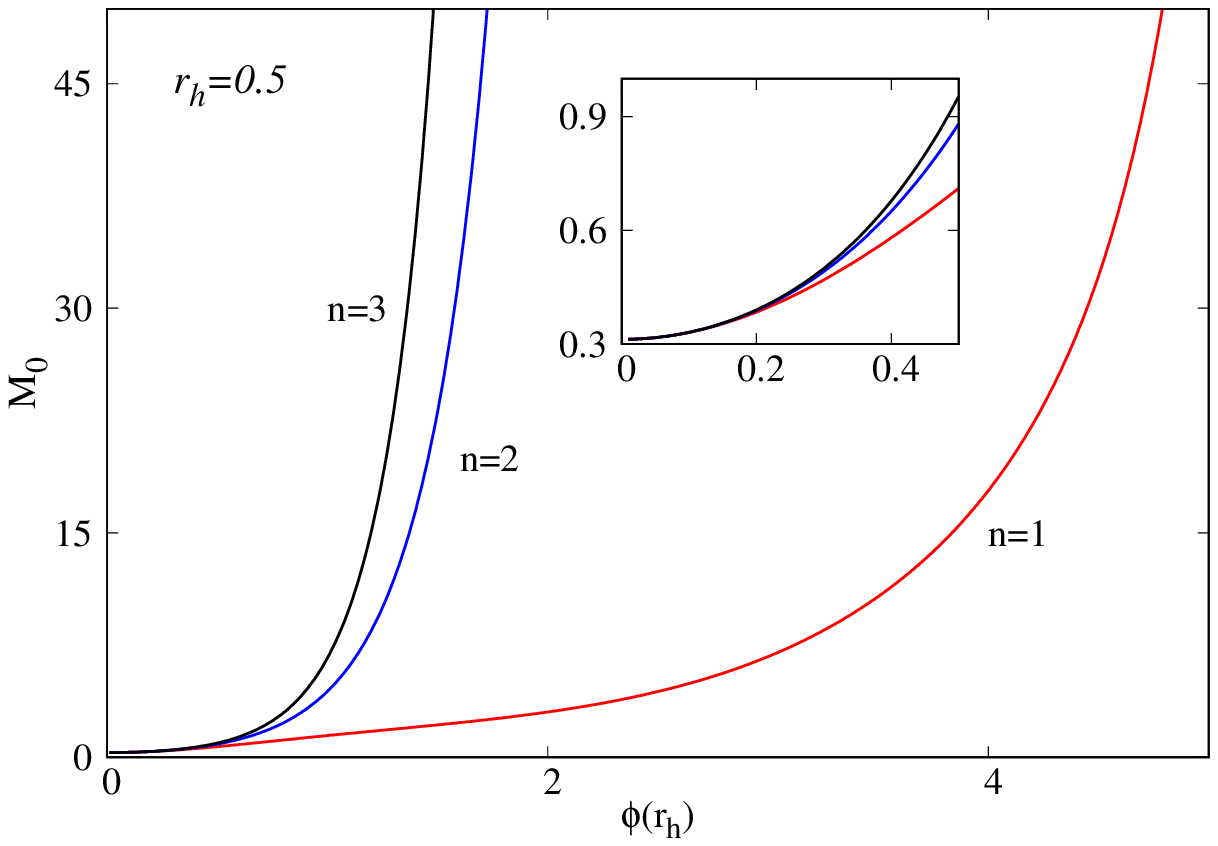} 
\includegraphics[height=.34\textwidth, angle =0 ]{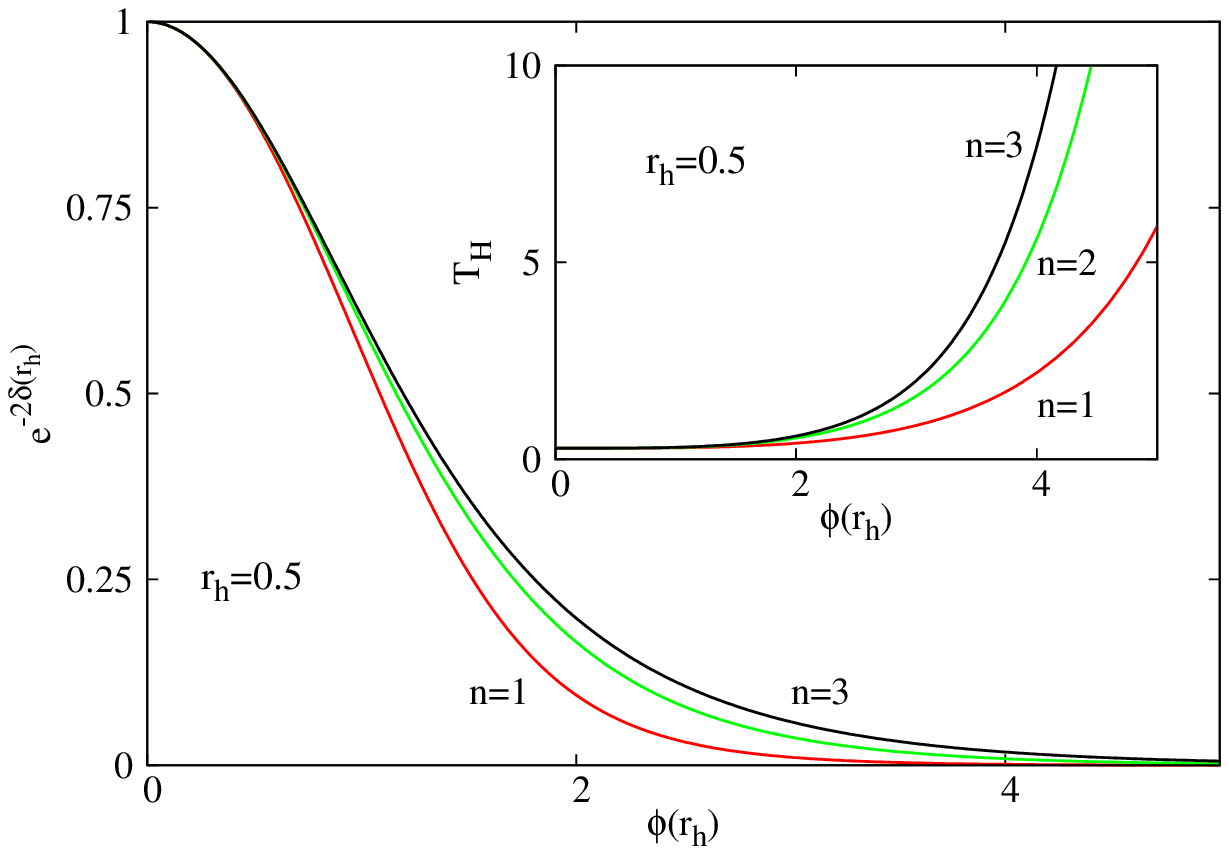}
%
\end{center}
\caption{ 
The mass-parameter $M_0$,  the value of the function $e^{-2\delta}$
at the horizon and the Hawking temperature are shown 
as a function of $\phi(r_h)$
for families of $n=1,2,3$ black hole solutions
with a fixed value of the horizon radius.
}
\label{MT-sugra123}
\end{figure}
%
\begin{figure}[ht!]
\centering
{ 
\includegraphics[height=.34\textwidth, angle =0 ]{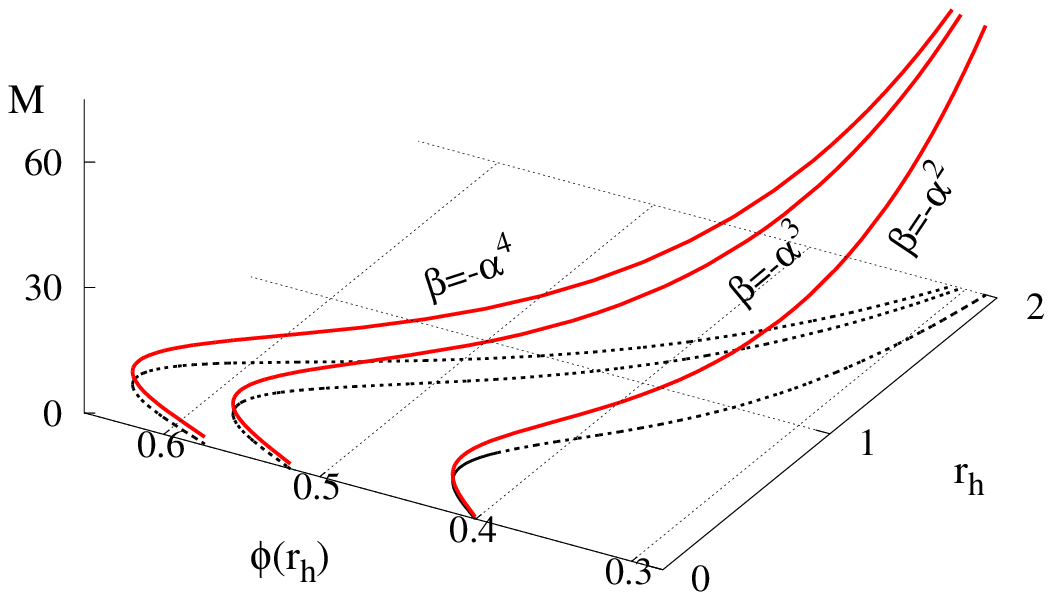}
\includegraphics[height=.34\textwidth, angle =0 ]{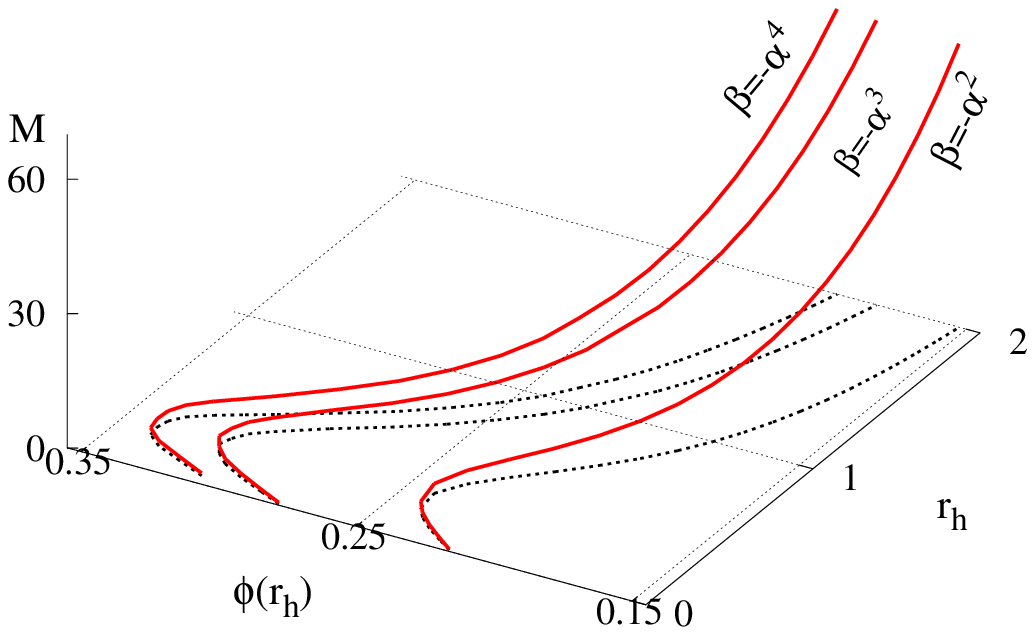}
}
\caption{ 
The mass of the solutions is shown as a function of the values of the event horizon
radius and  the scalar field at the horizon
for sugra solutions with $n=1$ (left panel) 
 and for $\lambda=-3$ solutions of the $\phi^4$ model  
(right panel),
with three
different choices of
the boundary condition 
 $\beta(\alpha)$.
}
\label{various-W}
\end{figure}

\subsubsection{ Nonperturbative results } 
 
The nonperturbative solutions are 
found by integrating 
 numerically
the equations 
(\ref{eomeq}).
In our approach,
suitable
initial conditions
resulting 
from
(\ref{r=0}), (\ref{r=rh}), 
are imposed  at $r=r_0+10^{-6}$
(with $r_0=(0,r_h)$ for solitons and BHs respectively),
for global tolerance $10^{-15}$,
the equations being integrated
towards $r\to \infty$.

In principle, the full set of solutions
can be scanned in this way by varying the
 boundary data at $r=r_0$ (which is provided by $\phi(r_0)$)
and extracting from the numerical output
the parameters $(\alpha,\beta,M_0)$ in the far field,
together with $\delta(r_0)$.

The profile of a typical soliton solution is
shown in Figure  \ref{profile-soliton-sugra13}
for $n=1$ (left panel)
and $n=3$ (right panel).
Both configurations have the value 
of the scalar field
at the origin, $\phi(0)=0.65$;
however,  some features of the solutions
depend on the value of
 $n$
(one finds $e.g.$
$\alpha=0.9294$,
$\beta=-0.5226$
for $n=1$
and
$\alpha=1.1263$,
$\beta=-1.0134$
for $n=3$).
A similar picture is found for BHs, as shown in Figure
\ref{profile-BH-sugra13}
for $n=1,3$ solutions
 with $\phi(r_h)=1.4$ and $r_h=2.3$
(in which case 
$e.g.$
$\alpha=5.234$,
$\beta=-2.919$
for $n=1$
and
$\alpha=11.3536$,
$\beta=-74.9321$
for $n=3$).
%
\begin{figure}[t]
\centering
{ 
\includegraphics[height=.34\textwidth, angle =0 ]{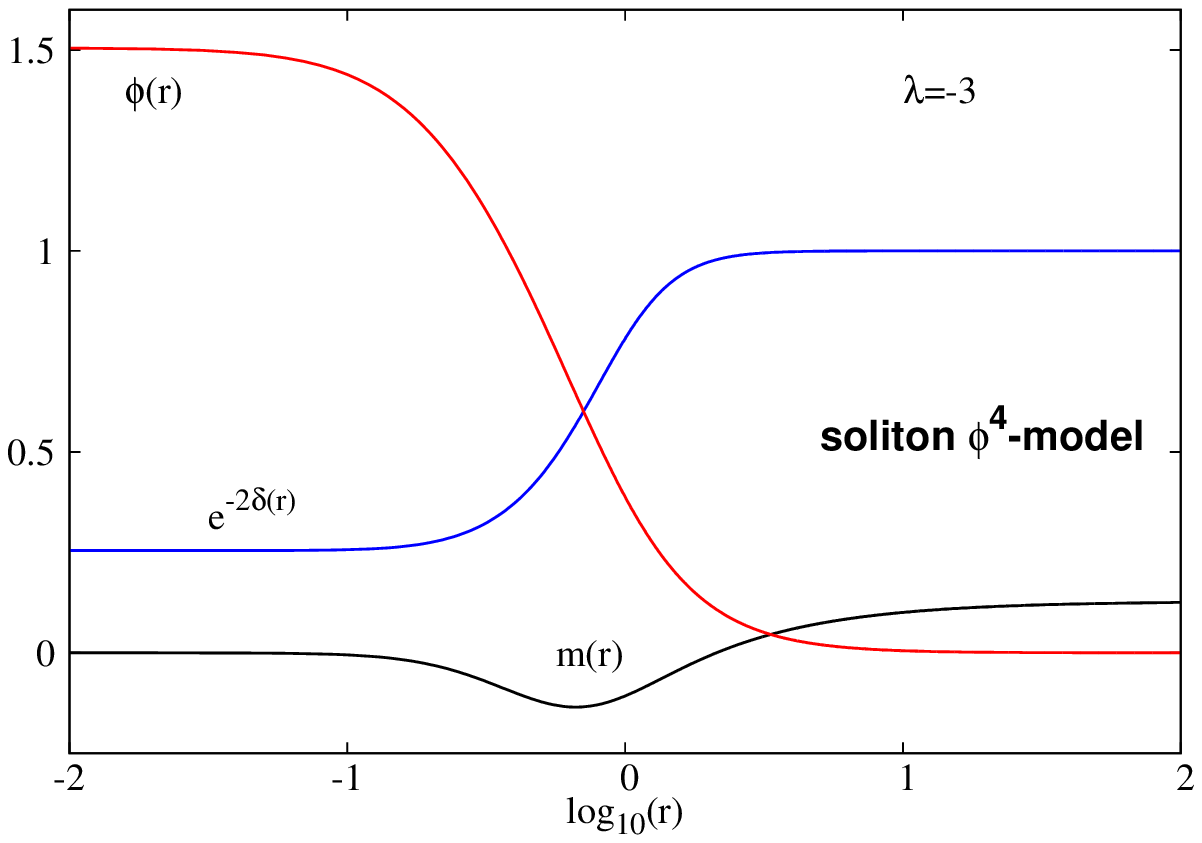}
\includegraphics[height=.34\textwidth, angle =0 ]{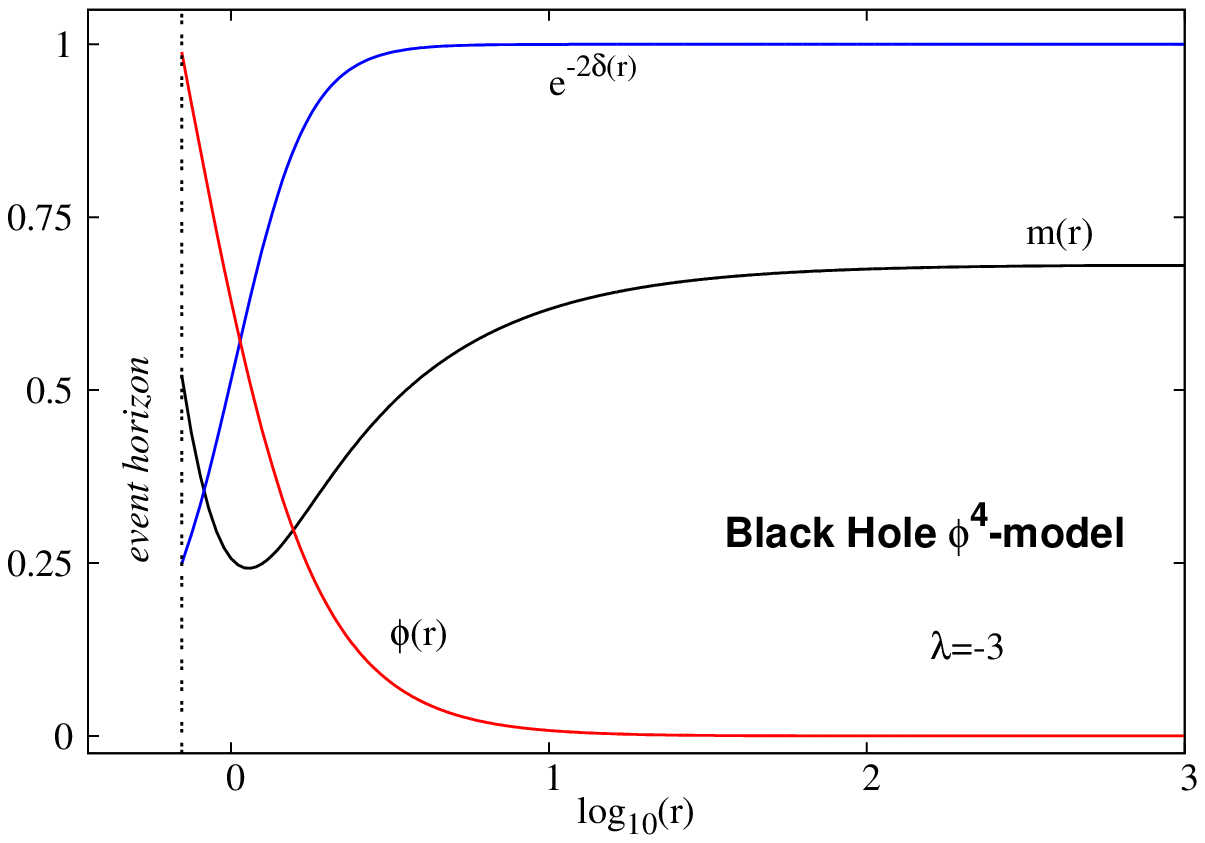}
}
\caption{ 
The profile of the nodeless soliton
of the $\phi^4$-model
is  shown together with a typical black hole  solution.
In both cases, the scalar field decays as $1/r^2$.
}
\label{profiles-phi4}
\end{figure}
%
Also,
for $n=1$
we have included the profile of the perturbative solution
with $\epsilon=0.65$; as one see,   this  provides a good approximation
of the non-perturbative result.
Moreover, 
the displayed profiles are typical and
 so far we could not find any indication 
for the existence of solutions with the function $\phi(r)$
changing sign ($i.e.$ with the existence of nodes).

In Figure \ref{quant123}
we show how the parameters 
$\alpha$,
$\beta$,
$M_0$
and $g_{tt}(0)$
vary with
$\phi(0)$
for soliton solutions with\footnote{We have considered as well solutions
with  $n=4,~5$  
and have found that they follow the $n=2,3$  pattern.}
$n=1,2,3$.
In particular, we remark
that $\alpha$
is always
strictly positive
(and increasing with $\phi(0)$)
while
$\beta<0$.
For $n=1$,
we have included also the corresponding perturbative results.
As one can see, they stop to be reliable when $\phi(0)$
becomes around one. 

Turning now to BH solutions,
we have considered a systematic 
scan of 
$n=1,2,3$
configurations 
by varying both $r_h$
and $\phi(r_h)$
in steps of $10^{-3}$.
The emerging picture is displayed in Figures 
\ref{alpha-beta-sugra123},
\ref{MT-sugra123}
and can be summarized as follows.
First, no configurations with
$\alpha \leq 0$
or 
$\beta \geq 0$
exist, at least for the considered
range of $(r_h,\phi(r_h))$
(note, however, the existence of  local extrema of these quantities).
Second, for any horizon size,
both the parameter $M_0$
and the Hawking temperature increase with $\phi(r_h)$.
Also, we mention that the  function $e^{-2\delta(r_h)}$
decreases monotonically with  $\phi(r_h)$
which makes the study of solutions
with large values of the scalar field at the horizon  difficult.

Finally, let us remark that the results in 
Figure 
\ref{alpha-beta-sugra123}
imply only rather weak restrictions on the
function
${\cal W}$
 which connects
$\alpha$ and $\beta$ 
in {\it designer gravity theories},
since 
all positive (negative) 
values of $\alpha$ 
$(\beta)$
are realized\footnote{
For example, 
the choice 
$\beta= f \alpha^k$ 
imposes only $f<0$.}.
In Figure (\ref{various-W})
we show the mass of $n=1$
BHs 
as a function of $(r_h,\phi(r_h))$
for several different functions
$W$ (which corresponds to
consider
 specific
slices in the general plots above).
As one can see, the minimal 
value of $M$
is achieved in the solitonic limit,
with the existence of two solitons for the same
value of the scalar field at the horizon.
A similar picture has been found for $n=2,3$ solutions.

\subsection{ Solutions in the $\phi^4$-model} 
\label{phi4}

Some of the features above are shared by
the  gravitating solutions in the $\phi^4$-model.
For example, a continuum of solutions
is found again when varying the values of 
$r_h$ 
and
$\phi(r_h)$
 (or $\phi(0)$),
without any indication for the existence of an upper bound for
these parameters.
As with the sugra case,
the generic $\phi^4$-solutions have nonzero parameters $\alpha$,
$\beta$.
In Figure \ref{various-W} 
(right panel) we show 
the mass of $\lambda=-3$ BH solutions with 
three different
 choices of
the condition $\beta(\alpha)$.
One can notice a (qualitatively) similar
picture to that found for solutions
with the $n=1$ potential (\ref{U}).

There are also a number of new specific properties, the
most interesting one being that
a scalar potential with  quartic self-interaction 
allows for solutions with 
$\alpha=0$
or 
$\beta=0$
in the asymptotic expansion (\ref{inf}).

\begin{figure}[ht!]
\centering
{ 
\includegraphics[height=.34\textwidth, angle =0 ]{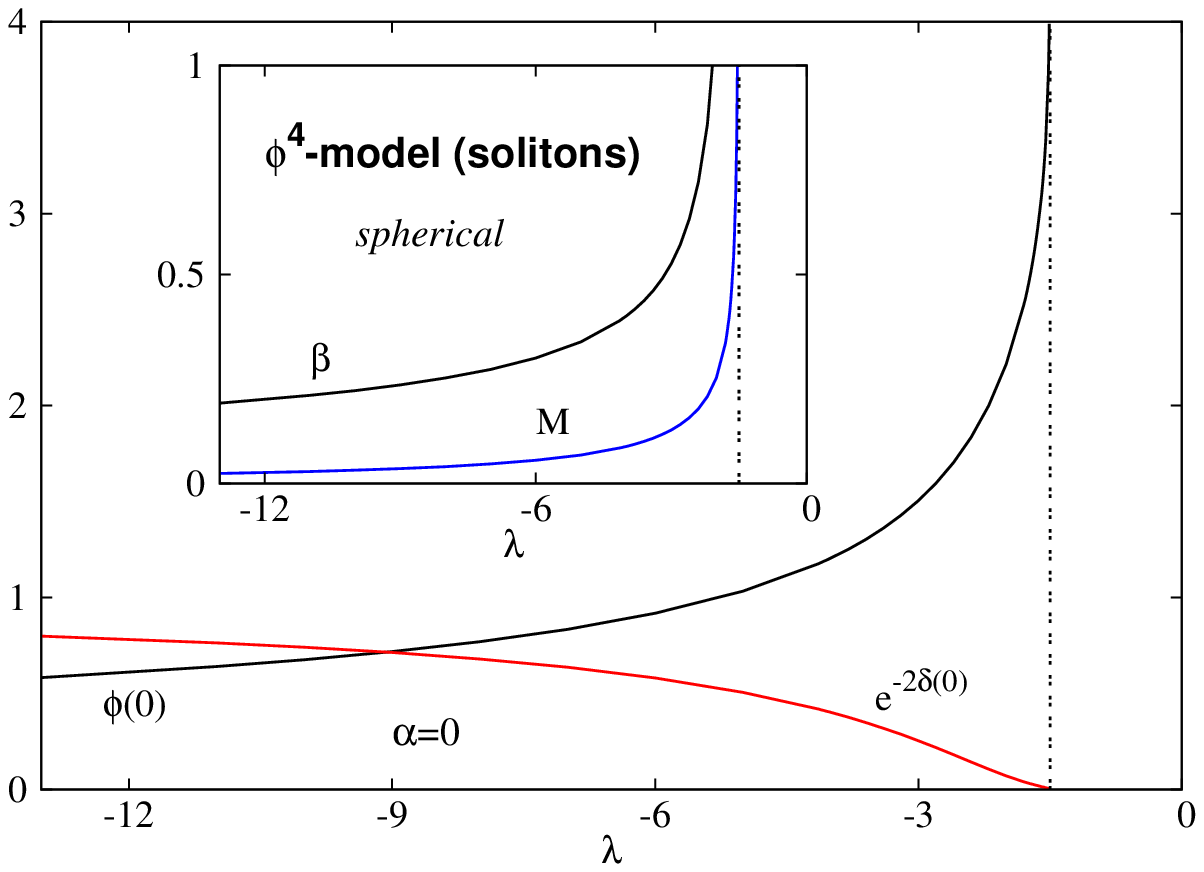}
\includegraphics[height=.34\textwidth, angle =0 ]{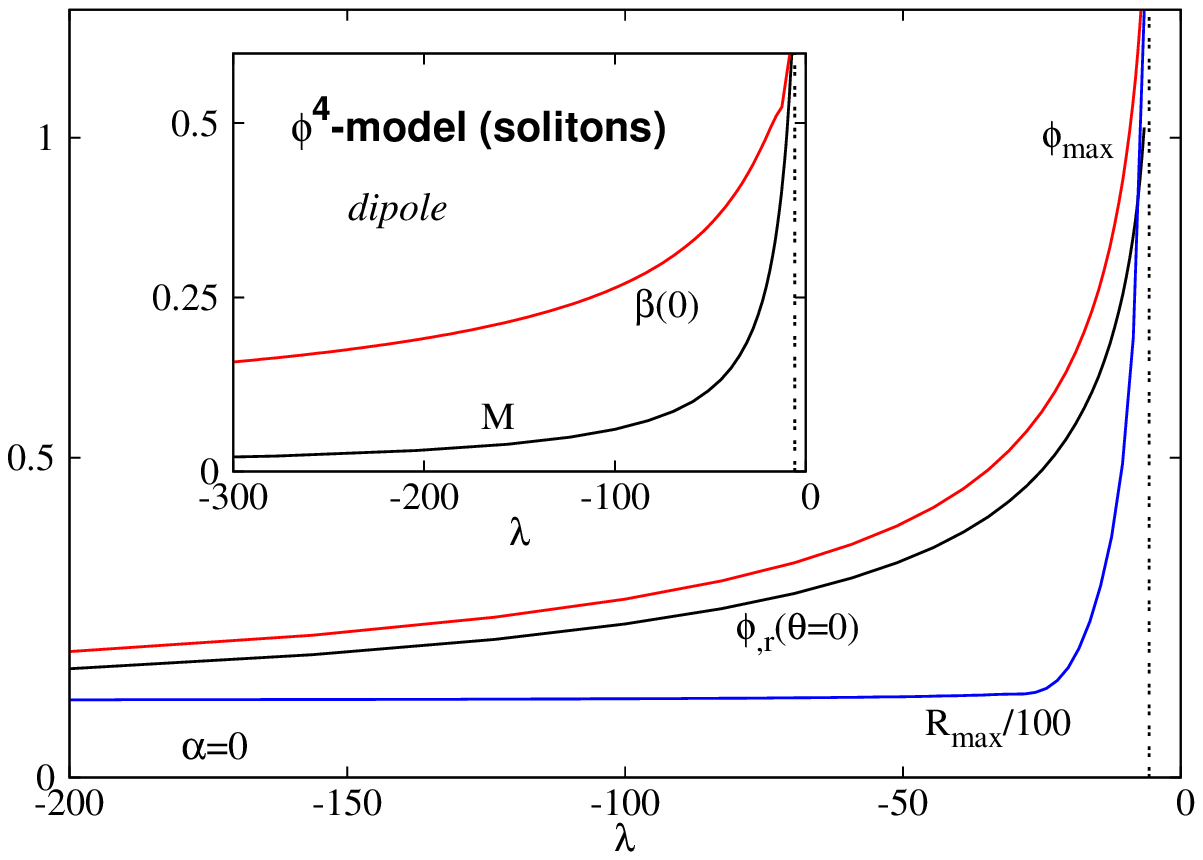}
}
\caption{ 
Several quantities of interest are shown for spherically
symmetric  
and axially symmetric (dipole) solitons  
of the $\phi^4$-model
as a function of the coupling constant $\lambda$.
 The scalar field decays
for these solutions
 as $1/r^2$.
}
\label{soliton-phi4-lambda}
\end{figure}
%
\begin{figure}[ht!]
\centering
{ 
\includegraphics[height=.34\textwidth, angle =0 ]{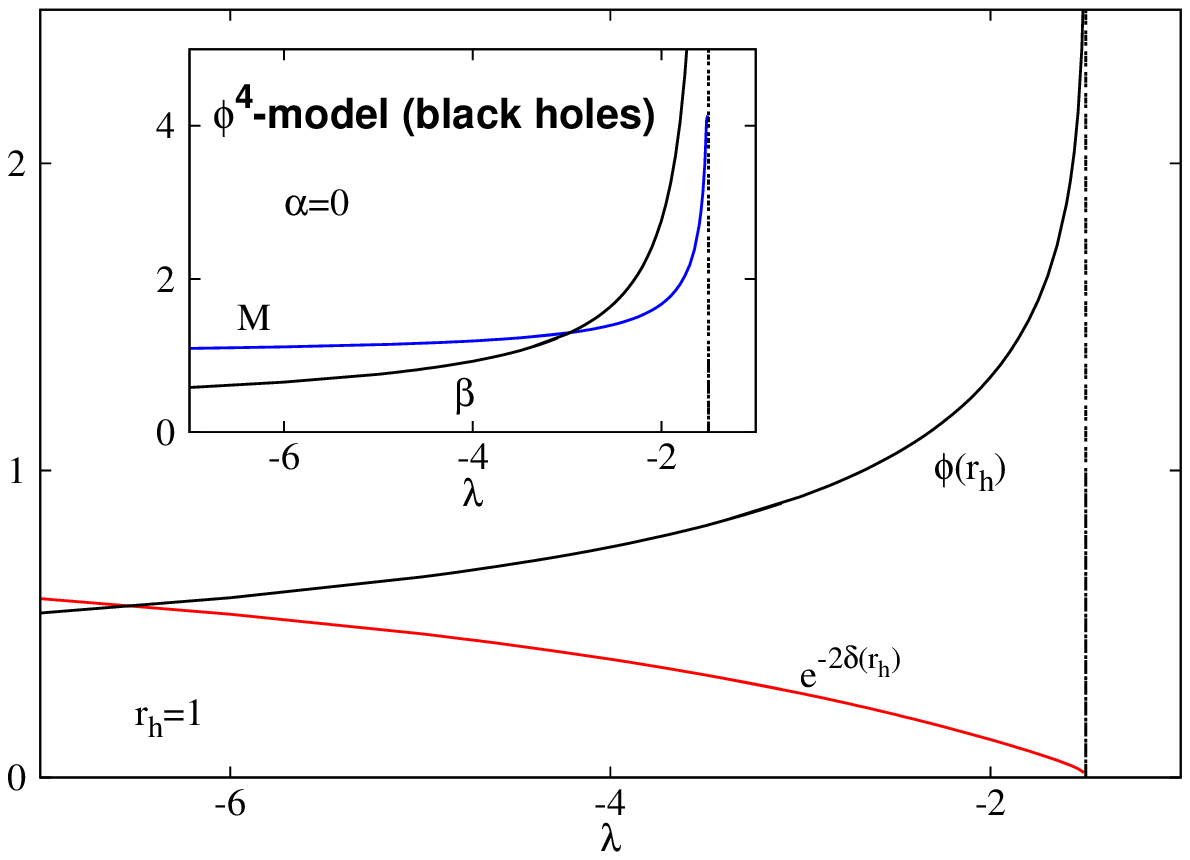}
\includegraphics[height=.34\textwidth, angle =0 ]{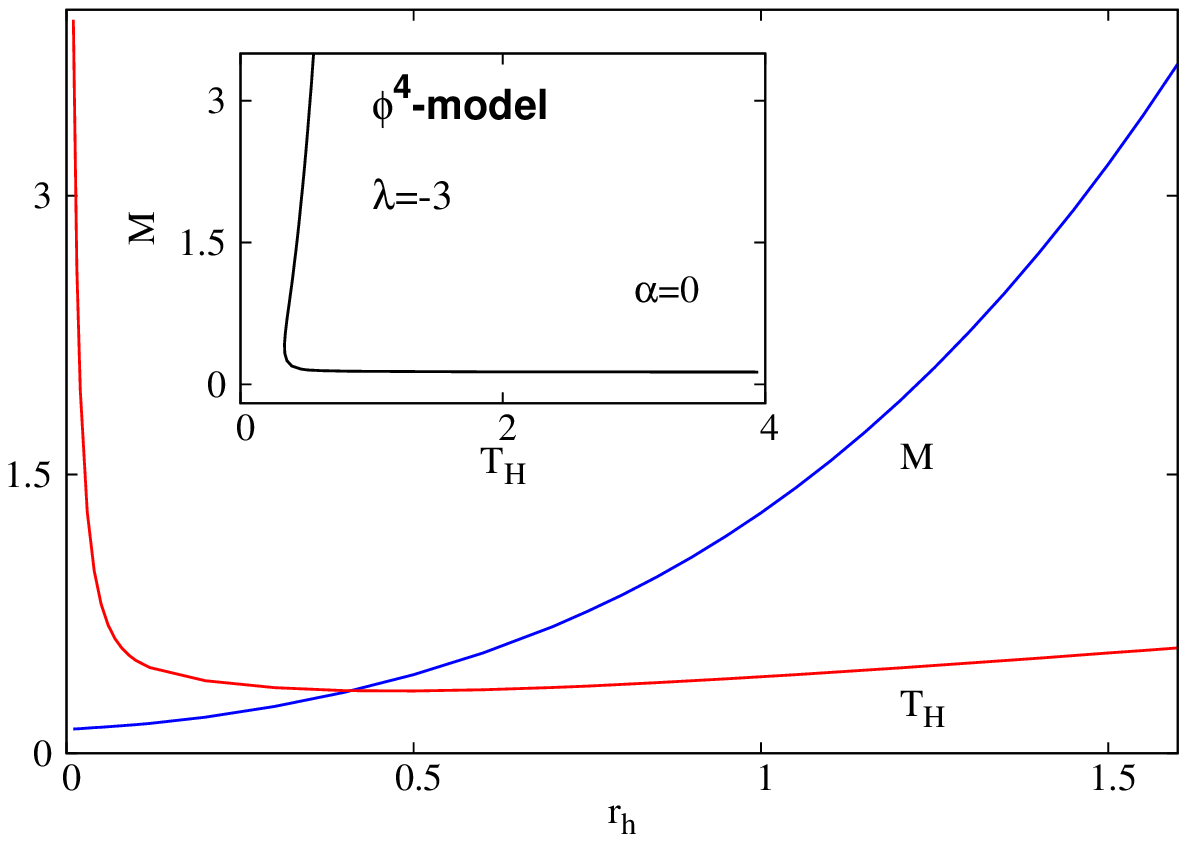} 
} 
\caption{ 
Several quantities of interest are shown for spherically
symmetric black hole solutions with $\alpha=0$  
of the $\phi^4$-model
as a function of the coupling constant $\lambda$
(left panel) and as a function of event horizon radius 
(right panel). 
}
\label{quant-BH-phi4}
\end{figure}

In what follows we shall restrict our study to configurations with $\alpha=0$,
in which case the mass function $m(r)$ approaches
a constant value at infinity.
The profile of two typical solutions are shown 
in Figure \ref{profiles-phi4}
(note that $m'(r)<0$ for some range of $r$,
such that solutions violate the weak energy conditions).
In Figure \ref{soliton-phi4-lambda}
(left panel)
we show how several quantities of interest
vary as a function of $\lambda$ for $\phi^4$-solitons.
No solutions with $\alpha=0$ were found 
for $\lambda>0$,
while
our results  suggest\footnote{Note that this
is smaller 
than the
 value  $\lambda=-1/(3n)$
found for the truncation of the sugra-potential (\ref{U});
 the absence of configurations
with
$\alpha=0$
or 
$\beta=0$
in the sugra-case
can presumably be attributed to the fact that
the quartic \textit{effective} term in the potential (\ref{U})
never becomes dominant.
} 
the existence of a maximal value of $\lambda$,
with
$|\lambda|\geq 1.5$.
As $|\lambda|_{min}$ is approached, the function 
$e^{-2\delta (0)}$
takes very small values close to zero,
and the Ricci scalar appears to diverge.

As seen in the left panel of Figure  
\ref{quant-BH-phi4}, 
a similar picture is found 
in the presence of a BH horizon,
without the existence of an upper bound on the horizon size
(we mention that the same picture was found  
for other values of $r_h$).
In Figure  
\ref{quant-BH-phi4} (right panel)
we show the result for solutions with a fixed value of 
$\lambda=-3$
and a varying horizon size.
One can see that the familiar SAdS thermodynamics
is recovered for $\alpha=0$
BHs with scalar hair,
with the existence of two branches of solutions which join for a minimal
value of the Hawking temperature.

\section{Beyond spherical symmetry: gravitating scalar dipoles } 
\label{dipole}

On general grounds, one expects that
each (linear) AdS scalar cloud
 with
given numbers $(\ell,m)$ 
would
possess nonlinear continuations in the full 
Einstein-scalar field model (and 
thus, the spherically symmetric ($\ell=0$ mode) 
case discussed above is not special).
In what follows we present  results for the
simplest case of (axially symmetric) scalar dipoles
(note, however the perturbative construction of the
quadrupole solution in the Appendix B).
Such configurations are first  constructed 
within a perturbative approach,
by considering the backreacting version 
of the $\ell=1$ (linear) mode in Section \ref{linear-clouds}.
Non-perturbative solutions of
the Einstein-scalar field equations
with a $1/r^2$
far field decay of the scalar field 
are constructed in the $\phi^4$-model.

\subsection{Perturbative results} 
 \label{pert-dipole}

In constructing perturbatively axially symmetric
solutions
it is convenient
to consider
a generalization of
 the pure $AdS$ line element~\eqref{ads} 
with three unknown functions $F_i$,  
 \be
\label{nex1}
ds^2=-F_1(r,\theta)N_0(r)dt^2+F_2(r,\theta)\frac{dr^2}{N_0(r)}+F_3(r,\theta) r^2 
\left(
d\theta^2+\sin^2\theta  d\varphi^2
\right) ,
\ee
with
\be
N_0(r)=1+\frac{r^2}{L^2}.
\ee
The scalar field only depends 
on $r,\theta$,
with the following perturbative 
ansatz
 up to order $\mathcal{O}(\epsilon^3)$: 
 \begin{eqnarray}
\label{nex2}
\phi(r,\theta) =\epsilon \phi^{(1)}(r,\theta) +  \epsilon^3 \phi^{(3)}(r,\theta)+\dots\ , 
\end{eqnarray}
where $\phi^{(1)}(r,\theta)$ is a linear scalar  
on $AdS$ studied in Section \ref{linear-clouds}
and $\epsilon$ is an infinitesimally small parameter.
 The backreaction of the scalar field on the geometry 
is taken into account by
defining 
(with $i=1,2,3$)
%
 \begin{eqnarray}
\label{nex3}
 F_i(r,\theta) =1+\epsilon^2 F_{i2}(r,\theta)
+\dots. 
\end{eqnarray}
Then the coupled Einstein--scalar field equations are solved order by order
in $\epsilon$,
the constants which enter the solution
being fixed by imposing regularity at 
$r=0$
and AdS asymptotics.

To illustrate this procedure, 
let us consider the backreaction on the geometry of a scalar dipole cloud
(similar results for the $\ell=2,m=0$ case
are given in the Appendix \ref{quadrupole}).
Thus the lowest order data is
 \begin{eqnarray}
\phi^{(1)}(r,\theta)= R_1(r)\cos \theta~~{\rm with}~~
R_1(r)=\frac{L}{r}(1-\frac{L}{r}\arctan(\frac{r}{L})) .
\end{eqnarray}
%
Then,
 the perturbed metric solution is constructed by considering 
 an angular expansion of $F_i$  in 
terms of
Legendre functions, with coefficients given by radial functions.
To lowest order one takes the consistent ansatz
 \begin{eqnarray}
\nonumber
F_{i2}(r,\theta)=\kappa^2 \left(a_{i}(r)+\mathcal{P}_2(\cos\theta)b_{i}(r) \right).
\label{nex31} 
\end{eqnarray}
In solving the Einstein equations, one uses a residual gauge freedom to set the radial function 
$a_3=0$,
the expressions of the other functions being 
(we recall ${\cal X}=\arctan (r/L)$)
\begin{eqnarray}
\nonumber
&&
a_1(r)=\frac{\pi^2}{6}-\frac{2}{3N_0(r)}
-\frac{4r}{L N_0(r)}(1+\frac{3L^2}{4r^2}){\cal X}(r)
-\frac{2}{3}({\cal X}(r))^2,
\\
\nonumber
&&
a_2(r)= -\frac{L^2}{3 r^2}(1+\frac{1}{N_0(r)})
+\frac{4L^3}{3 r^3}{\cal X} (r)
\left(
(1+\frac{3r^2}{4 L^2})\frac{1}{N_0(r)}
-\frac{L}{2r}{\cal X}(r)
\right),~~
\\
\label{dipole-exact}
&&
b_1(r)= \frac{2L^2}{9 L^2}
\left(
\frac{1}{N_0(r)}-7 +3{\cal X}(r)
({\cal X}(r)+\frac{2L}{r})
\right),
\\
\nonumber
&&
b_2(r)=\frac{2}{9 N_0(r)}+\frac{4L^3}{3r^3}{\cal X}(r)
\left(1-\frac{L}{2r}(1+N_0(r)){\cal X}(r)\right),
\\
\nonumber
&&
b_3(r)=-\frac{10}{9}+\frac{L^2}{r^2}+\frac{2L }{3r}(1-\frac{L^2}{r^2}){\cal X}(r)
+\frac{1}{3}(1-\frac{L^4}{r^4})({\cal X}(r))^2.
\end{eqnarray}
Moving now to the next order in $\epsilon$,
we shall restrict again to the case $n=1$
in the potential (\ref{U}).
The correction induced by
the metric corrections to the scalar field are found by taking  
a (consistent) ansatz for
$\phi^{(3)}$
 with two unknown functions,
 \begin{eqnarray}
\label{dipole-nlo}
\phi^{(3)}(r,\theta)= 
\phi_{31}(r) \mathcal{P}_{1}(\cos\theta) +\phi_{33}(r)  \mathcal{P}_{3}(\cos\theta).
\end{eqnarray} 
The explicit form of the functions 
$\phi_{31}(r)$ and 
$\phi_{33}(r)$
is given in the Appendix  
\ref{dipole-appendix}.
In deriving it,
we impose them to be regular  at $r=0$ 
and to decay as $1/r^2$ in $r\to\infty$. 
As such,
the expansion parameter $\epsilon$
can be identified with the function 
$\alpha$
(evaluated at $\theta=0$)
that enters the far field 
expansion
(\ref{inf})
of the scalar field,
and thus
\begin{eqnarray}
\alpha (\theta)= \epsilon L \cos \theta. 
\end{eqnarray}
The corresponding expression of $\beta$ is 
\begin{eqnarray}
\beta (\theta)
= \left(\frac{L^2 \pi}{2} \epsilon + \bar \beta_1  \epsilon^3  \right) \cos \theta   +
\bar \beta_3 \epsilon^3  L_3(\cos \theta) ,
\end{eqnarray}
where we denote
\begin{eqnarray}
&&
\nonumber
\bar \beta_1=\frac{8L^2 \pi}{25}
\left(
1+ \frac{\pi^2}{32}(-4+\log (256)-\frac{9}{8}\zeta(3)
+\frac{6403 \kappa^2}{2016} 
(-1+\frac{\pi^2}{6403}(801-1272 \log (2))+\frac{5724}{6403} \zeta(3))
\right),
\\
\nonumber
&&
\bar \beta_3= \frac{ L^2 \pi}{7000}
\left(
61+72 \zeta(3)-\frac{29 \pi^2}{12}(1+\frac{192}{29} \log(2) )
+\kappa^2 (-\frac{94}{3}+\frac{\pi^2}{8} (-47+384 \log(2)-216 \zeta(3)) )
 \right).
\end{eqnarray}
The presence of the term proportional 
with
$\mathcal{P}_{3}(\cos\theta)$
in the far field expansion of $\beta$ above 
indicates that,
 to order $1/r^2$,
the asymptotic behaviour of the scalar field
deviates from that of a dipole.

Different from the spherical case,
we were not able to solve the equations
to higher order in $\epsilon$.
However, likely the above solution  already captures
same basic features of the general configurations.
One finds, $e.g.$,
 \begin{eqnarray}
-g_{tt}(0)=1-\frac{1}{6}(10-\pi^2)\kappa^2 \epsilon^2,
\end{eqnarray}
while the leading order terms in the large-$r$ expressions 
of the metric potentials are
 \begin{eqnarray}
\nonumber
&&
 g_{rr} =\frac{L^2}{r^2}-
\left(
1+\frac{1}{24}(\pi^2+\frac{20}{3}+(3\pi^2-4)\cos 2\theta)\kappa^2 \epsilon^2
\right)
\frac{L^4}{r^4}+\dots,~~
\\
\nn
&&
g_{\varphi \varphi}=
\sin^2 \theta g_{\theta \theta}
=
\left(
1+\frac{\kappa^2 \epsilon^2}{24}(\frac{40}{3}-\pi^2)(1-3\cos^2\theta))
\right) r^2
+\frac{1}{6} L^2 (1+3\cos 2\theta)\kappa^2 \epsilon^2+\dots,
\\
\label{xcz}
&&
g_{tt}=-\frac{r^2}{L^2}-
\left(
1+\frac{1}{72}
(3\pi^2-28)(1+3\cos 2\theta)
\kappa^2 \epsilon^2
\right)
+\frac{\kappa^2 \epsilon^2 \pi L}{18 r}+\dots.
\end{eqnarray}

Also, 
 the non-vanishing components of
the 
boundary stress tensor, as computed 
by using the prescription in Section \ref{mass}
are  
 \begin{eqnarray}
{\rm T}_\theta^\theta
={\rm T}_\varphi^\varphi=
-\frac{\pi L^2 (2+3\cos 2\theta) }{24}\frac{\epsilon^2}{r^3}+\dots,\qquad
{\rm T}_t^t=
 -\frac{\pi L^2 }{12}\frac{\epsilon^2}{r^3}+\dots.
\end{eqnarray}
Then, to order 
$\epsilon^2$, 
the (cubic) term 
(which is multiplied with the function $W$)
in the scalar counterterm (\ref{Tsup})
 does not show up,
and one finds the following simple
expression for the
 mass of the gravitating dipole solution 
 \begin{eqnarray}
M=\frac{L  \pi^2}{3}\epsilon^2 ,
\end{eqnarray}
(where we choose $\partial M$
to be a   surface at constant $r$,
while $n_{\nu}=\delta_{\nu r}\sqrt{F_2/N_0}$).

\subsection{Non-perturbative solitons in the $\phi^4$-model }

\subsubsection{The framework}

The (axially symmetric)
non-perturbative solutions 
are constructed by employing the Einstein-De Turck approach
\cite{Headrick:2009pv}, 
\cite{Adam:2011dn}.
 Therefore, instead of the Einstein equations, 
we solve the so called Einstein-DeTurck (EDT) equations
%
\begin{eqnarray}
\label{EDT}
R_{ab}-\nabla_{(a}\xi_{b)}=-\frac{3}{L^2}g_{ab}+
2\kappa^2 (T_{ab}-\frac{1}{2}T  g_{ab})~,
~~{\rm with}~~\xi^a=g^{bc}(\Gamma_{bc}^a-\bar \Gamma_{bc}^a), 
\end{eqnarray} 
 $\Gamma_{bc}^a$ being the Levi-Civita connection associated to the
spacetime metric $g$ that one wants to determine.
Also, a  reference metric $\bar g$ is introduced, with
$\bar \Gamma_{bc}^a$ the corresponding Levi-Civita connection.
Solutions to (\ref{EDT}) solve the Einstein equations
iff $\xi^a \equiv 0$ everywhere on
${\cal M}$.
To achieve this,
we shall impose boundary conditions  which are compatible with
$\xi^a = 0$
on the boundary of the domain of integration. 

Within this approach, the (static, axially symmetric)
metric Ansatz is more complicated than the perturbative one, eq. (\ref{nex1}), 
with five metric functions 
\begin{eqnarray}
\label{metric}
ds^2=-f_0(r,\theta)N(r)dt^2
+f_1(r,\theta)\frac{dr^2}{N(r)}+S_1(r,\theta)(rd\theta+S_2(r,\theta)dr)^2
+f_2(r,\theta)r^2\sin^2\theta  d\varphi^2.\nn\\
\end{eqnarray}
For solitons with a $1/r^2$ decay of the scalar field
(the only considered case),
the obvious reference metric is AdS  spacetime, while
the numerics is done with
a scalar field Ansatz  
\begin{eqnarray}
 \phi=\frac{\psi(r,\theta)}{r},
\end{eqnarray}
such that a vanishing $\psi$
as $r\to \infty$ corresponds to $\alpha=0$
in (\ref{inf}).

Then 
the EDT equations (\ref{EDT}) together with scalar field equation 
result in a set of  six elliptic partial differential equations,
 which are solved numerically
as a boundary value problem.
Following the standard approach  \cite{Dias:2015nua}, 
the boundary conditions are found by constructing an approximate form
of the solutions on the boundary of the domain
of integration  compatible with the requirement $\xi^a = 0$. 
They read
\begin{eqnarray}
\nonumber
&&
\partial_r f_1\big|_{r=0}=\partial_r f_2\big|_{r=0}=\partial_r f_0\big|_{r=0}
=\partial_r S_1\big|_{r=0}=\partial_r S_2\big|_{r=0}=0, ~~\psi|_{r=0}=0,
\\
\nn
&&
\partial_\theta f_1\big|_{\theta=0,\pi}=\partial_\theta f_2\big|_{\theta=0,\pi}=
\partial_\theta f_0\big|_{\theta=0,\pi}=\partial_\theta S_1\big|_{\theta=0,\pi}
=S_2\big|_{\theta=0,\pi}=0,~~\partial_\theta\psi \big|_{\theta=0,\pi}~,
\\
\nn
&&
 f_1\big|_{r=\infty}= f_2\big|_{r=\infty}=
 f_0\big|_{r=\infty}=1,~~ S_1\big|_{r=\infty}=
=S_2\big|_{r=\infty}=0,~~
\psi \big|_{r=\infty}=0.
\end{eqnarray} 
Moreover, we shall assume again that the solutions are
symmetric $w.r.t.$ a reflection in the
equatorial plane, which implies
that 
the functions $f_1,f_2,f_0,S_1$
satisfy Neumann boundary conditions at $\theta=\pi/2$
while $S_2$ 
and
$\psi$ 
vanish there.  
It is also of interest to display the
  far field behaviour of solution
\begin{eqnarray}
\label{far-field}
&&
\phi=\frac{\beta(\theta)}{r^2}+O(1/r^4),
~~
 f_0=1+\frac{f_{03}(\theta)}{r^3}+ O(1/r^4),~~
f_1= 1+\frac{24 \pi G \beta(\theta)^2}{r^4}+O(1/r^5),~~
 \\
 &&
 \nonumber
 f_2=1+\frac{f_{23}(\theta)}{r^3}+ O(1/r^4),~~
 S_1=1-\frac{f_{03}(\theta)+f_{23}(\theta)}{r^3}+ O(1/r^4),~~
 S_2= O(1/r^5),
\end{eqnarray}
the functions 
$\beta(\theta)$
and $f_{03}(\theta)$,
 $f_{23}(\theta)$,
 $s_{13}(\theta)$
being determined from the numerics. 
 
One finds in this way the following large-$r$ expressions of
the non-vanishing components of the boundary stress tensor
(note the absence of a contribution from the scalar counterterm (\ref{Tsup})):
\begin{eqnarray}
\label{boundary-Tij}
T_{\theta}^\theta=-\frac{3}{4\kappa^2 L}\frac{(f_{03}(\theta)+f_{23}(\theta))}{r^3}+\dots,
~~
T_{\varphi}^\varphi = \frac{3}{4\kappa^2 L}\frac{ f_{23}(\theta)}{r^3}+\dots,
~~
T_{t}^t= \frac{3}{4\kappa^2  L}\frac{ f_{03}(\theta)}{r^3}+\dots,\nn\\
\end{eqnarray}
which is  traceless, as expected.
Then a straightforward computation leads to the following expression for
the  mass:
\begin{eqnarray}
\nonumber
M=\frac{3\pi }{2\kappa^2 L^2}
\int_0^\pi d\theta   \sin \theta  f_{03}(\theta) .
\end{eqnarray}

\subsubsection{Numerical results}

In this approach, the only input parameter is 
$\lambda$,
the constant of the quartic self-interaction.
Instead of $r$, the numerics is done by using a compactified
radial coordinate $x=r/(1+r)$,
 the  equations being discretized on a ($x,~\theta$) grid with around $250\times 50$ points.
Then the resulting system is solved iteratively until convergence is achieved.
The typical  numerical error
for the solutions reported in this work is estimated to be of the order of $10^{-4}$
(also, the order of the difference formulae  was 6).

The profile of 
the typical scalar field,
the function
$\beta(\theta)$ and 
the energy density
$\rho=-T_t^t$
are 
(qualitatively)
similar to those displayed in Figure \ref{profile-phi4-probe}
for solutions in the probe limit.
As expected, the $\phi^4$-(AdS probe) solution 
with $\alpha=0$ found in Section \ref{phi4-probe}
possesses gravitating generalizations.
The resulting picture shares the basic features
found for spherically symmetric solitons,
see Figure \ref{soliton-phi4-lambda}.
The solutions with a $1/r^2$ decay
 exist up to a minimal value of 
$|\lambda|$
(while again no such solutions are found for $\lambda>0$).
As the minimal value of
$|\lambda|$
is approached,  both the mass and $\beta(0)$
increase, while the numerics become increasingly challenging,
with large numerical errors.

Also, the solutions appear to exist for arbitrary large 
values of 
$|\lambda|$.
To understand this limit, 
one notes that these
Einstein-scalar field
solutions can 
also be constructed
 by using 
an alternative scaling, with
$
\lambda \to \lambda c^2,
$
$
\phi \to \phi/c,
$
and
$
\kappa^2 \to \kappa^2 c^2,
$
with $c$ an arbitrary nonzero constant.
This can be used to set $\lambda=-1$,
and work instead
with the  following form of the EDT equations
$
R_{ab}-\nabla_{(a}\xi_{b)}={3}g_{ab}/{L^2}+
2\bar \kappa^2 (T_{ab}-\frac{1}{2}T  g_{ab})~,
$
with $\bar \kappa^2=\kappa^2/|\lambda|$.
As such, $\bar \kappa^2 \to 0$
corresponds 
 to solutions in the probe limit
(being approached for large values of $|\lambda|$).

We mention that the preliminary results
indicate the existence of BH generalizations
of these solutions,
with the presence, as in the probe limit in  Section \ref{phi4-probe},
of two branches of solutions 
which merge for a maximal value for the
horizon area.  
However, their study is more involved, being beyond the
purposes of this work. 
 
Returning to the solitonic case,
such solutions should exist 
as well
for
a pure $1/r$ asymptotic decay of the scalar field,
or, more generally,
with nonzero $\alpha$ and $\beta$ in Eq. (\ref{inf}).
However, so far we did not manage to adapt our numerical scheme 
to these cases. The obstacle  
is that
the EDT approach requires the choice of a suitable 
background metric $\bar g$, 
which is not obvious for a $1/r$ decay of the scalar field.
For example, when choosing AdS for $\bar g$,
we could not find a consistent far field expression 
of the solutions 
which is
 compatible with the requirement
$\xi^a = 0$.
This  obstacle has also prevented us to find 
non-perturbative solutions 
in the ${\cal N}=8$ model.

We also mention that no results were found when modifying
the code used in the $\phi^4$-model
 for a scalar potential given by (\ref{U}),
while keeping the same set of boundary condition,
which strongly suggests 
the absence of solutions with $\alpha=0$
in that case.

\section{ Discussion} 
\label{final}
 
The Einstein-scalar field system with 
mass $\mu^2=-2/L^2$
 in  AdS$_4$ spacetime 
provides an interesting toy model
to investigate the issues of  asymptotics and possible
boundary conditions,
together with the  existence of
scalar multipolar solutions. 
Moreover, for a suitable scalar potential,
 this model is a consistent truncation 
of ${\cal N}=8$ $D=4$
gauged supergravity \cite{Duff:1999gh},
this being the main case studied in this work.
Apart from this case, we have considered also a model with
a quartic self-interaction of the scalar field.

The main results can be summarized as follow.
First, both the perturbative and the numerical results
for the ${\cal N}=8$  model
strongly suggest that 
 no (soliton or BH) 
solutions can be found subject to  
the `standard' boundary conditions
$\alpha=0$
or
$ \beta=0$ 
(with
$\alpha$ and $\beta$
the parameters which enter
 the asymptotic scalar field expansion (\ref{inf})). 
As such, all solutions  of (the considered truncation)
of
the ${\cal N}=8$
model belong to 
{\it designer gravity} theories 
\cite{Hertog:2004ns}. 
 Then the existence of the relation
between $\alpha$ and $\beta$
 of the form (\ref{W}) 
is essential, from a physical point of view, 
for obtaining an integrable mass for the solutions.
The fact that the scalar self-interaction potential 
in the ${\cal N}=8$  gauged supergravity 
supports only mixed boundary conditions implies
that the bulk solution is consistent 
with RG flows generated in the dual field theory
by multi-trace deformations. 
In particular, the (marginal) triple-trace deformation 
is consistent with mixed boundary conditions 
that preserve the conformal symmetry, in which case $\beta \sim \alpha^2$.

However,  
this result depends on the precise form 
of the scalar field self-interaction.
As shown in this work, a different picture is
found for a scalar field with quartic self-interaction,
with the existence 
as well
of 
spherically symmetric solitons and BHs  with a 
 $1/r$
or
 $1/r^2$
asymptotic decay of the scalar field. 
 
 \medskip

In a different direction,
our results suggest that the 
spherically symmetric Einstein-scalar field  solitons 
are only the first member of
a discrete family of solutions,
which can be viewed as non-linear
continuations of the 
the linear scalar clouds in a fixed AdS background.
Moreover, similar configurations should
exist when adding a
BH horizon at the center of these solitons.
The main case studied in our work was that of (axially symmetric)
dipoles, where we have found both perturbative and  
non-perturbative results. 
However, we emphasize that the multipole structure in AdS is quite different than in flat spacetimes because all the multipoles come at the same order in AdS. 

 \medskip
As avenue for future research, we mention first
 the possible existence of 
configurations {\it without} isometries,
which would be the backreacting version
of the $m\neq 0$ scalar multipoles.
Moreover, already in the dipole case,
similar solutions were shown to exist in a model with U(1) fields.
Also, it would be interesting to consider a similar study for 
other AdS parametrizations (here we mention the existence
in the $n=3$  sugra-model
 of an
exact solution describing a BH with scalar hair\footnote{See also \cite{Anabalon:2017yhv}
for exact solutions in an extended supergravity model.},
whose event horizon is a surface of negative constant
curvature 
\cite{Martinez:2004nb}).
Finally, we conjecture the existence of 
(qualitatively) similar results 
for any value of the scalar field mass
above
the Breitenlohner-Freedman 
 bound \cite{Breitenlohner:1982jf}.

\section*{Acknowledgements}
D.A. was supported during this work by the Fondecyt grant 1200986. H.H. is grateful for support by the National Natural Science Foundation of China (NSFC)
grants No. 12205123 and by the Sino-German (CSC-DAAD) Postdoc Scholarship Program,2021 (57575640).
The work of E. R. is supported by the Fundacao para a Ci\^encia e a Tecnologia (FCT) project UID/MAT/04106/2019 (CIDMA) and by national funds (OE), through FCT, I.P., in the scope of the framework contract foreseen in the numbers 4, 5 and 6 of the article 23, of the Decree-Law 57/2016, of August 29, changed by Law 57/2017, of July 19. We acknowledge support from the project PTDC/FIS-OUT/28407/2017 and PTDC/FIS-AST/3041/2020.  
E.R. gratefully acknowledges the support of the Alexander von Humboldt Foundation.
We are also grateful to the DFG RTG 1620 \textit{Models of Gravity}.
This work has further been supported by  the  European  Union's  Horizon  2020  research  and  innovation  (RISE) programmes H2020-MSCA-RISE-2015 Grant No.~StronGrHEP-690904 and H2020-MSCA-RISE-2017 Grant No.~FunFiCO-777740. 
The authors would like to acknowledge networking support by the
COST Actions CA15117 {\sl CANTATA} 
and CA16104 {\sl GWverse}.

\appendix
 \setcounter{equation}{0}
\renewcommand{\theequation}{A.\arabic{equation}}

\section{The ${\cal N}=8$
$D=4$
gauged supergravity action: the Einstein-scalar field(s) truncation }

Among other results,
Ref.~\cite{Duff:1999gh}
shows the existence of a consistent truncation
of the bosonic sector 
of the gauged $N=8$ supergravity,
which, apart from the Einstein term,
contains
three scalar fields
$\phi^{(12)}$,
$\phi^{(13)}$,
$\phi^{(14)}$
and
four U(1) gauge fields
$F^{(C)}_{\mu \nu}$ ($C=1,\dots,4$).
Its (bulk) action reads (eq.~(2.11) in Ref.\cite{Duff:1999gh}):
\begin{eqnarray}
\label{L1}
I
&=&\frac{1}{4 \kappa^2} 
\int d^4 x
\sqrt{-g}\Bigl[R
-\frac{1}{e}\left(
(\partial_\mu\phi^{(12)})^2+
(\partial_\mu\phi^{(13)})^2+
(\partial_\mu\phi^{(14)})^2\right)-U(\phi)\\
&&\qquad-2\left(
e^{-\lambda_1}(F_{\mu\nu}^{(1)})^2+
e^{-\lambda_2}(F_{\mu\nu}^{(2)})^2+
e^{-\lambda_3}(F_{\mu\nu}^{(3)})^2+
e^{-\lambda_4}(F_{\mu\nu}^{(4)})^2
\right)\Bigr],\nonumber
\end{eqnarray}
with
the scalar potential 
\begin{equation}
U=-4g^2\left(\cosh{\phi^{(12)}}+\cosh{\phi^{(13)}}+\cosh{\phi^{(14)}}\right).
\end{equation}
(with 
$4\pi G=\kappa^2$
and
$2g^2=1/L^2$ for the notation in this work)
while  $\lambda_i$ are linear combination of the scalar fields, as given by
\bea
&&\lambda_1=-\phi^{(12)}-\phi^{(13)}-\phi^{(14)},~\qquad
\lambda_2=-\phi^{(12)}+\phi^{(13)}+\phi^{(14)},\nn\\
&&\lambda_3=\phi^{(12)}-\phi^{(13)}+\phi^{(14)},~\qquad
\lambda_4=\phi^{(12)}+\phi^{(13)}-\phi^{(14)}.\nn
\eea

Let us remark that
one can take consistently  
$F^{(C)}_{\mu\nu}=0$ 
and thus we are left 
with a model with
three gravitating scalar fields
 \begin{equation}
\phi^{(12)}\equiv 2\kappa \phi^{(1)},~
\phi^{(13)}\equiv 2\kappa \phi^{(2)},~
\phi^{(14)}\equiv 2\kappa \phi^{(3)}.
\end{equation}
The case of  only one nonzero scalar field 
$\phi^{(a)}$
results in the action (\ref{action})
with $n=1$
in  the potential (\ref{U}).
Let us assume now  that
two scalar fields
(for example $a=1,2$)
 are equal,
while the third one vanishes.
Then  the redefinition
 \begin{equation}
 \phi^{(1)}=\phi^{(2)}=\frac{\phi}{\sqrt{2}} ~ ,
\end{equation}
leads to the $n=2$ 
case in 
 (\ref{action}), (\ref{U}).
Finally, when all scalars are equal, 
  the sugra-model in Section \ref{sec_eq_motion_action}
	with $n=3$
is recovered via the redefinition
 \begin{equation}
 \phi^{(1)}=\phi^{(2)} =\phi^{(3)}=\frac{\phi}{\sqrt{3}} .
\end{equation}

Also,
it was pointed out in 
\cite{Martinez:2004nb}
 that, for 
a scalar field potential  (\ref{U})  with
$n=3$,
the model (\ref{action}) can be obtained via the field redefinition
from 
the action of a scalar field
conformally coupled to Einstein gravity with a negative
cosmological constant.

To clarify if this result holds for the general $n$-case, 
we consider the following 
tranformation in  (\ref{action})
\be
\hat{g}_{\mu\nu}=(1-\ft{\kappa^2}{n}\psi^2)^{-1}g_{\mu\nu}, \qquad \psi=\sqrt{\ft{n}{\kappa^2}}\tanh(\sqrt{\ft{\kappa^2}{n}}\phi).
\ee
Then the original action \eqref{action} becomes
\bea
\label{Sgen}
 S 
=\int \sqrt{-\hat{g}}
           \bigg(
\frac{1}{4\kappa^2}  (\hat{R}+\frac{6}{L^2})
+\ft{3\kappa^2\psi^2-n^2}{2n(n-\kappa^2\psi^2)}
\hat{\nabla}^a \psi \hat{\nabla}_a \psi
-\ft{\hat{R}}{4n}\psi^2
+\ft{(n-3)n }{n^2 L^2}\psi^2
+\ft{ (3-2n)\kappa^2}{2n^2 L^2}\psi^4
             \bigg) .\nn\\
\eea
It is obvious that the case   $n=3$ is special, 
with a simple form of the above expression 
\be\label{taction}
S=
\int \sqrt{-\hat{g}}\bigg(\ft{1}{4\kappa^2}(\hat{R}+\frac{6}{L^{2}})-
\ft{1}{2}\hat{g}^{\mu\nu}\nabla_a \psi \nabla_a \psi 
-\ft{1}{12} \hat{R}\psi^2-\ft{\kappa^2}{6L^2}\psi^4\bigg)~.
\ee
Also, 
this is the only case where the matter part in 
(\ref{Sgen}) (which includes also the $\hat{R}\psi^2$ term)
is conformally invariant.

\setcounter{equation}{0}
\renewcommand{\theequation}{B.\arabic{equation}}
\section{Details on the perturbative axially symmetric solutions }
\label{details}

\subsection{The general equations}
For the metric Ansatz (\ref{nex1}),
and $\phi\equiv \phi(r,\theta)$,
the 
Einstein-scalar field equations
\eqref{eqs} 
reduce to the following equations:
\bea
&&-\frac{3}{L^2}-\ft{2\kappa^2\sinh^2(\phi)}{L^2}+\ft{N+rN'-\kappa^2r^2N\phi'^2}{r^2F_2}+\ft{F_3(\cot(\theta)\dot{F_1}+\ddot{F_1})+F_1(\cot(\theta)\dot{F_3}+\ddot{F_3})}{2r^2F_1F_3^2}+\ft{r^2NF_1^2F_3'^2-F_2F_3\dot{F_1}^2}{4r^2F_1^2F_2F_3^2}
\nn\\
&&+\ft{rF_1N'F_3'+N(2F_3F_1'+(2F_1+rF_1')F_3')}{2rF_1F_2F_3}=0,
\nn\\
&&\ft{2NF_2\dot{F_1}+rN'(F_1\dot{F_2}-F_2\dot{F_1})}{4r^3NF_1F_2F_3}+\ft{\dot{F_2}-4\kappa^2r F_2\dot{\phi}\phi'}{2r^3F_2F_3}+\ft{(F_2\dot{F_1}+F_1\dot{F_2})(F_3F_1'+F_1F_3')}{4r^2F_1^2F_2F_3^2}+\ft{-F_3^2\dot{F_1}'+F_1(\dot{F_3}F_3'-F_3\dot{F_3}')}{2r^2F_1F_3^3}=0,
\nn\\
&&-\frac{3}{L^2}-\ft{2\kappa^2\sinh^2(\phi)}{L^2}+\ft{\dot{F_1}(2\cot(\theta)F_2+\dot{F_2})+F_1(-4\kappa^2F_2\dot{\phi}^2+2\cot(\theta)\dot{F_2})}{4r^2F_1F_2F_3}+\ft{(F_2\dot{F_1}+F_1\dot{F_2})\dot{F_3}}{4r^2F_1F_2F_3^2}+\ft{2N'+rN''}{2rF_2}
\nn\\
&&+\ft{3r N' F_1'+2N(2\kappa^2rF_1\phi'^2+F_1')}{4rF_1F_2}+\ft{N(F_1^2F_3F_2'F_3'+F_2(F_3^2F_1'^2+F_1^2F_3'^2))}{4F_1^2F_2^2F_3^2}-\ft{(rF_1N'+N(2F_1+rF_1'))F_2'}{4rF_1F_2^2}
\nn\\
&&+\ft{(2rF_1N'+N(4F_1+rF_1'))F_3'}{4rF_1F_2F_3}+\ft{N(F_3F_1''+F_1F_3'')}{2F_1F_2F_3}=0,
\nn\\
&&-\frac{3}{L^2}-\ft{2\kappa^2\sinh^2(\phi)}{L^2}+\ft{2N'+rN''+2\kappa^2 rN\phi'^2}{2rF_2}+\ft{\dot{F_1}(F_3\dot{F_1}+F_1\dot{F_3})}{4r^2F_1^2F_3^2}+\ft{\dot{F_1}\dot{F_2}+2F_2\ddot{F_1}+2F_1\ddot{F_2}}{4r^2F_1F_2F_3}-\ft{\dot{F_2}(F_3\dot{F_2}+F_2\dot{F_3})}{4r^2F_2F_3^2}\nn\\
&&+\ft{\kappa^2\dot{\phi}^2}{r^2F_3}-\ft{NF_1'(F_2F_1'+F_1F_2')}{4F_1^2F_2^2}+\ft{(2N+rN')(-F_3F_2'+2F_2F_3')}{4rF_2^2F_3}-\ft{NF_3'(F_3F_2'+F_2F_3')}{4F_2^2F_3^2}+\ft{N(F_1'F_3'+2F_3F_1''+2F_1F_3'')}{4F_1F_2F_3}=0,
\nn\\
&&-\frac{3}{L^2}-\ft{2\kappa^2\sinh^2(\phi)}{L^2}+\ft{2NF_3+2rF_3N'+2F_2(\kappa^2\dot{\phi}^2-1)+\cot(\theta)\dot{F_2}}{2r^2F_2F_3}-\ft{r^2NF_2F_3'^2+F_3(\dot{F_2}^2+2r^2NF_2'F_3')}{4r^2F_2^2F_3^2}\nn\\
&&+\ft{F_3(\cot(\theta)\dot{F_3}+\ddot{F_3})-\dot{F_3}}{2r62F_3^3}+\ft{\ddot{F_2}+r(rN'F_3'+2N(3F_3'+rF_3''))}{2r^2F_2F_3}+\ft{N(\kappa^2 r F_2\phi'^2-F_2')}{rF_2^2}=0,
\eea
where a prime denotes the derivative with respect to $r$ and a dot denotes the derivative with respect to $\theta$.

\subsection{The dipole solution: the ${\cal O}(\epsilon)^2$
term for the scalar field  }
\label{dipole-appendix}

The expression of the functions 
$\phi_{21}(r)$, $\phi_{23}(r)$
which enter the $n.l.o.$ expression 
(\ref{dipole-nlo})
of the scalar field reads
(with
$
 N_0=1+ r^2/L^2$,
$
{\cal X}=\arctan(r/L) 
$):
\bea
\label{g21}
\phi_{31}&=&\ft{1}{31500r^6L N_0}\times \bigg(-\kappa^2 r^2 \big(r(5700L^4+424i L^3\pi^4r N_0+5(3520+\gamma_1)L^2r^2+5c_1r^4)\nn\\
&&-5L(3420L^4+(2536+\gamma_1)L^2r^2+(\gamma_1-1934)r^4{\cal X}\big)+6L^2N_0\big(-120i\gamma_2 Lr^4 \text{Li}_4(x)\nn\\
&&+7\pi^2 r^4(4iL\pi^2-45r(\ln(4)-1))+60\gamma_2 r^4 \text{Li}_3(x)(r+3L{\cal X})+5((-570\kappa^2L^4r+28(-9+22\kappa^2)L^2r^3\nn\\
&&+6(-42+151\kappa^2+2i\gamma_2\pi)r^5+24\gamma_2r^4(r{\cal Y}+i L\text{Li}_2(x)-r\ln(2))){\cal X}^2+2(95\kappa^2L^5+6(7-6\kappa^2)L^3r^2\nn\\
&&+L(\kappa^2(81-106i\pi)+42i(i+\pi))r^4+4i\gamma_2r^5+4\gamma_2L r^4(-{\cal Y}+\ln(2))){\cal X}^3+9\gamma_2r^5\zeta(2)\nn\\
&&+4r^4{\cal X}(8i\gamma_2r\text{Li}_2(x)+7L(8+\pi^2(-3+\ln(63)))-3\gamma_2L\zeta(3)))\big)\bigg),
\nonumber
\eea
\bea
\label{g23}
\phi_{33}&=&\ft{L}{126000r^6}\times\bigg(\ft{8400\kappa^2L^2r^3}{N_0}-240{\cal X}^2\big(315\kappa^2L^4r+L^2r^3(34+3\kappa^2(71-60i\pi)+60i\pi)\nn\\
&&+(34+\kappa^2(-67-48i\pi)+16i\pi)r^5+8\gamma_3 r^3(15L^2+4r^2)({\cal Y}-\ln(2))\big)+80{\cal X}^3\big(\kappa^2(315L^5-507L^3r^2)\nn\\
&&-8\gamma_3\gamma_6r^2{\cal Y}+2r^2(60i\gamma_3 L^2 r+\gamma_3(\gamma_7+16i r)r^2+L^3(137+30\gamma_3(i\pi+\ln(4))))\big)
\nn\\
&&+15r^2{\cal X}\big(-15L^3(-88+208\kappa^2+\Gamma_2)+L(376+1392\kappa^2-9\Gamma_2)r^2+32\gamma_3(4i\Gamma_1\text{Li}_2(x)+4\gamma_6 \text{Li}_3(x)\nn\\
&&-3\gamma_6\zeta(2))\big)
+3r^2\big(6i\gamma_3\gamma_6\pi^4+25L^2(1184\kappa^2+3(-88+\Gamma_2))r+20\Gamma_2 r^3+80\gamma_3(4\Gamma_1 \text{Li}_3(x)\nn\\
&&-8i \gamma_6 \text{Li}_4(x)+3\Gamma_1\zeta(3))\big)\bigg),
\nonumber
\eea
where we define
\bea
&&\gamma_1=9\pi^2(151-212\ln(2)),\qquad
\gamma_2=21-53\kappa^2,\qquad
\gamma_3=3\kappa^2-1,\qquad
\gamma_4=(17+16\ln(2))\pi^2,\nn\\
&&\gamma_5=(67+96\ln(2))\pi^2,\qquad
\gamma_6=3(5L^2+3r^2)L,\qquad
\gamma_7=3L(6i\pi-31+\ln(4096)),
\nn
\\
&&\Gamma_1=15L^2r+4r^3+\gamma_6 {\cal X},\qquad
\Gamma_2=2\gamma_4-\gamma_5\kappa^2,\qquad
{\cal Y}=\ln(L+ir)-\ln(r),
\nonumber
\eea
where $Li_n(x)$ is the poly-logarithm function and $\zeta(x)$ 
is the Riemann zeta function\footnote{ 
Both $\phi_{31}$ and $\phi_{33}$,
 are  real functions, although $i$ appears in their expressions. }.

\subsection{The perturbative quadrupolar solution }
\label{quadrupole}

In principle, the computation presented in 
Section  \ref{pert-dipole}
can be repeated starting with any (axisymmetric) scalar 
$\ell-$mode.
Here we present some results for the
$\ell=2$, $i.e.$ a scalar quadrupole.

The general equations
(\ref{nex1})-(\ref{nex3})
are still valid; 
however, for $\ell=2$ 
the expression of the perturbed metric functions is more complicated, with
 \begin{eqnarray}
\nonumber
F_{i2}(r,\theta)=\kappa^2 \left(a_{i}(r)
+\mathcal{P}_2(\cos\theta)b_{i}(r) 
+\mathcal{P}_4(\cos\theta)c_{i}(r) \right) .
\label{nex31q} 
\end{eqnarray}
Then a straightforward but cumbersome computation
leads to the following expressions 
of the functions $a_{i},b_{i},c_{i}$:
\begin{eqnarray}
\nonumber
&&
a_1(r)=\ft{4}{5\pi^2}\bigg(-13\pi^2+\ft{12L^2}{r^2}(1+\ft{10r^2}{3L^2 N_0(r)})+\ft{40L^3}{r^3}{\cal X}(r)\big(-1+\ft{13 r^2}{5L^2}+\ft{2}{5N_0(r)}+\ft{3L}{10r}(1+\ft{13r^4}{3L^4}){\cal X}(r)\big)\bigg),
\\
\nonumber
&&
a_2(r)=-\ft{432L^4}{5\pi^2 r^4 N_0(r)}\bigg(1-\ft{L{\cal X}(r)}{r}(1+\ft{r^2}{3L^2})\bigg)\bigg(1+\ft{7r^2}{9L^2}-\ft{LN_0(r){\cal X}(r)}{r}(1+\ft{r^2}{9L^2})\bigg),
\\
\nonumber
&&
a_3(r)=0,
\\
&&
\nonumber
b_1(r)=\ft{64L^2}{21\pi^2r^2N_0(r)}\bigg(-1+34N_0(r)-\ft{42L}{r}N_0(r){\cal X}(r)+9{\cal X}^2(r)(\ft{L^2}{r^2}-\ft{r^2}{L^2})\bigg),
\\
&&
\nonumber
b_2(r)=-\ft{1728L^4}{7\pi^2r^4N_0(r)}\bigg(1+\ft{38r^2}{27L^2}+\ft{34r^4}{81L^4}-\ft{2L}{r}N_0(r){\cal X}(r)\big(1+\ft{16r^2}{27L^2}-\ft{L{\cal X}(r)}{2r}(1+\ft{7r^2}{9L^2})\big)\bigg),
\\
\nonumber
&&
b_3(r)=-\ft{176L^2}{7\pi^2r^2}\bigg(1-\ft{68r^2}{33L^2}+\ft{10L{\cal X}(r)}{11r}(1+\ft{11r^2}{5L^2})+(1-\ft{21L^2}{11r^2})N_0(r){\cal X}^2(r)\bigg),
\\
\nonumber
&&
c_1(r)=-\ft{768L^4}{5\pi^2 r^4 N_0(r)}\bigg(1+\ft{125r^2}{84L^2}+\ft{17r^4}{35L^4}-\ft{LN_0(r){\cal X}(r)}{2r}\big(1+N_0(r)+\ft{2r{\cal X}(r)}{7L}(1+\ft{5}{4}N_0(r))\big)\bigg),
\\
\nonumber
&&
c_2(r)=\ft{384L^4}{7\pi^2r^4N_0(r)}\bigg(1+\ft{71r^2}{30L^2}+\ft{34r^4}{25L^4}-\ft{4L}{5r}N_0(r){\cal X}(r)\big(-1+\ft{r^2}{4L^2}+\ft{9L{\cal X}(r)}{4r}(1+\ft{7r^2}{6L^2}+\ft{7r^4}{18L^4})\big)\bigg),
\\
\nonumber
&&
c_3(r)=\ft{96L^4}{\pi^2r^4}\bigg(1+\ft{r^2}{35L^2}-\ft{136r^4}{1575L^4}+\ft{2L{\cal X}(r)}{5r}\big(-1+\ft{2r^2}{21L^2}+\ft{r^4}{7L^4}+\ft{r^3N_0(r){\cal X}(r)}{14L^3}
(1-\ft{21L^4}{r^4})\big)\bigg),
\end{eqnarray}
which are found subject to the assumption of regularity at $r=0$
and AdS asymptotics.

To find the $\epsilon^3$-corrections to 
 the scalar field, we  consider 
an expansion similar to (\ref{nex3}), 
with
 \begin{eqnarray}
\label{quadrupole-nlo}
\phi^{(3)}(r,\theta)= 
\phi_{32}(r)   \mathcal{P}_{2}(\cos\theta) 
+\phi_{34}(r)  \mathcal{P}_{4}(\cos\theta)
+\phi_{35}(r)  \mathcal{P}_{6}(\cos\theta).
\end{eqnarray} 
Although an exact solution for 
$\phi_{32}(r)$,  
$\phi_{34}(r)$,
$\phi_{36}(r)$
can be found,
its expression is too complicated to include here. 
As with the dipole case,
we impose these functions to be regular  at $r=0$ 
and to decay as $1/r^2$ in $r\to\infty$. 
Then
the expansion parameter $\epsilon$
can be identified with the function 
$\alpha$
(evaluated at $\theta=0$) 
in eq.~(\ref{inf}),
\begin{eqnarray}
\alpha= \epsilon L_2(\cos \theta), 
\end{eqnarray}
while the expression for 
$\beta$ is 
\begin{eqnarray}
\beta= \left(-\frac{8L^2 }{\pi} \epsilon + \bar \beta_{32}  \epsilon^3  \right) {\cal P}_2(\cos \theta) 
  +
(\bar \beta_{34}    {\cal P}_4(\cos \theta) 
+ \bar \beta_{36}  {\cal P}_6(\cos \theta) )\epsilon^3,
\end{eqnarray}
where we denote
\bea
\nonumber
&&
\bar \beta_{32}=\frac{256 L^2}{245 \pi^3}
\left(132-
\pi^2(19-48 \ln(2))-216 \zeta(3)
-\ft{\kappa^2}{11}
\left(
\ft{340742}{15}-\pi^2(\ft{11973}{4}-5008 \ln(2))+22536 \zeta(3)
\right)
\right),
\\
\nonumber
&&
\bar \beta_{34}=-\frac{3477504 L^2}{148225 \pi^2}
\bigg(
-\ft{864\zeta(3)}{283}+\ft{1676482\kappa^2}{99333}\big(-1+3\pi^2(\ft{349357}{6705928}-\ft{90528\ln2}{838241})+\ft{1222128\zeta(3)}{838241}\big)\nn\\
&&\qquad\quad+1+\pi^2\big(\ft{192\ln 2}{283}-\ft{1579}{6792}\big)
\bigg),
\eea

\bea
&&\bar \beta_{36}=-\ft{192L^2}{1926925\pi^3}\bigg(-5\pi^2\big(6389+18432\ln2\big)+1296\big(303+320\zeta(3)\big)+\kappa^2\big(\pi^2(382237+153600\ln2)\nn\\
&&\qquad\quad-(\ft{59050384}{15}+691200\zeta(2)\big)\bigg).
\eea
To lowest order, the mass of the gravitating quadrupole, as computed within the same approach
as other solutions in this work,
is
\begin{eqnarray}
M=\frac{64 L \epsilon^2}{5}.
\end{eqnarray}

 \begin{small}
 
 \end{small}
\end{document}